\documentclass[useAMS,usenatbib]{mn2e}

\usepackage[T1]{fontenc}

\usepackage{enumerate}
\usepackage{amssymb}
\usepackage{graphicx}
\usepackage{natbib}
\usepackage{fixltx2e}
\usepackage{mathptmx}
\usepackage{pdflscape}
\usepackage{lscape}

\begin{document}
\title[Massive Cluster Galaxies in XMMUJ2235-2557]{Sizes, Colour gradients and Resolved Stellar Mass Distributions for the Massive Cluster Galaxies in XMMUJ2235-2557 at z = 1.39}

\author[J. C.C. Chan et al.]{Jeffrey C.C. Chan$^{1,2}$\textsuperscript{\thanks{E-mail: jchan@mpe.mpg.de}}, Alessandra Beifiori$^{2,1}$,  J. Trevor Mendel$^{1}$,  Roberto P. Saglia$^{1,2}$, \newauthor Ralf Bender$^{1,2}$, Matteo Fossati$^{2,1}$, Audrey Galametz$^{1}$, Michael Wegner$^{2}$, \newauthor David J. Wilman$^{2,1}$, Michele Cappellari$^{3}$, Roger L. Davies$^{3}$, Ryan C. W. Houghton$^{3}$, \newauthor Laura J. Prichard$^{3}$, Ian J. Lewis$^{3}$, Ray Sharples$^{4}$ and John P. Stott$^{3}$ 
\\
$^{1}$Max-Planck-Institut f\"ur extraterrestrische Physik (MPE), Giessenbachstr. 1, D-85748 Garching, Germany\\
$^{2}$Universit\"ats-Sternwarte, Ludwig-Maximilians-Universit\"at, Scheinerstrasse 1, D-81679 M\"unchen, Germany\\
$^{3}$Sub-department of Astrophysics, Department of Physics, University of Oxford, Denys Wilkinson Building, Keble Road, Oxford OX1 3RH, UK\\
$^{4}$Centre for Advanced Instrumentation, Department of Physics, Durham University, South Road, Durham DH1 3LE, UK\\}

\date{Accepted 2100 December 3. Received 2050 December 3; in original form 2014 November 13}
\pubyear{2015}

\label{firstpage}
\pagerange{\pageref{firstpage}--\pageref{lastpage}}
\maketitle

\begin{abstract}
We analyse the sizes, colour gradients, and resolved stellar mass distributions for 36 massive and passive galaxies in the cluster XMMUJ2235-2557 at $z=1.39$ using optical and near-infrared Hubble Space Telescope imaging. We derive light-weighted S\'ersic fits in five HST bands ($i_{775},z_{850},Y_{105},J_{125},H_{160}$), and find that the size decreases by $\sim20\%$ going from $i_{775}$ to $H_{160}$ band, consistent with recent studies.  We then generate spatially resolved stellar mass maps using an empirical relationship between $M_{*}/L_{H_{160}}$ and $(z_{850}-H_{160})$ and use these to derive mass-weighted S\'ersic fits: the mass-weighted sizes are $\sim41\%$ smaller than their rest-frame $r$-band counterparts compared with an average of $\sim12\%$ at $z\sim0$.  We attribute this evolution to the evolution in the $M_{*}/L_{H_{160}}$ and colour gradient.  Indeed, as expected, the ratio of mass-weighted to light-weighted size is correlated with the $M_{*}/L$ gradient, but is also mildly correlated with the mass surface density and mass-weighted size.  The colour gradients $(\nabla_{z-H})$ are mostly negative, with a median value of $\sim0.45$ mag dex$^{-1}$, twice the local value.  The evolution is caused by an evolution in age gradients along the semi-major axis ($a$), with $\nabla_{age} = d \log(age) / d \log(a)$ $\sim-0.33$, while the survival of weaker colour gradients in old, local galaxies implies that metallicity gradients are also required, with $\nabla_{Z} = d \log(Z) / d \log(a)$ $\sim-0.2$. This is consistent with recent observational evidence for the inside-out growth of passive galaxies at high redshift, and favours a gradual mass growth mechanism, such as minor mergers.
\end{abstract}

\begin{keywords}
galaxies: clusters: general  -- galaxies: elliptical, lenticular, cD -- galaxy: evolution -- galaxies: formation -- galaxies: high-redshift -- galaxies: fundamental parameters.
\end{keywords}


\section{Introduction}
The study of galaxy clusters at high redshift has attracted a lot of attention over the last decade, as these large structures provide a unique environment for understanding the formation and evolution of massive galaxies we see in the present day Universe.  Massive galaxies in clusters especially in the cluster cores are preferentially in the red passive population, have regular early-type morphology and are mainly composed of old stars \citep[e.g.][]{Dressler1980, Rosatietal2009, Meietal2009}. Nevertheless, in higher redshift clusters ($z \gtrsim1.5$) a substantial massive population are recently found to be still actively forming stars \citep[e.g.][]{Hayashietal2011, Gobatetal2013, Strazzulloetal2013, Baylissetal2014}.  The member passive galaxies reside on a well-defined sequence in colour-magnitude space, namely the red sequence which is seen in clusters up to redshift $z\sim2$ \citep[e.g.][]{KodamaArimoto1997, Stanfordetal1998, Gobatetal2011, Tanakaetal2013, Andreonetal2014}.  Previous works have shown that the stars in these galaxies have formed (and the star formation was quenched) early, the stellar mass is largely assembled before $z \sim1$ \citep[e.g.][]{Lidmanetal2008, Manconeetal2010, Strazzulloetal2010, Fassbenderetal2014}. These galaxies then evolve passively in the subsequent time \citep[e.g.][]{Andreonetal2008, DeProprisetal2013}.  Nonetheless, the details of how these massive passive cluster galaxies formed and evolved, in particular the physical processes involved, remains a matter of debate.

An important component of the above question is the evolution of the structure of these passive galaxies over time.  It has now been established that galaxies at high redshift are much more compact: those with stellar masses $M_{*} \geq 10^{11} M_{\odot}$ at $z \sim2$ have an effective radius of only $\simeq1$ kpc \citep[e.g.][]{Daddietal2005, Trujilloetal2006a}.  At $z\sim0$ such massive dense objects are believed to be relatively rare \citep{Trujilloetal2009}, yet the exact abundance is still under debate \citep{Valentinuzzietal2010a, Trujilloetal2012, Poggiantietal2013}.  Previous studies suggest that massive passive galaxies have grown by a factor of $\sim2$ in size since $z \sim1$ \citep[e.g.][]{Trujilloetal2006b, Longhettietal2007, Cimattietal2008, vanderWeletal2008, Sagliaetal2010, Beifiorietal2014}, and a factor of $\sim4$ since $z \sim2$ \citep[e.g.][]{Trujilloetal2007, Buitragoetal2008, vanDokkumetal2008, Newmanetal2012, Szomoruetal2012,  Barroetal2013, Vanderweletal2014}.  This progressive growth appears to happen mainly at the outer envelopes, as several works have shown that massive ($M_{*} \ga 1 \times 10^{11} M_{\odot}$) passive galaxies at high-redshift have comparable central densities to local ellipticals, suggesting the mass assemble took place mainly at outer radii over cosmic time \citep[i.e. the ``inside-out'' growth scenario,][]{Bezansonetal2009, vanDokkumetal2010, Pateletal2013}.

To explain the observed evolution, the physical processes invoked have to result in a large growth in size but not in stellar mass, nor drastic increase in the star formation rate.  Most plausible candidates are mass-loss driven adiabatic expansion (``puffing-up'') \citep[e.g.][]{Fanetal2008, Fanetal2010, RagoneFigueroaetal2011} and dry mergers scenarios \citep[e.g.][]{Bezansonetal2009, Naabetal2009, Trujilloetal2011}.  In the former scenario, galaxies experience a mass loss from wind driven by active galactic nuclei (AGN) or supernovae feedback, which lead to an expansion in size due to a change in the gravitational potential. In the latter, mergers either major involving merging with another galaxy of comparable mass, or minor that involves accretion of low mass companions, have to be dry to keep the low star formation rate \citep{Trujilloetal2011}.  Nevertheless, major mergers are not compatible with the observed growth in mass function in clusters as well as the observed major merger rates since $z \sim 1$ \citep[e.g.][]{Nipotietal2003, Bundyetal2009}.  On the other hand, minor mergers are able to produce an efficient size growth \citep[see e.g.][]{Trujilloetal2011, Shankaretal2013}. The rates of minor mergers are roughly enough to account for the size evolution only up to $z\lesssim1$ \citet{Newmanetal2012}, at z $\sim$ 2 additional mechanisms are required \citep[e.g. AGN feedback-driven star formation][]{Ishibashietal2013}.  In addition, the effect of continual quenched galaxies onto the red sequence as well as morphological mixing (known as the ``progenitor bias'') further complicates the situation \citep[e.g][]{vanDokkumetal2001}.  Processes that are specific in clusters such as harassment, strangulation and ram-pressure stripping \citep[e.g.][]{Treuetal2003, Moranetal2007} might play an important role in quenching and morphologically transforming galaxies.  Several studies have already shown that the progenitor bias has a non-negligible effect on the size evolution \citep[e.g.][]{Sagliaetal2010, Valentinuzzietal2010b, Carolloetal2013, Poggiantietal2013, Beifiorietal2014, Delayeetal2014, Bellietal2015, Shankaretal2015}.

In addition to size or structural parameter measurements, colour gradients also provide valuable information for disentangling the underlying physical processes involved in the evolution of passive galaxies, and have been used as tracers of stellar population properties and their radial variation.  In local and intermediate redshift passive galaxies, colour gradients are mainly attributed to metallicity gradients \citep[e.g.][]{Sagliaetal2000, LaBarberaetal2005, Tortoraetal2010}, although also affected by age and dust \citep[see, e.g.][]{Vulcanietal2014}.  Measuring the colour gradients at high redshift is more challenging due to compact galaxy sizes and limitations on instrumental angular resolution.  Passive galaxies at high redshift appear to show negative colour gradients, in the sense that the core is redder than the outskirts \citep[e.g.][]{Wuytsetal2010, Guoetal2011, Szomoruetal2011}, implying a radial variation in the stellar mass-to-light ratio (hereafter $M_{*}/L$).

Due to $M_{*}/L$ gradients within the galaxies, the size of the galaxies measured from surface brightness profiles (i.e. luminosity-weighted sizes) is not always a reliable proxy of the mass distribution, especially at high redshifts when the growth of the passive galaxies is more rapid.  Hence, measuring characteristic sizes of the mass distribution (i.e. mass-weighted sizes) is preferable over the wavelength dependent luminosity-weighted sizes.  Recently a number of works attempted to reconstruct stellar mass profiles taking into account the $M_{*}/L$ gradients primarily using two techniques: resolved spectral energy distribution (SED) fitting \citep[e.g.][]{Wuytsetal2012, Langetal2014} and the use of a scaling $M_{*}/L$ - colour relation \citep[e.g.][]{BelldeJong2001, Belletal2003}.  In the former, stellar population modeling is performed on resolved multi-band photometry to infer spatial variations in the stellar population in 2D.  While this is a powerful way to derive resolved properties, deep and high-resolution multi-band imaging are required to well constrain the SED in each resolved region, which is not available for most datasets.  The latter method, demonstrated by \citet{Zibettietal2009} and \citet{Szomoruetal2013}, relies on a $M_{*}/L$ - colour relation to determine the spatial variation of $M_{*}/L$.  Although this method cannot disentangle the degeneracy between age, dust and metallicity, it provides a relatively inexpensive way to study the mass distribution of galaxies.

In this study, we analyse a sample of 36 passive galaxies in the massive cluster XMMUJ2235-2557 at $z\sim1.39$.  We focus on their light-weighted sizes (in rest-frame optical, from the near-IR \textit{HST}/WFC3 images), resolved stellar mass distribution, as well as mass-weighted sizes and colour gradients.  This paper is organised as follows.  The sample and data used in this study are described in Section~\ref{sec:Data}.  Object selection, photometry, structural analysis, and the procedure to derive resolved stellar mass surface density maps are described in Section~\ref{sec:Analysis}. We also examine the reliability of our derived parameters with simulated galaxies and present the results in the same section.  In Section~\ref{Local Comparison Sample} we describe the local sample we used for comparison.  In Section~\ref{sec:Results} we present the main results,  including both light-weighted and mass-weighted structural parameters derived from the stellar mass surface density maps, colour and $M_{*}/L$ gradients. The results are then compared with the local sample, and discussed in Section~\ref{sec:Discussion}.  Lastly, in Section~\ref{sec:Conclusion} we draw our conclusions.

Throughout the paper, we assume the standard flat cosmology with $H_{0} = 70$~km~s$^{-1}$~Mpc$^{-1}$, $\Omega_{\Lambda} = 0.7 $ and $\Omega_{m} = 0.3$.  With this cosmological model at redshift 1.39, 1 arcsec corresponds to 8.4347 kpc.  Magnitudes quoted are in the AB system \citep{OkeGunn1983}.  The stellar masses in this paper are computed with a \citet{Chabrier2003} initial mass function (IMF).  Quoted published values are transformed to Chabrier IMF when necessary.

\begin{table}
  \caption{HST imaging of XMMUJ2235-2557 used in this study.}
  \label{tab_data}
  \begin{tabular}{cccc}
  \hline
  \hline
  Name & Filter &  Rest pivot wavelength  & Exposure time \\
              &           &  at $z = 1.39$ (\AA) & (s) \\
  \hline
  $i_{775}$ & ACS F775W &  3215.2  &  8150  \\
  $z_{850}$ & ACS F850LP  & 3776.1  &  14400  \\   
  $Y_{105}$ & WFC3 F105W  & 4409.5  &  1212  \\   
  $J_{125}$ & WFC3  F125W & 5217.7  &  1212   \\   
  $H_{160}$ & WFC3  F160W & 6422.5  &  1212  \\   
  \hline
\end{tabular}
\end{table}


\section{Data}
\label{sec:Data}
\subsection{Sample}
The cluster XMMUJ2235-2257 was serendipitously detected in an X-ray observation of a nearby galaxy by XMM-Newton and discovered by \citet{Mullisetal2005}.  Subsequent VLT/FORS2 spectroscopy confirmed the redshift of the cluster to be $z \sim 1.39$.  \citet{Rosatietal2009} confirmed the cluster membership of 34 galaxies.  Among them 16 within the central 1 Mpc are passive.  \citet{Jeeetal2009} performed a weak-lensing analysis on the cluster and estimated the projected mass of the cluster to be $\sim 8.5 \times 10^{14}$ $M_{\odot}$, making it one of the most massive clusters seen at high-redshift.  \citet{Grutzbauchetal2012} studied the star formation in this cluster out to a projected radius of 1.5 Mpc and found that all massive galaxies have low specific star formation rates, and galaxies in the cluster centre have lower specific star formation rates than the rest of the cluster galaxies at fixed stellar mass.  For the galaxy structural properties, this cluster has been investigated by \citet{Strazzulloetal2010} and was also included in the cluster sample of \citet{Delayeetal2014} and \citet{DeProprisetal2015}.  

\subsection{\textit{HST} imaging}
We make use of the deep optical and IR archival imaging of the cluster XMMUJ2235-2557, obtained with \textit{HST}/ACS WFC and \textit{HST}/WFC3 in June 2005 (PID 10698), July 2006 (PID 10496) and April 2010 (PID 12051).  The ACS data are mostly from a program designed to search for Type Ia supernovae in galaxy clusters \citep{Dawsonetal2009}, while the WFC3 data are from a calibration program aiming at cross-calibrating the zero point of WFC3 and NICMOS.  The \textit{HST}/ACS data consists of F775W and F850LP bands (hereafter $i_{775}$ and $z_{850}$), while the WFC3 data comprises four IR bands, F105W, F110W, F125W and F160W (hereafter $Y_{105}$, $YJ_{110}$, $J_{125}$ and $H_{160}$).  The $YJ_{110}$ data is not used in this study as it has a shorter exposure time.  The WFC3 data has a smaller field of view than the ACS data, $145'' \times 126''$.  A summary of the observational setup can be found in Table \ref{tab_data}.

Data in each band are reduced and combined using Astrodrizzle, an upgraded version of the Multidrizzle package in the $\tt{PyRAF}$ interface \citep{Gonzagaetal2012}.  Relative WCS offsets between individual frames are first corrected using the \texttt{tweakreg} task before drizzling.  The ACS and WFC3 images have been drizzled to pixel scales of 0.05 and 0.09 arcsec pixel$^{-1}$ respectively.  The full-width-half-maximum (FWHM) of the PSF is $\sim$0.11 arcsec for the ACS data and $\sim$0.18 arcsec for the WFC3 data, measured from median stacked stars.  We produce weight maps using both inverse variance map (\texttt{IVM}) and error map (\texttt{ERR}) settings for different purposes.  The \texttt{IVM} weight maps, which contain all background noise sources except Poisson noise of the objects, are used for object detection, while the \texttt{ERR} weight maps are used for structural analysis as the Poisson noise of the objects is included.  Due to the nature of the drizzle process, the resulting drizzled images have correlated pixel-to-pixel noise. To correct for this we follow \citet{Casertanoetal2000} to apply a scaling factor to the weight maps.  Absolute WCS calibrations of the drizzled images are derived using GAIA (Graphical Astronomy and Image analysis Tool) in the Starlink library \citep{Berryetal2013} with Guide Star Catalog II (GSC-II) \citep{Laskeretal2008}.


\section{Analysis}
\label{sec:Analysis}
\subsection{Object detection, sample selection, and photometry}
\label{sec:Object Detection, Sample Selection and Photometry}
\subsubsection{Method}
The WFC3 $H_{160}$ image, the reddest available band, is used for object detection with SExtractor \citep{BertinArnouts1996}.  The multi-band photometry is obtained with SExtractor in dual image mode with the $H_{160}$ image as the detection image.  {\tt MAG\_AUTO} magnitudes are used for galaxy magnitudes and aperture magnitudes are used for colour measurements.  We use a fixed circular aperture size of $1''$ in diameter.  The effective radii of most galaxies in the cluster are generally much smaller than the aperture size.
Galactic extinction is corrected using the dust map of \citet{Schlegeletal1998} and the recalibration E(B-V) value from \citet{Schlaflyetal2011}.

\begin{figure}
  \includegraphics[scale=0.55]{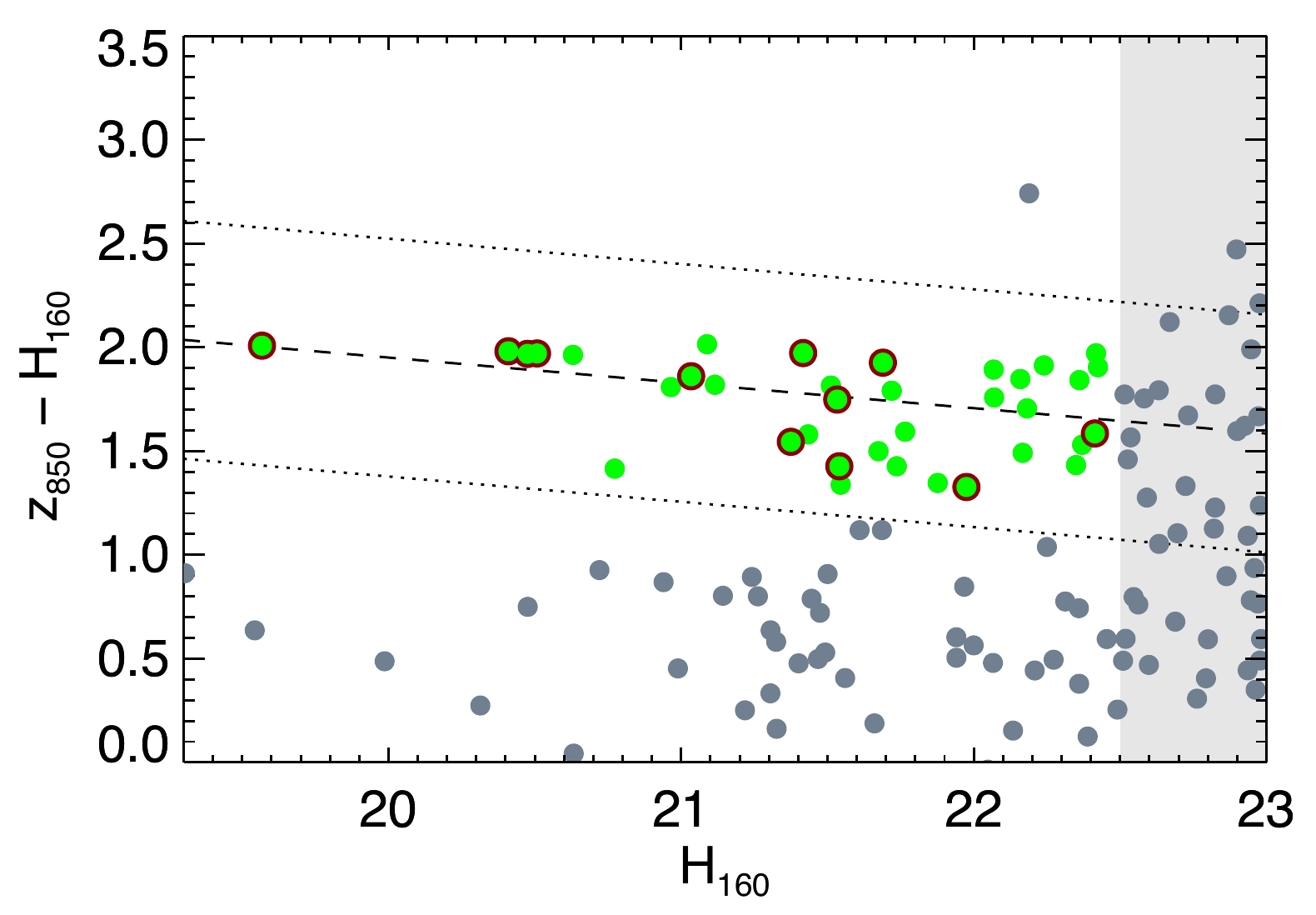}
  \caption{Colour-magnitude diagram of the cluster XMMUJ2235. WFC3 $H_{160}$ magnitudes are \texttt{MAG\_AUTO} magnitudes while the $z_{850} - H_{160}$ colour are $1''$ aperture magnitudes.  The dashed line corresponds to the fitted red sequence and the dotted lines are $\pm 2\sigma$.  Green circles correspond to objects that are included in our sample, which are within the dotted line and are not in the shaded area (i.e. $H_{160} < 22.5$). Objects that are spectroscopically confirmed cluster members from the catalogue of \citet{Grutzbauchetal2012} are circled in dark red.}
  \label{fig_colourmag}
\end{figure}

As described in the introduction, \citet{Grutzbauchetal2012} studied the star formation in the cluster out to a projected radius of 1.5 Mpc.  We cross-match our SExtractor catalogue to theirs to identify spectroscopically confirmed cluster members from previous literature \citep{Mullisetal2005, Lidmanetal2008,Rosatietal2009}.  12 out of 14 spectroscopically confirmed cluster members are within the WFC3 FOV and identified.   Figure \ref{fig_colourmag} shows the colour-magnitude diagram of the detected sources within the WFC3 FOV.  We identify passive galaxies through fitting the red sequence from the colour-magnitude diagram.  We measure the scatter through rectifying the $z_{850} - H_{160}$ colour with our fitted relation, then marginalise over the $H_{160}$ magnitude to obtain a number distribution of the galaxies.  The dotted lines correspond to $\pm 2\sigma$ derived from a Gaussian fit to the number distribution.

Objects that are within 2$\sigma$ from the fitted red-sequence are selected as the passive sample.  We trim the sample by removing point sources indicated by SExtractor (i.e. those with \texttt{class\_star} $ \geq 0.9$) and applying a magnitude cut of $H_{160} < 22.5$, which corresponds to a completeness of $\sim$95\% (see below).  This selection results in a sample of 36 objects in the cluster XMMUJ2235-2557.

\subsubsection{Quantifying the uncertainties on the photometry}
\label{sec:Quantifying the uncertainties on the photometry}
Since the photometric uncertainties are folded directly into our mass estimates as well as the structural parameters measurements, a realistic estimate of the photometric uncertainties is required.  Previous works have shown that SExtractor tends to underestimate the photometric uncertainties and there can be a small systematic shift between {\tt MAG\_AUTO} output and the true magnitudes \citep{Haussleretal2007}.  Hence, we perform an extensive galaxy magnitude and colour test with a set of 50000 simulated galaxies with surface brightness profiles described by a S\'ersic profile on the ACS $z_{850}$ and WFC3 $H_{160}$ band images.  This set of galaxies is also used for assessing the completeness and accuracy of the light and mass structural parameter measurements.  Details of the simulations can be found in Appendix~\ref{Details of the simulations}.  Here we focus on the photometric uncertainties estimates.

The detection rate above a certain magnitude reflects the completeness of the sample at that particular magnitude cut.  We find that a magnitude cut of $H_{160} < 22.5$ corresponds to a completeness of $\sim$95\%.  We then assess the accuracy of the recovered magnitudes and colours.  Since the accuracies depend strongly on both input magnitude ($mag_{in}$) as well as the effective semi-major axis ($a_e$) of the galaxies, we assess the accuracy in terms of input mean surface brightness ($\Sigma = mag_{in} + 2.5 \log(2\pi a_e^{2})$ in mag arcsec$^{-2}$) rather than input magnitudes.  Below we quote the results at a mean surface brightness of 23.5 mag arcsec$^{-2}$ in $H_{160}$ (or 24.5 mag arcsec$^{-2}$ in $z_{850}$) as a benchmark, as most objects we considered are brighter than 23.5 mag arcsec$^{-2}$.

For ACS $z_{850}$, the typical $1\sigma$ uncertainty for the \texttt{MAG\_AUTO} output at mean surface brightness of 24.5 mag arcsec$^{-2}$ is $\sim$0.33 mag.  For WFC3 $H_{160}$, at a mean surface brightness of 23.5 mag arcsec$^{-2}$ the typical $1\sigma$ uncertainty is $\sim$0.19 mag.  Previous studies have shown that SExtractor {\tt MAG\_AUTO} misses a certain amount of flux especially for the faint objects \citep[e.g.][]{BertinArnouts1996,Labbeetal2003,Tayloretal2009b}.  We find a systematic shift for both filters towards low surface brightness, the shifts are on average $\sim0.42$ mag for a $H_{160}$ mean surface brightness of 23.5 mag arcsec$^{-2}$ or $\sim0.50$ mag for a $z_{850}$ mean surface brightness of 24.5 mag arcsec$^{-2}$.  

\begin{figure}
  \includegraphics[scale=0.385]{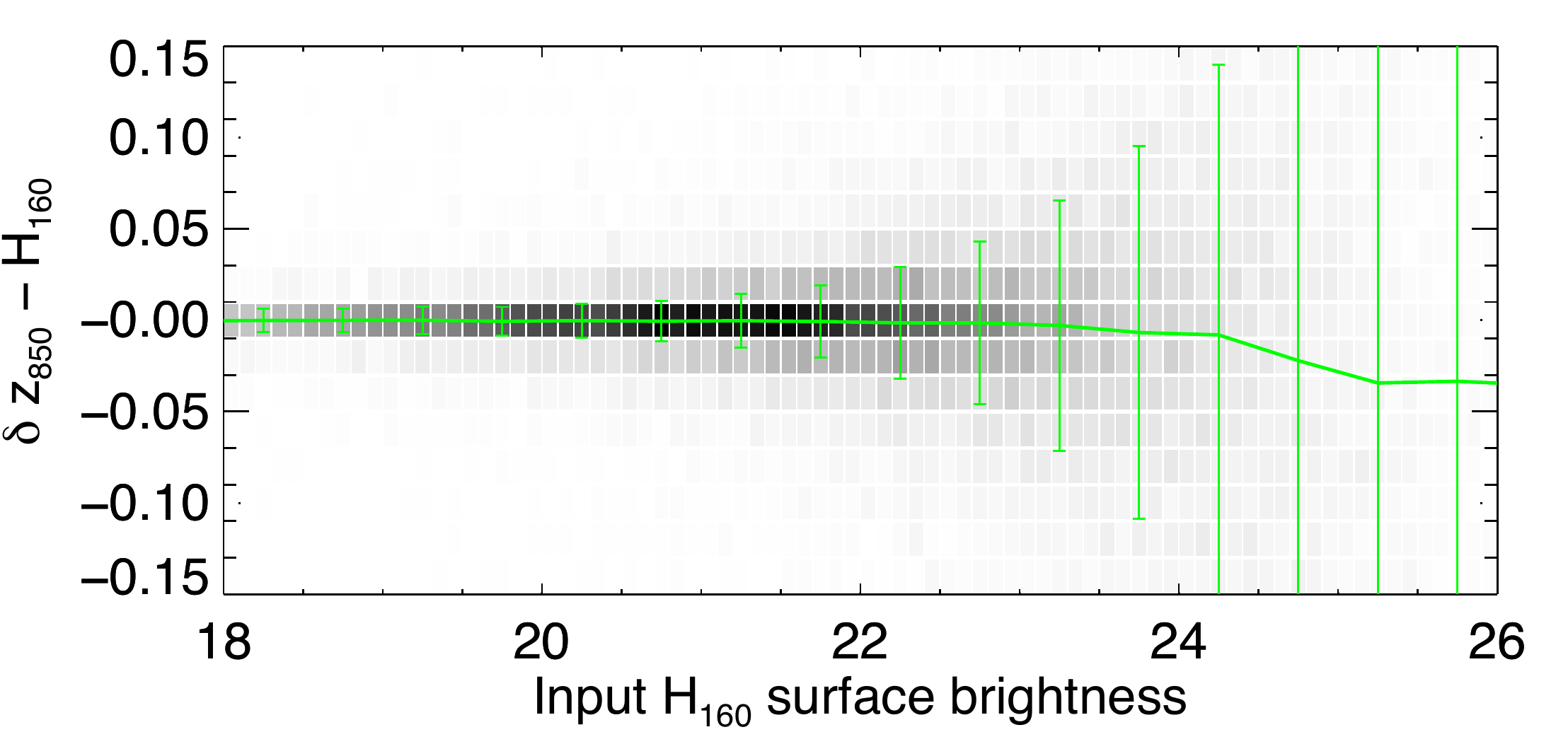}
  \caption{Differences between recovered and input aperture colour $\delta z_{850} - H_{160} = (z_{850} - H_{160})_{out} - (z_{850} - H_{160})_{in}$ as a function of input mean $H_{160}$ surface brightness. The green line indicates the median and $1\sigma$ dispersion in different bins (0.5 mag arcsec$^{-2}$ bin width), and the grey-shaded 2D histogram shows the number density distribution of the simulated galaxies.}
  \label{fig:photouncertainty_mod}
\end{figure}

On the other hand, we find no systematics between the input and recovered aperture colour.  Figure \ref{fig:photouncertainty_mod} shows the result for the $z_{850} - H_{160}$ colour from simulated galaxies.  The uncertainties on colour are small i.e. $\sim0.07$ mag for a $H_{160}$ mean surface brightness of 23.5 mag arcsec$^{-2}$. The uncertainty in colour tends to be larger for objects with redder $z_{850} - H_{160}$ colour, solely due to the fact that the $z_{850}$ aperture magnitude has a larger uncertainty for a redder colour.

\subsection{Light-weighted structural parameters}
\label{sec:Light-weighted structural parameters}
\subsubsection{Method}
\label{sec:Light-weighted Methods}
We measure the light-weighted structural parameters of the passive galaxies in five HST bands ($i_{775}$, $z_{850}$, $Y_{105}$, $J_{125}$ and $H_{160}$) using a modified version of GALAPAGOS \citep{Bardenetal2012}.  For each object detected by SExtractor, GALAPAGOS generates a postage stamp and measures the local sky level around the object using an elliptical annulus flux growth method.  This local sky level is then used by GALFIT \citep[v.3.0.5,][]{Pengetal2002}, in order to model the galaxy surface brightness profile.  We examine different settings of the sky estimation routine in GALAPAGOS to ensure the robustness of the results.  Since the ACS and WFC3 images have a different spatial resolution, we modify GALAPAGOS to allow the use of a single detection catalogue (in our case, the $H_{160}$ band) in all bands.  The code is further adjusted to use the RMS maps derived from the \texttt{ERR} weight maps output by Astrodrizzle.

As shown in \citet{Haussleretal2007}, contamination by neighbouring objects has to be accounted while fitting galaxy surface brightness profiles, especially in regions where the object density is high.  To deal with this issue, adjacent sources are identified from the SExtractor segmentation map and are masked out or fitted simultaneously if their light profiles have a non-negligible influence to the central object.
We fit a two-dimensional S\'ersic profile \citep{Sersicetal1963} to each galaxy, which can be written as

\begin{equation}
  I(a)= I_{e} ~\exp\left[-b_n\left ((\frac{a}{a_e})^{1/n} -1 \right ) \right]
\end{equation}
where the effective intensity $I_{e}$ can be described by
\begin{equation}
  I_{e} = \frac{L_{tot}}{2\pi nqa_e^2~b_n^{-2n}~\Gamma(2n)}
\end{equation}
where $\Gamma (2n)$ is the complete gamma function.

The S\'ersic profile of a galaxy can be characterised by five independent parameters: the total luminosity $L_{tot}$, the S\'ersic index $n$, the effective semi-major axis $a_e$, the axis ratio $q$ ($=b/a$, where $a$ and $b$ is the major and minor axis respectively) and the position angle $P.A.$. The parameter $b_n$ is a function of the S\'ersic index ($\Gamma(2n) = 2\gamma(2n, b_n)$, where $\gamma$ is the incomplete gamma function) and can only be solved numerically \citep{Ciotti1991}.  All five parameters as well as the centroid ($x,y$) are left to be free parameters in our fitting process with GALFIT. The constraints of each parameter for GALFIT are set to be: $0.2 < n < 8$, $0.3 < a_{e} < 500$ (pix), $0 < mag < 40$, $0.0001 < q < 1$, $-180^{\circ} < P.A. < 180^{\circ}$.  The sky level on the other hand, is fixed to the value determined by GALAPAGOS.

The S\'ersic model is convolved with the PSF constructed from stacking bright unsaturated stars in the images.  Note that we have also tried to derive a \texttt{TinyTim} PSF composite by adding PSF models using the \texttt{TinyTim} code \citep{Krist1995} into the raw data and drizzling them as science images.  Nevertheless, we notice that the \texttt{TinyTim} drizzled PSF does not match well the empirical PSF in the outer part: a much stronger outer envelope (as well as diffraction spikes) can be seen in the empirical PSF \citep[see also, Appendix A in][for a similar description]{Bruceetal2012}.  On the other hand, \citet{Vanderweletal2012} produced hybrid PSF models by replacing the central pixels of the median-stacked star by the \texttt{TinyTim} PSF.  We do not employ this correction as we find that the median-stacked star matches the TinyTim PSF reasonably well in the inner part.

\begin{figure}
  \includegraphics[scale=0.44]{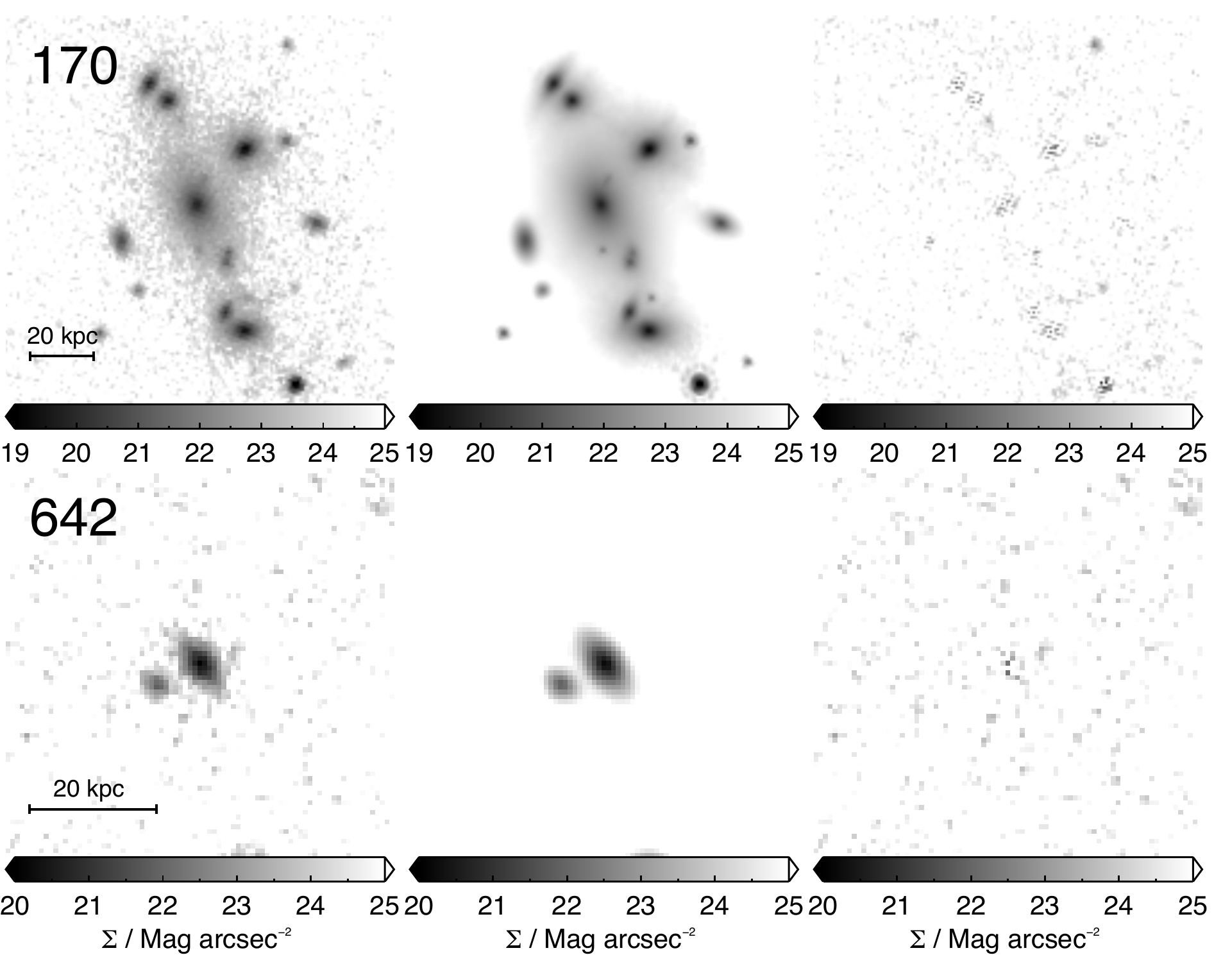}
  \caption{Examples of surface brightness profile fitting of two passive galaxies (ID 170, 642) in cluster XMMUJ2235-2557. From left to right: $H_{160}$ galaxy image cut-out centered on the primary object, GALFIT best-fit models and residuals.  The two examples are selected to demonstrate the clustering of sources.  Galaxy 170, the BCG of this cluster is located in the central region of the cluster with high object density.  Galaxy 642 is located in a more outer region of the cluster, yet is still affected by an extremely close neighbour.  Multiple objects are fitted simultaneously as described in Section~\ref{sec:Light-weighted Methods}.}
  \label{fig_lightfitex}
\end{figure}

The best-fitting light-weighted parameters are listed in Table~\ref{tab_para} in Appendix~\ref{parametertablesection}. Two galaxies (ID 170, 642) and their best-fits are shown in Figure~\ref{fig_lightfitex} for illustrative purposes.  These two objects have been chosen to show the impact of clustering of sources in dense regions.  Even in the cluster centre where there are multiple neighbouring objects,  GALFIT can do a good job in determining the structural parameters by fitting multiple object simultaneously.  Below we discuss the reliability and uncertainties in these light-weighted structural parameters.

\subsubsection{Reliability of the fitted structural parameters}
GALAPAGOS coupled with GALFIT performs well in most cases.  However in some exceptions, it is rather tricky to obtain a good-quality fit due to various issues.  We are not referring here to the global systematics and uncertainties (which are addressed in the next section), but on stability and quality control of individual fits.  We find that using an inadequate number of fitting components for the neighbouring sources (due to inadequate deblending in the SExtractor catalogue or appearance of extra structures / sources in bluer bands, e.g. $z_{850}$ band, compare to our $H_{160}$ detection catalogue) can lead to significant residuals that adversely affect the fit of the primary object.  Similarly, since GALAPAGOS fits sources with a single S\'ersic profile by default, GALFIT will likely give unphysical outputs for unresolved sources / stars in the field (with $a_e$ hitting the lower boundary of the constraint $a_e = 0.3$ pix, or S\'ersic index hitting the upper boundary $n = 8$) or even not converging in these cases, which again affects the result of the object of primary interest.  Moreover, the best-fit output can vary if we use a different treatment for neighbouring sources.  We notice that in a few cases the results can be very different depending upon whether neighbouring sources are masked or are fitted simultaneously.

To ensure high reliability, we perform the following checks for each galaxy: 1) We visually inspect the fits as well as the segmentation maps (output by SExtractor) in each band to ensure adjacent sources are well-fitted.  Extra S\'ersic components are added to poorly fitted neighbouring objects iteratively if necessary.
2) For neighbours for which GALFIT gives ill-constrained results (i.e. hitting the boundaries of the constraints), we replace the S\'ersic model with a PSF model and rerun the fit, which often improves the convergence and the quality of the best-fit model.  Regarding this, \citet{Bardenetal2012} explained the need of fitting S\'ersic profiles to saturated stars instead of PSF model in GALAPAGOS, since the PSF often lacks the dynamic range to capture the diffraction spikes of the bright saturated stars.  In our case this is not necessary since there are only a few bright saturated stars in the field, for which we can safely mask their diffraction spikes.
3) We compare the results of masking and simultaneously fitting neighbouring objects.  In most cases the two methods give results that are within 1$\sigma$.  For galaxies with close neighbours (e.g. within 5 $a_e$) we prefer to fit them simultaneously as any inadequate or over-masking can result in problematic fits, judging by examining the residual map output by GALFIT.  On the other hand, masking is more suitable when the neighbouring object are not axisymmetric or show certain substructures, which causes the single S\'ersic fit to not reach convergence.

\subsubsection{Quantifying the uncertainties in light-weighted structural parameters}
We quantify the systematic uncertainties using the set of 50000 simulated galaxies inserted on the images.  In this section we focus on the result of the test; details of the simulations can be found in Appendix~\ref{Details of the simulations-lightstruct}.  Note that the uncertainties quoted here are more likely to represent lower limits to the true uncertainties, as the simulated galaxies are also parametrised with a Se\'rsic profile.
 
\begin{figure}
  \includegraphics[scale=0.46]{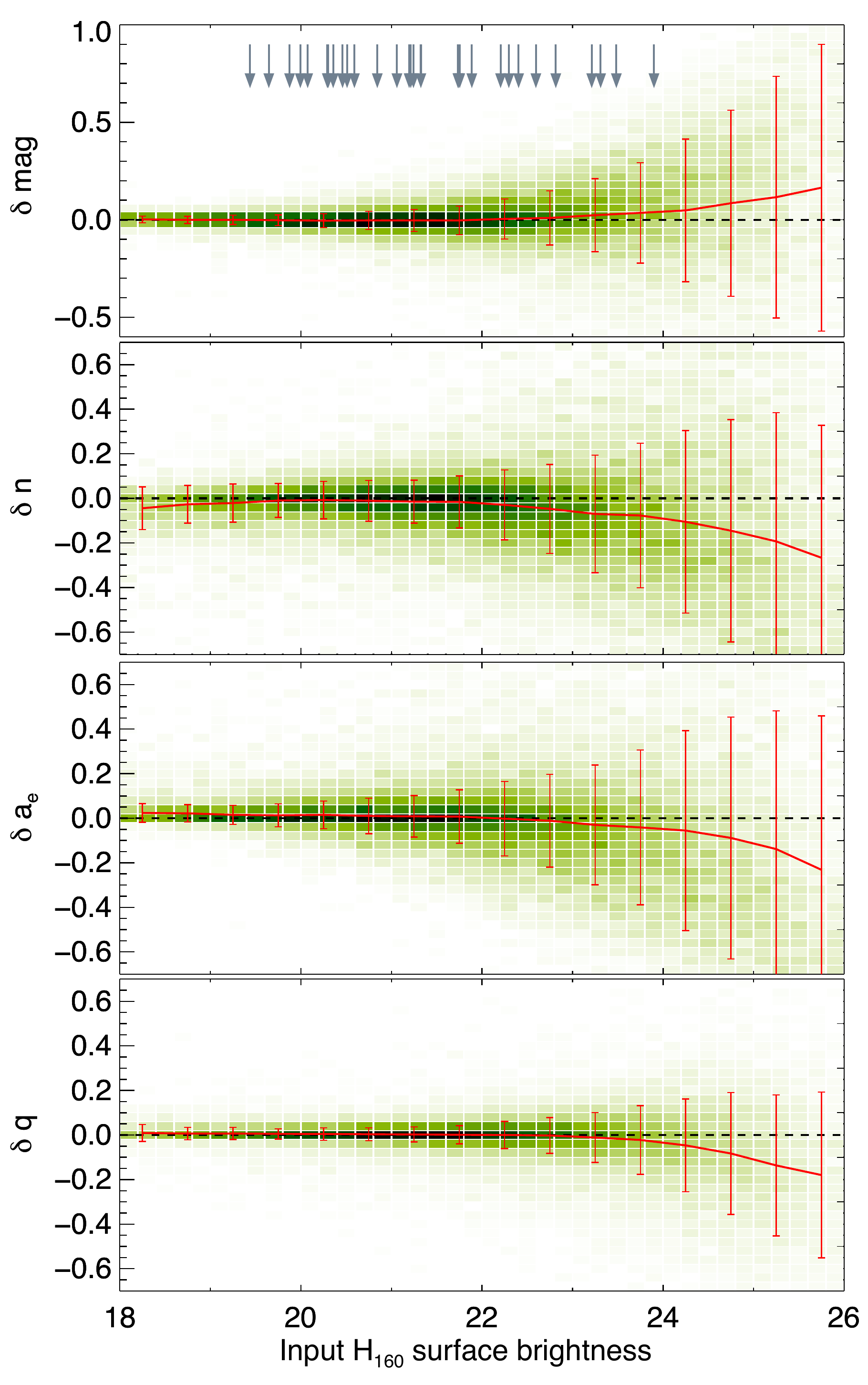}
  \caption{Differences between recovered and input structural parameters by GALFIT in function of input mean $H_{160}$ surface brightness. From top to bottom: magnitude $\delta mag = mag_{out} - mag_{in}$, S\'ersic indices $\delta n = (n_{out} - n_{in})/n_{in}$, effective semi-major axes $\delta a_e = (a_{e-out} - a_{e-in})/a_{e-in}$  and axis ratio $\delta q = (q_{out} - q_{in})/q_{in}$.  Red line indicates the median and $1\sigma$ dispersion in different bins (0.5 mag arcsec$^{-2}$ bin width) and green-shaded 2D histogram shows the number density distribution of the simulated galaxies. The grey arrows indicate the $H_{160}$ surface brightness of the galaxies in our cluster sample.}
  \label{fig:lightuncertainty}
\end{figure}

Figure~\ref{fig:lightuncertainty} shows the comparison between the input and recovered magnitudes and structural parameters for the $H_{160}$ band.   The magnitudes recovered by S\'ersic profile fitting are accurate with almost no systematics and a $1\sigma$ dispersion less than 0.25 for objects having mean $H_{160}$ surface brightness brighter than 23.5 mag arcsec$^{-2}$.  The S\'ersic index, effective radius and axis ratio measurements are generally robust for objects brighter than a mean $H_{160}$ surface brightness of 23.5 mag arcsec$^{-2}$.  The bias between the recovered and input S\'ersic indices is less than $8\%$ and the $1\sigma$ dispersion is lower than $30\%$.  Effective radii have a bias less than $4\%$ and a $1\sigma$ dispersion lower than $30\%$ for objects brighter than $H_{160}$ surface brightness of 23.5 mag arcsec$^{-2}$.  For objects with high mean surface brightness (i.e. $<19$ mag arcsec$^{-2}$) in our simulated sample, the effective radii are slightly overestimated ($\sim 2\%$) and the S\'ersic indices are underestimated ($\sim -4\%$) by GALFIT.  We find out that this bias is due to unresolved objects in our simulations.  A related discussion can be found in Appendix~\ref{Details of the simulations-lightstruct}.

We have also performed the same test on a simulated background similar to the actual images (where the main difference is that the simulated background has no issue of neighbour contamination), and find that the uncertainties on the effective radius are on average $\sim 15 - 20\%$ lower compared to those derived from real images.  

For each galaxy in the sample, we compute the mean $H_{160}$ surface brightness and add the corresponding dispersion in quadrature to the error output by GALFIT.

\subsection{Elliptical aperture photometry and color gradients}
\label{sec:Elliptical aperture photometry and color gradients}
In addition to structural parameters, we derive $z_{850}-H_{160}$ colour profiles for the passive sample with PSF-matched elliptical annular photometry.  We first convert the 2D image in both bands into 1D radial surface brightness profiles.  \citet{morishitaetal2015} demonstrated that deriving 1D profiles with elliptical apertures has certain advantages over circular apertures. Profiles derived with concentric circular apertures are biased to be more centrally concentrated.  We perform an elliptical annular photometry on the PSF-matched $z_{850}$ and $H_{160}$ images at the galaxy centroid derived from GALFIT.  The GALFIT best-fit axis ratios and position angles of individual galaxies (in $H_{160}$ band) are used to derive a set of elliptical apertures for each galaxy.  

Due to the proximity of objects in the cluster, it is necessary to take into account (as in 2D fitting) the effect of the neighbouring objects.  The neighbouring objects are first removed from the image by subtracting their best S\'ersic fit (or PSF fit in some cases) in both bands.  While the fit might not be perfect, we find that this extra step can remove the majority of the flux of the neighbouring objects contributing to surface brightness profiles. For some galaxies the colour profiles show substantial change after we apply the correction.

We then measure the colour gradients of individual galaxies by fitting the logarithmic slope of their $z_{850}-H_{160}$ colour profiles along the major axis, which are defined as follows:

\begin{equation}
    z_{850} - H_{160}  = \nabla_{z_{850} - H_{160}} \times \log(a) + Z.P.
\end{equation}
  
\begin{figure}
  \includegraphics[scale=0.45]{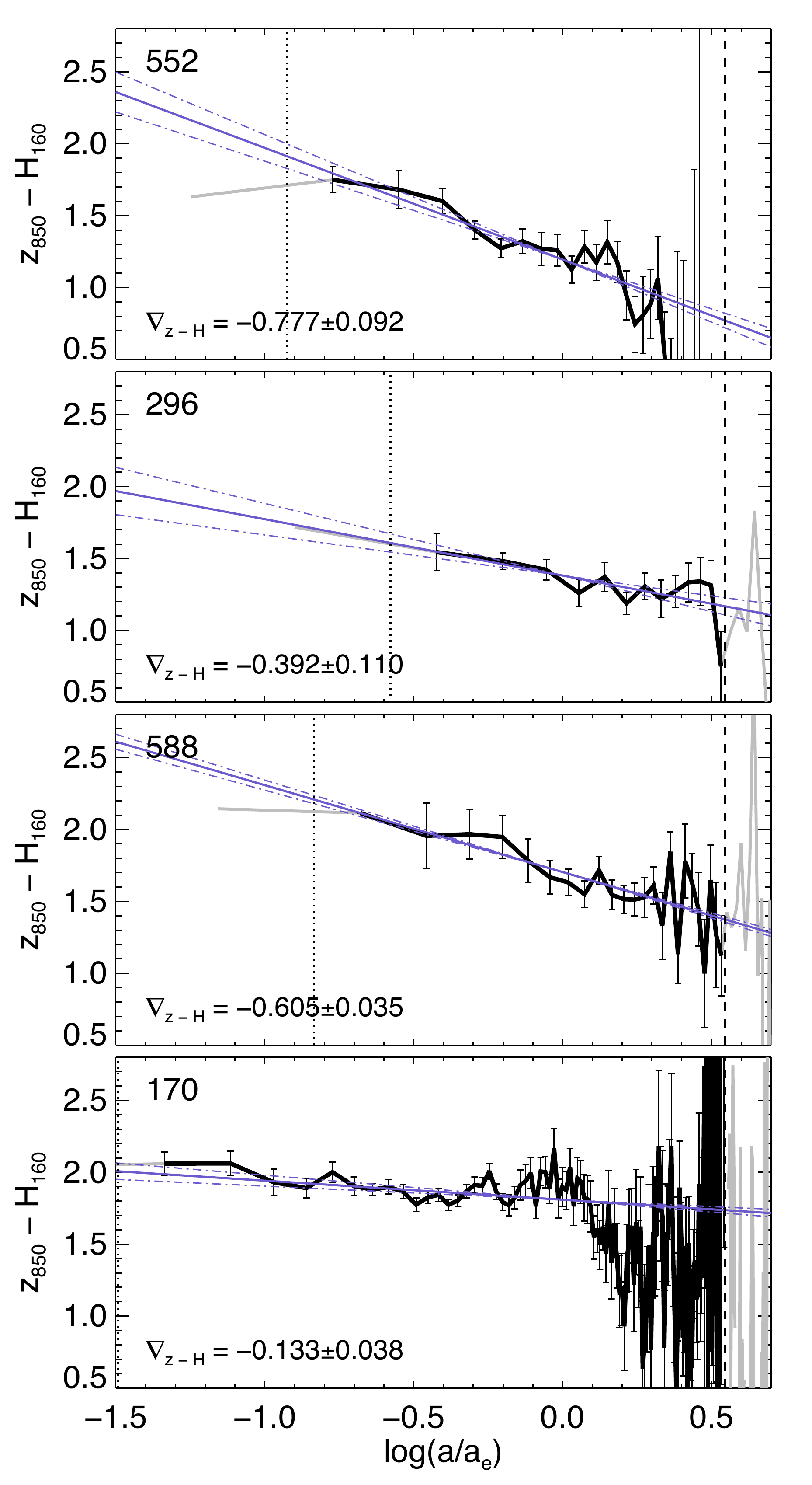}
  \caption{Examples of colour profile fitting of four passive galaxies in the cluster XMMUJ2235-2557.  From top to bottom:  colour profiles for galaxies ID 552, 296, 588 and 170 (with $\log(M_{*}/M_\odot) =$ 10.46, 10.54, 10.81, 11.81) along the logarithmic major axis ($\log(a/a_{e})$.  The grey line in each panel is the elliptical-averaged $z_{850} - H_{160}$ colour profile.  Regions that are fitted (PSF HWHM $< a < 3.5~a_{e}$) are over-plotted in black.  The vertical black dotted and dashed line show the minimum (PSF HWHM) and maximum radial distance for fitting ($3.5~a_{e}$).  The error bars show the error on the mean of the $z_{850} - H_{160}$ colour at each distance.  The blue solid line is the best logarithmic gradient fits, and the blue dotted-dashed lines are the $\pm 1\sigma$ error of the slope.}
  \label{fig_colourgradexample}
\end{figure}

\noindent At redshift 1.39 this corresponds roughly to the rest-frame $(U-R)$ colour gradient.  The depth and angular resolution of our WFC3 data allow us to derive a 1D colour profile accurately to $\sim3-4~a_{e}$, hence the colour gradient is fitted in the radial range of PSF half-width-half-maximum (HWHM) $< a < 3.5~a_{e}$.  We note that the colour gradients of most galaxies, as well as the median colour gradient, do not strongly depend on the adopted fitting radial range.  Figure~\ref{fig_colourgradexample} shows the colour profiles and logarithmic gradient fits of four passive galaxies as an example. The colour profiles are in general well-described by logarithmic fits.

\subsection {Stellar mass-to-light ratio -- colour relation}
\label{sec:Stellar Mass-to-light Ratio-Colour relation}
We estimate the stellar mass-to-light ratios of the cluster galaxies in XMMUJ2235-2557 using an empirical relation between the observed $z_{850} - H_{160}$ colour and stellar mass-to-light ratio ($M_{*}/L$).  At redshift 1.39, the $z_{850} - H_{160}$ colour (rest-frame $U - R$) straddles the $4000 \AA$ break.  Hence, this colour is sensitive to variations in the properties of the stellar population (i.e. stellar age, dust and metallicity).  In addition, the effects of these variations are relatively degenerate on the colour - $M_{*}/L$ plane \citep[almost parallel to the relation,][]{BelldeJong2001, Belletal2003, Szomoruetal2013}, which makes this colour a useful proxy for the $M_{*}/L$.

We derive the relation using the NEWFIRM medium band survey (NMBS) catalogue, which combines existing ground-based and space-based UV to mid-IR data, and new near-IR medium band NEWFIRM data in the AEGIS and COSMOS fields \citep{Whitakeretal2011}.  The entire catalogue comprises photometries in 37 (20) bands, high accuracy photometric redshifts derived with \texttt{EAZY} \citep{Brammeretal2008} and spectroscopic redshifts for a subset of the sample in COSMOS (AEGIS).  Stellar masses and dust reddening estimates are also included in the catalogue, and are estimated by SED fitting using \texttt{FAST} \citep{Krieketal2009}. 

To derive the $M_{*}/L$-colour relation, we use the stellar masses from the NMBS catalogue in COSMOS estimated with stellar population models of \citet{BruzualCharlot2003}, an exponentially declining SFHs, and computed with a \citet{Chabrier2003} IMF.  We do not use the sample in AEGIS as it contains photometries with fewer bands.  We derive the relation in the observer frame and compute the observed $z_{850} - H_{160}$ colour for all NMBS galaxies.  Note that we do not adopt the typical approach to interpolate the cluster data to obtain a rest-frame colour \citep[e.g. with InterRest,][]{Tayloretal2009a} due to limited availability of bands, which would likely lead to degeneracy in choices of templates.  Firstly, we rerun \texttt{EAZY} for all NMBS galaxies to obtain the best-fit SED template, these SEDs are then integrated with the ACS $z_{850}$ and WFC3 $H_{160}$ filter response for the $z_{850} - H_{160}$ colour.  Similarly we obtain the luminosity $L_{H_{160}}$of each galaxy in the observed $H_{160}$ band, from which we calculate the stellar mass-to-light ratio $M_{*}/L_{H_{160}}$.  We select NMBS galaxies within a redshift window of 0.1 of the cluster redshift, i.e. $1.29 < z < 1.49$, and apply the magnitude cut ($H_{160} < 22.5$) and a chi-square cut ($\chi^2 < 2.0$, from template fitting in \texttt{EAZY}) to better match the cluster sample.

\begin{figure}
  \includegraphics[scale=1.16]{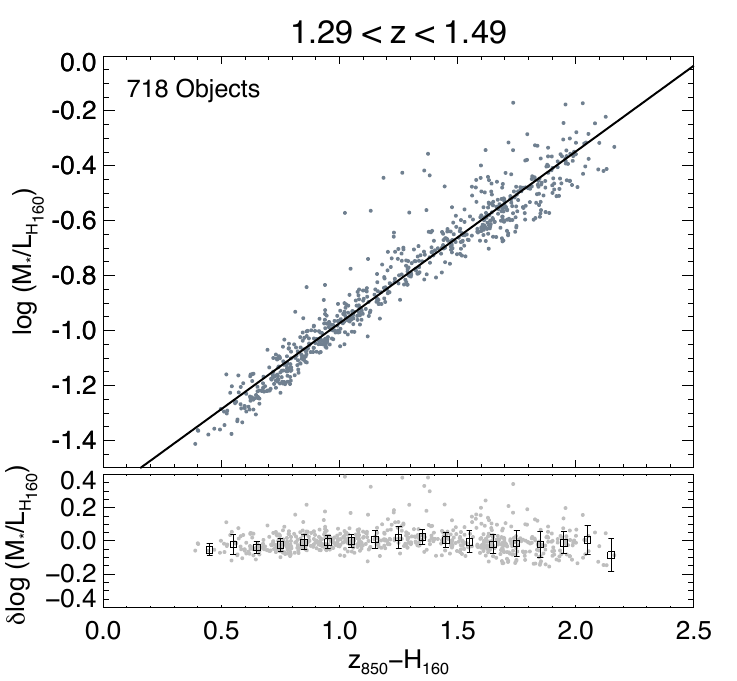}
  \caption{Relation between stellar mass-to-light ratio and z-H colour at redshift  $\sim$1.39 using the public NMBS catalogue.  Gray points are 718 galaxies from the NMBS catalogue that satisfy the selection criteria.  Black line is the best-fit linear relation.  Bottom panel shows the residuals of the relation $\delta log(M_{*}/L_{H_{160}}) = $ data - linear fit in colour bins of 0.1.}
  \label{fig:molcolourxmmu}
\end{figure}

A total of 718 objects are selected by this criterion.  A redshift correction is applied to these 718 galaxies to redshift their spectra to the cluster redshift (i.e. similar to k-correction in observer frame).  We then measure their $z_{850} - H_{160}$ colour and $L_{H_{160}}$ in the observer frame. This redshift correction is effective in reducing the scatter of the relation, indicating that some of (but not all) the scatter is simply due to difference in redshifts.  Figure~\ref{fig:molcolourxmmu} shows the fitted relation between $\log(M_{*}/L_{H_{160}}$) and $z_{850} - H_{160}$ colour.  The black line is the best-fit linear relation with:

\begin{equation}
   \log((M_{*}/L_{H_{160}}) / (M_\odot/L_\odot)) =  0.625\>(z_{850} - H_{160}) - 1.598 
\end{equation}

The relation is well-defined within a colour range of $0.4 < z_{850} - H_{160} <2.2$ (hence we choose the same range for our simulated galaxies, see Appendix~\ref{Details of the simulations}).  The global scatter of the fit is $\sim 0.06$ dex.  In the lower panel of Figure \ref{fig:molcolourxmmu} we plot the residuals of the fit in colour bins of 0.1.  The uncertainty in log($M_{*}/L$) is generally $<0.1$ in each bin and the bias is negligible.  The remaining scatter results from redshift uncertainties and stellar population variations (age, dust and metallicity), as their effects are not exactly parallel to the relation.  Note that this can lead to small systematics in measuring mass-to-light ratios and the mass-to-light ratio gradients. For example in metal-rich or old regions the mass-to-light ratio will likely be systematically slightly underestimated, and overestimated in metal-poor or young regions \citep{Szomoruetal2013}.

\subsection{Integrated stellar masses}
\label{sec:Integrated Stellar Masses}
We estimate the integrated stellar masses ($M_{*}$) of the cluster galaxies using our $M_{*}/L$-colour relation, the $z_{850} - H_{160}$ aperture colours and the total luminosity $L_{H_{160}}$ from best-fit S\'ersic models.  The uncertainties in stellar mass comprise photometric uncertainties in the colour and $H_{160}$ luminosity, as well as the scatter in the derived colour - $M_{*}/L$ relation.  The typical uncertainty of the masses is $\sim0.1$ dex, comparable to the uncertainties obtained from SED fitting.

Previous literature computed SED mass with multi-band \texttt{MAG\_AUTO} photometry obtained with SExtractor \citep[e.g. for this cluster,][]{Strazzulloetal2010,Delayeetal2014}.  Nevertheless, as we have shown in Section~\ref{sec:Quantifying the uncertainties on the photometry}, it is known that \texttt{MAG\_AUTO} can be systematically biased, due to the assumption in SExtractor that the sky background comprises only random noise without source confusion \citep{Brownetal2007}.  Hence, more recent studies use the total luminosity from best-fit S\'ersic models to correct the masses to account for the missing flux in \texttt{MAG\_AUTO} \citep{Bernardietal2013, Benzansonetal2013}.  In our case, we have demonstrated that total luminosity from the best-fit S\'ersic models can recover input galaxy magnitudes to a high accuracy.  Hence, we scale our masses with the total luminosity $L_{H_{160}}$ from best-fit S\'ersic models rather than $H_{160}$ \texttt{MAG\_AUTO} magnitudes.  We also compute masses with $H_{160}$ \texttt{MAG\_AUTO};  the difference between the two is small for our sample, with $\langle M_{*, MAG\_AUTO} - M_{*, Sersic} \rangle = -0.039$ dex.

For this particular cluster, \citet{Delayeetal2014} estimated the galaxies masses through SED fitting with four bands (HST/ACS $i_{775}$, $z_{850}$, HAWK-I $J$, $Ks$), which also gave an uncertainty of $\sim$0.1 dex in mass.  The masses derived with our method are consistent with the SED masses in \citet{Delayeetal2014} within the uncertainties.  A comparison of masses estimated using $M_{*}/L$-colour relation with masses computed using SED fitting can be found in Appendix~\ref{app:masscomparison}.  The uncertainty of the absolute stellar masses is of course larger (as in the case of SED fitting), depending on the details of NMBS SED fitting and e.g. choice of IMF.

\subsection{Resolved stellar mass surface density maps}
\label{sec:Resolved Stellar Mass Surface Density Maps}
We further exploit the $M_{*}/L$-colour relation to derive stellar mass surface density maps.  This allows us to study the mass distribution within each galaxy, at the same time eliminating the effect of internal colour gradient which influences the light-weighted size measurements.  Below we describe the main steps involved in deriving stellar mass surface density maps with the $M_{*}/L$-colour relation.

\subsubsection{PSF matching}
We first match the resolution of the ACS $z_{850}$ image ($\sim0.1''$) to the WFC3 $H_{160}$ image ($\sim0.18''$).  PSF matching is critical in this kind of study as the measured colour has to come from the same physical projected region.  We stack the unsaturated stars for each band to obtain characteristic PSFs, then generate a kernel that matches the $z_{850}$ to $H_{160}$ PSF using the \texttt{psfmatch} task in IRAF.  The difference between the resultant $z_{850}$ PSF and the $H_{160}$ PSF is less than 2.5\%.  Details of the PSF matching can be found in Appendix~\ref{app:psfmatching}.
We then apply the kernel to the ACS $z_{850}$ image. The PSF matched $z_{850}$ image is resampled to the same grid as the $H_{160}$ image using the software SWarp \citep{Bertinetal2002}.  We then generate postage stamps of each galaxy in both $H_{160}$ and PSF matched $z_{850}$ images for deriving resolved stellar mass surface density maps.

\subsubsection{From colour to stellar mass surface density}
\label{sec:From Colour to Stellar Mass Surface Density}
The next step is to convert the $z_{850} - H_{160}$ colour information into mass-to-light ratios with the $M_{*}/L$-colour relation described in Section~\ref{sec:Stellar Mass-to-light Ratio-Colour relation}.  Nevertheless, a direct pixel-to-pixel conversion is not possible for our data.  The conversion requires a certain minimum signal-to-noise (S/N) level because:  a) significant biases or massive uncertainties may arise if colours are not well measured. b) our relation is only calibrated within the colour range of $0.4 < z_{850} - H_{160} <2.2$. Any low S/N colour that falls outside the calibrated range could convert to an unphysical $M_{*}/L$.

Therefore, we adopt the Voronoi binning algorithm as described by \citet{Cappellarietal2003}, grouping pixels to a target S/N level of 10 per bin.  For each galaxy, we run the Voronoi binning algorithm on the sky-subtracted PSF-matched $z_{850}$ band postage stamps as a reference, as it has a lower S/N compared to the $H_{160}$ image.  The same binning scheme is then applied to the sky-subtracted $H_{160}$ image.  The subtracted sky levels are determined by GALAPAGOS.  The two images are then converted into magnitudes.  Binned $z_{850} - H_{160}$ colour maps are obtained by subtracting the two.  We then construct a binned $M_{*}/L$ map by converting the colour in each bin to a mass-to-light ratio with the derived colour - $M_{*}/L$ relation.

An extrapolation scheme is implemented to determine the $M_{*}/L$ in regions or bins with insufficient S/N, for example in the galaxy outskirts and the sky regions.  We first run an annular average to derive a 1-dimensional S/N profile in $z_{850}$ for individual galaxies using the light-weighted galaxy centroid, axis ratio and position angle determined in Section~\ref{sec:Light-weighted structural parameters}.  For the area outside the elliptical radius that has a S/N less than half of our target S/N (i.e. S/N $\sim 5$), we fix the $M_{*}/L$ to the annular median of $M_{*}/L$ bins at the last radius with sufficient S/N.  We find that this extrapolation is crucial for the following structural analysis as the sky noise is preserved (see the discussion in Appendix~\ref{Details of the simulations-massstruct}).

\begin{figure*}
  \includegraphics[scale=0.610]{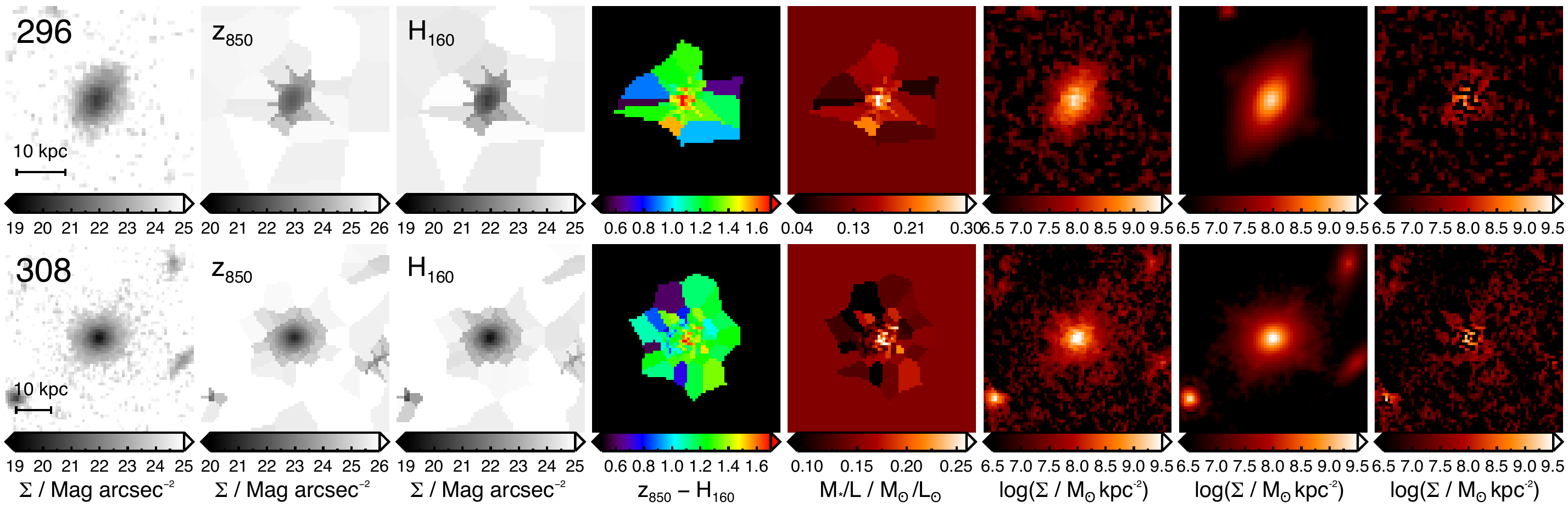}
  \caption{Examples of mass map derivation and fitting of two passive galaxies (ID 296, 308) in cluster XMMUJ2235-2557. From left to right: $H_{160}$ galaxy image cut-outs centred on the primary object, Voronoi-binned $z_{850}$ images, Voronoi-binned $H_{160}$ images, $z_{850} - H_{160}$ colour maps, $M_{*}/L$, the surface mass density maps $\Sigma_{mass}$, the GALFIT best-fit models and residuals in mass.  Bins that are extrapolated are masked out (shown in black) in the colour maps. The procedure is described in detail in Section~\ref{sec:From Colour to Stellar Mass Surface Density}.}
  \label{fig_massfitexample}
\end{figure*}

We construct resolved stellar mass surface density maps (hereafter referred to as mass maps) by directly combining the extrapolated $M_{*}/L$ map and the original (i.e. unbinned) $H_{160}$ images.  Figure~\ref{fig_massfitexample} illustrates the procedure of deriving mass maps from the $z_{850}$ and $H_{160}$ images.  Using the original $H_{160}$ image instead of the binned one allows us to preserve the WFC3 spatial resolution in the mass maps.  Note that in theory combining a binned (i.e. spatially discrete) $M_{*}/L$ map with a smooth luminosity image would result in a discrete mass profile in low S/N region, in order words, induce an ``discretization effect" in the mass maps.  This effect is more severe in low S/N regions, i.e. the galaxy outskirts where the bins are larger (hence less smooth).  For bright galaxies, since there are more bins with sufficient S/N and the dynamical range of the light distribution (surface brightness gradient) is much larger than the $M_{*}/L$ gradient,  this appears to have minimal effect and does not largely affect our result.  For fainter galaxies this issue is non-negligible.  To tackle this, for each galaxy we perform the above binning procedure 10 times, each with a slightly different set of initial Voronoi nodes.  This ends up with a set of $M_{*}/L$ maps which are then median-stacked to create the final mass map.  This extra step alleviates the discretization effect.

\subsection{Mass-weighted structural parameters}
\subsubsection{Method}
\label{sec:Mass-weighted Methods}
We measure mass-weighted structural parameters from the resolved stellar mass surface density maps.  We follow a similar procedure as with the light-weighted structural parameters, using GALFIT to model the mass profiles with two-dimensional S\'ersic profiles.  All five parameters of the S\'ersic profile ($M_{*, tot}$, $n$, $a_e$, $q$ and $P.A.$) and the centroid (x,y) are left to be free parameters in the fit.  We use the same GALFIT constraints as for the light-weighted structural parameters, except for allowing a larger range for the S\'ersic indices: $0.2 < n < 15.0$.  This is because the mass profiles are expected to be more centrally peaked compared to light profiles \citep{Szomoruetal2013}.  As the $H_{160}$ images are background subtracted before being converted into mass maps, the sky level (i.e. the mass level) is fixed to zero in the fitting process. The best-fitting mass-weighted parameters are given in Appendix~\ref{tab_para}.

\subsubsection{Quantifying the uncertainties in mass-weighted parameters}
We further assess the accuracy of our mass conversion procedures as well as the reliability of the mass-weighted structural parameter measurements.  The details of the test are described in Appendix \ref{Details of the simulations-massstruct}.  Similar to the uncertainty in the light-weighted parameters, the uncertainties quoted here are more likely to represent lower limits to the true uncertainties.

\begin{figure}
  \includegraphics[scale=0.46]{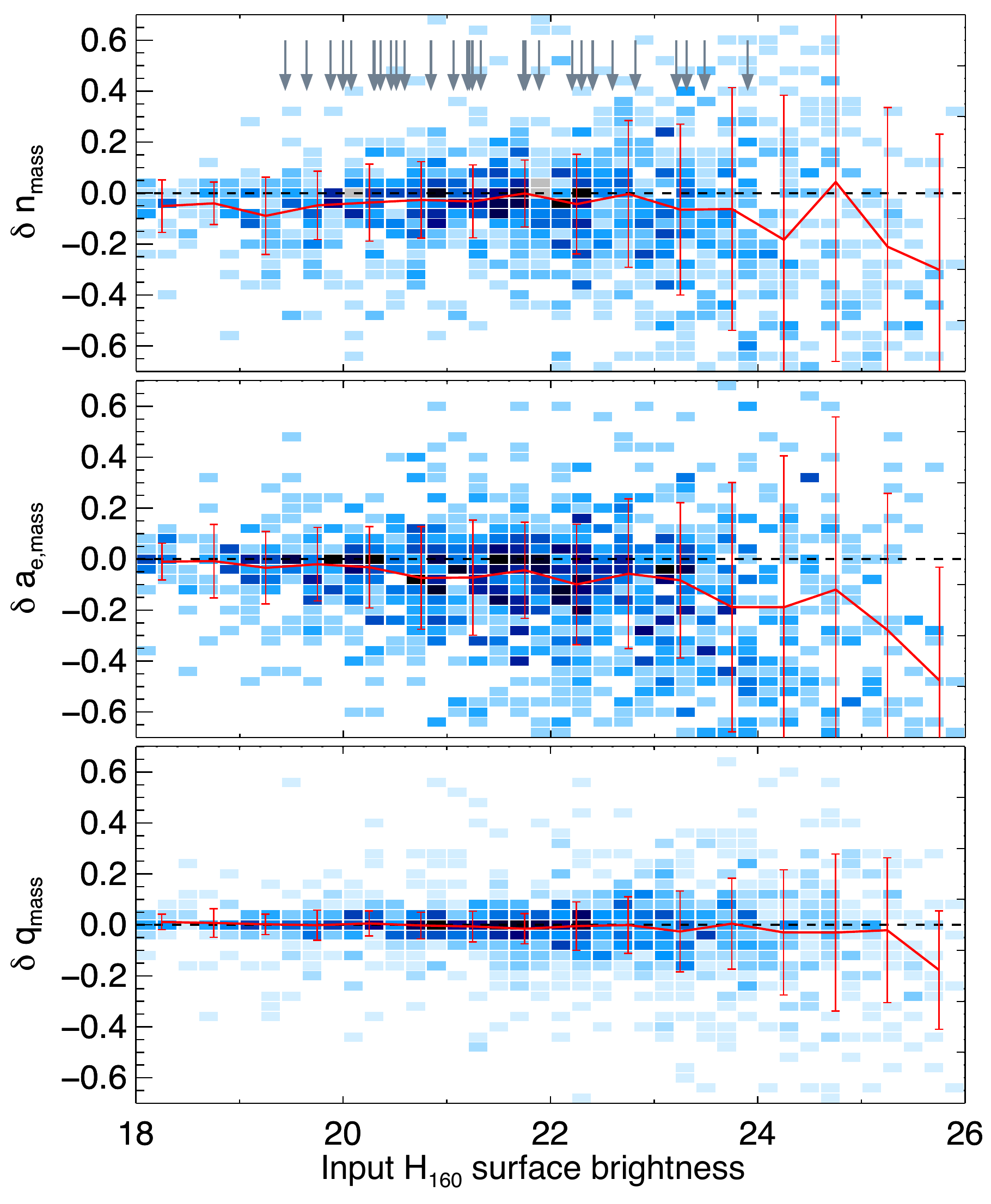}
  \caption{Differences between recovered and input mass-weighted structural parameters by GALFIT as a function of input $H_{160}$ surface brightness. Similar to Figure~\ref{fig:lightuncertainty}, but for mass-weighted structural parameters. From top to bottom: S\'ersic indices $\delta n = (n_{out} - n_{in})/n_{in}$, effective semi-major axes $\delta a_e = (a_{e-out} - a_{e-in})/a_{e-in}$  and axis ratio $\delta q = (q_{out} - q_{in})/q_{in}$.  Red line indicates the median and $1\sigma$ dispersion in different bins (0.5 mag arcsec$^{-2}$ bin width) and blue-shaded 2D histogram shows the number density distribution of the simulated galaxies. The grey arrows indicate the $H_{160}$ surface brightness of the galaxies in our cluster sample.}
  \label{fig:massuncertainty}
\end{figure}

Figure~\ref{fig:massuncertainty} shows the difference between input and recovered mass-structural parameters as a function of $H_{160}$ surface brightness.  The S\'ersic index, effective radius and axis ratio measurements are generally robust for objects brighter than $H_{160}$ surface brightness of 23.5 mag arcsec$^{-2}$.  This is important, as it demonstrates that our mass conversion procedure does not significantly bias the result.  The bias between the recovered and input S\'ersic indices is less than $7\%$ and the $1\sigma$ dispersion is lower than $40\%$, and effective radii have a bias less than $10\%$ and a $1\sigma$ dispersion lower than $40\%$.  Among the three parameters, the axis ratio can be recovered most accurately. Compared with the light uncertainties (Figure \ref{fig:lightuncertainty}), the mass uncertainties in all parameters are $\sim2$ times higher.  Similar to the light-weighted parameters, for each galaxies we add the corresponding dispersion in quadrature to the error output by GALFIT.

We find that for a couple of objects the fits do not converge, or have resultant sizes smaller than the PSF size.  To avoid biases and wrong conclusions we remove these objects that are not well-fitted from the mass parameter sample.  6 objects (out of 36) are discarded, among them one object is spectroscopically confirmed.  Three of them initially have small light-weighted sizes and their fitted mass-weighted sizes become smaller than half of the PSF HWHM, which are unreliable (see the discussion in Appendix~\ref{Details of the simulations-lightstruct}).  Most of them are low mass galaxies (i.e. $\log(M_{*}/M_\odot) < 10.5$).

\subsubsection{Deviation of mass-weighted parameters - 1D vs. 2D}
\label{deviation1Dvs2D}
\citet{Szomoruetal2013} derived 1D mass profiles from 1D radial surface brightness profiles and measured mass-weighted structural parameters.  In theory, fitting in 1D and in 2D should give identical results, as statistically fitting an averaged smaller group of points and fitting all the points without averaging are equivalent \citep[see,][for a detailed discussion.]{GalFAQ}.  Nevertheless, in practice deriving maps and fitting in 2D have certain advantages: a) It does not rely heavily on the S\'ersic profile fitting in light.  Deriving elliptical averaged profiles require a predetermined axis ratio and position angles, which, in our case, come from the light S\'ersic profile fitting.  This will of course fold in the uncertainties of these two parameters into the 1D profiles, which complicates the propagation of uncertainties in the mass-weighted parameters. b) In the cluster region, the object density is high and many galaxies have very close neighbours. Hence it will be more appropriate to fit all the sources simultaneously to take into account the contribution from the neighbouring objects, rather than deriving 1D profile without deblending the neighbouring contamination.  A possible way to solve this is to first subtract the best-fit 2D models of the neighbours from the 2D images before generating the 1D profiles, but of course this depends strongly on how well the neighbours can be subtracted, and still suffer from a).

\section{Local Comparison Sample}
\label{Local Comparison Sample}
In order to study the evolution of mass-weighted sizes over redshift, we compare our cluster sample at $z\sim1.39$ to a local sample of passive galaxies from the Spheroids Panchromatic Investigation in Different Environmental Regions (SPIDER) survey \citep{LaBarberaetal2010a}.  The publicly available SPIDER sample includes 39993 passive galaxies selected from SDSS Data Release 6 (DR6), among them 5080 are in the near-infrared UKIRT Infrared Deep Sky Survey-Large Area Survey Data release (UKIDSS-LAS DR4) in the redshift range of 0.05 to 0.095.  \citet{LaBarberaetal2010a} derived structural parameters in all available bands ($grizYJHK$) with single S\'ersic fitting with \texttt{2DPhot} \citep{LaBarberaetal2008}.

We use the structural parameters in $g$-band and $r$-band from the publicly available multiband structural catalogue from \citet{LaBarberaetal2010a} to derive mass-weighted structural parameters.  For the galaxy selection, we follow similar criteria as \citet{LaBarberaetal2010a}: we apply a magnitude cut at the 95\% completeness magnitude ($M_{r} \leq -20.55$), a $\chi^{2}$ cut from the S\'ersic fit for both $g$-band and $r$-band ($\chi^{2} < 2.0$), and a seeing cut at $\leq 1.5$''.  This results in a sample of 4050 objects.  We compute integrated masses for the sample as in Section~\ref{sec:Integrated Stellar Masses} with aperture $g - r$ colour.  The colours are obtained from direct numeral integration of the $g$-band and $r$-band S\'ersic profiles to 5 kpc instead of using GALFIT total magnitudes.  Extending the integration limit to larger radius (e.g. 10 kpc) does not change largely the derived masses.  With the $g - r$ colour we derive and select red-sequence galaxies within $2\sigma$ following the same method discussed in Section~\ref{sec:Object Detection, Sample Selection and Photometry}; we end up with a sample of 3634 objects (hereafter the SPIDER sample).  On top of that we use the group catalogue from \citet{LaBarberaetal2010b} to select a subsample of galaxies residing in high density environments.   Applying a halo mass cut to the SPIDER sample of $\log(M_{200}/M_{\odot}) \geq 14$, we end up with a subsample of 627 objects (hereafter the SPIDER cluster sample), which we will use as the main comparison sample for our high-redshift cluster galaxies.

2D S\'ersic model images in $g$-band and $r$-band are then generated with fitted parameters from the structural catalogue.  Given the large number and relatively low object density of local galaxies compared to our high redshift cluster sample, using fitted parameters from the structural catalogue is statistically reliable and issues mentioned in Section~\ref{deviation1Dvs2D} do not contribute substantially here.  We construct mass maps for individual galaxies using the procedure described in Section~\ref{sec:From Colour to Stellar Mass Surface Density} without Voronoi binning and stacking.  A $M_{*}/L$-colour relation is again derived from the NMBS sample as in Section~\ref{sec:Stellar Mass-to-light Ratio-Colour relation}, but in $g$-band and $r$-band at $0 < z < 0.27$, a window of 0.2 in redshift around the median redshift of the SPIDER sample.  A total of 1315 NMBS objects are selected.  The mass maps are then fitted with GALFIT to obtain mass-weighted structural parameters.


\section{Results}
\label{sec:Results}
\subsection{Wavelength dependence of light-weighted galaxy sizes}
The measured size of a galaxy depends on the observed wavelength, as different stellar populations are being traced at different wavelength (e.g., the ``morphological k-correction'', Papovich et al. \citeyear{Papovichetal2003}).  
With our multi-band measurements of light-weighted structural parameters of the cluster passive galaxies, we first investigate the wavelength dependence of galaxy sizes at this redshift. 
This wavelength dependence of sizes (or the size-wavelength relation) has been quantified for local passive galaxies in a number of studies \citep[e.g.][]{Bardenetal2005, Hydeetal2009, LaBarberaetal2010a, Kelvinetal2012, Vulcanietal2014, Kennedyetal2015}.  The dependence shown by the above mentioned studies is quite strong, in the sense that galaxy sizes can decrease up to $\sim38\%$ from $g$ through $K$ band in the GAMA sample \citep{Kelvinetal2012}, or $\sim32\%$ across the same range in SPIDER \citep{LaBarberaetal2010a}.  Nevertheless, different authors disagree on the extent of the reduction in sizes in various datasets.  For example, in a recent study \citet{Langeetal2015} revisited the GAMA sample with deeper NIR imaging data and found a smaller size decrease, $\sim13\%$ from $g$ to $K_s$ band.

At higher redshift, study of wavelength dependence of sizes is scarce in clusters. The star formation history and age gradient may contribute significantly to the size-wavelength dependence, for example the inside-out growth scenario suggests that younger stellar population are more widespread compared to the older population in the core of passive galaxies. Various authors have shown that measured sizes in the observed optical and NIR (i.e. rest-frame UV vs rest-frame optical for high-redshift galaxies) show a difference of $\sim$20--25\% \citep[e.g.][]{Trujilloetal2007,Cassataetal2010,Damjanovetal2011,Delayeetal2014}, although some find no difference \citep{Morishitaetal2014}.  The comparisons are usually done with only two bands, hence it is unclear whether this dependence can change with redshift.  Recent works from CANDELS studied the wavelength dependence of sizes for 122 early-type galaxies (ETG) in the COSMOS field in three HST bands (F125W, F140W and F160W) at redshift $0<z<2$, and found an average gradient of $d \log(a_{e}) / d \log(\lambda) = -0.25$ independent of mass and redshift \citep{Vanderweletal2014}.

\begin{figure}
  \includegraphics[scale=0.66]{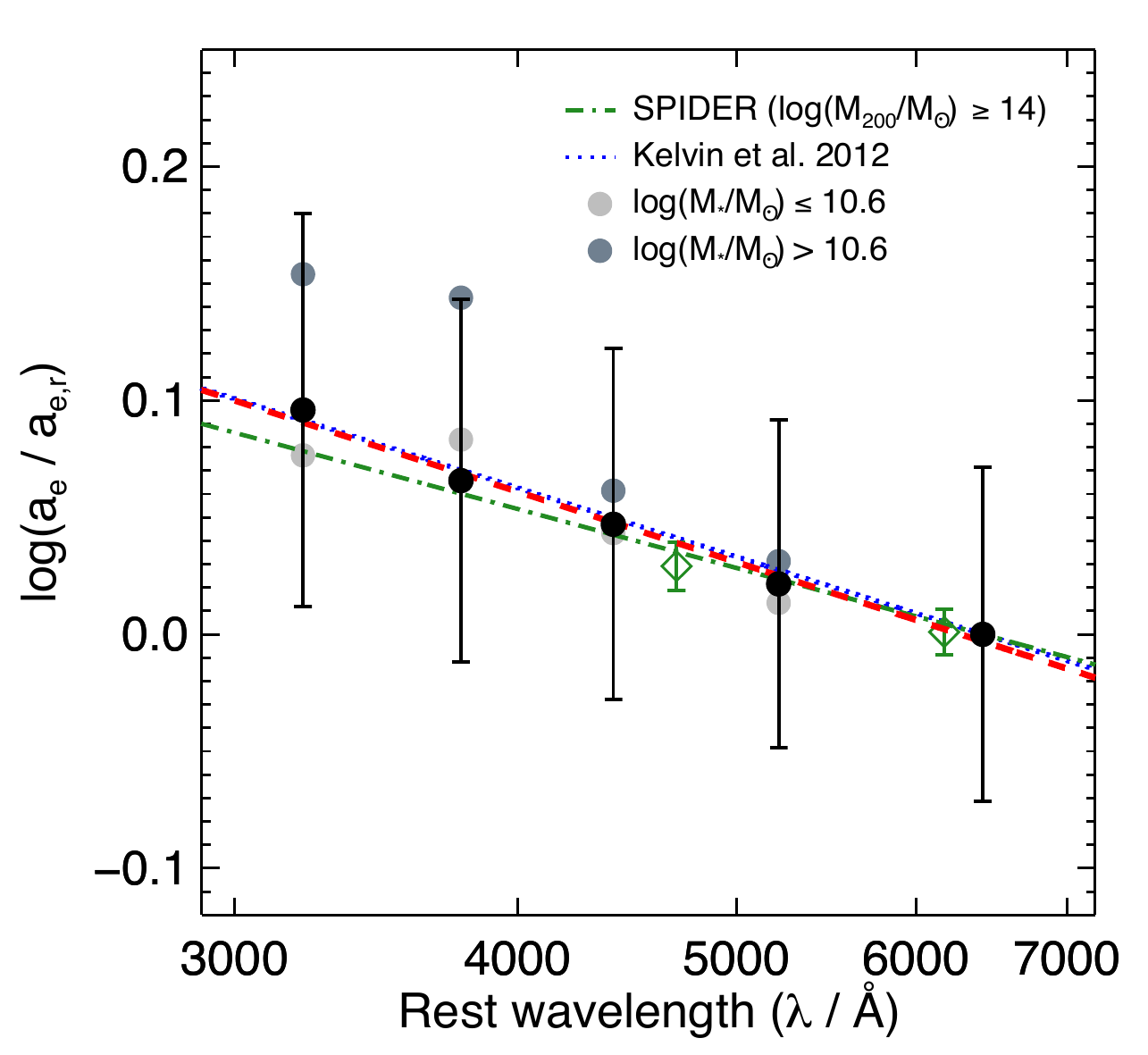}
  \caption{Size-wavelength relation of the passive galaxies in the cluster XMMUJ2235-2557.  Black circles show the median sizes of the sample in each band positioned at the rest-frame pivot wavelength, normalised with the median $H_{160}$-band sizes (approximately rest-frame $r$-band, $a_{e,r}$).  Error bars show the uncertainty of the median in each band, estimated as $1.253\sigma/\sqrt{N}$, where $\sigma$ is the standard deviation and $N$ is the size of the sample.  The best-fit power law to the sizes in our sample is shown as a red dashed line.  The green dot-dashed line is the best-fit relation for the SPIDER cluster sample (from $g$-band to $K_s$-band), while the green diamonds are the median size of the sample in $g$-band and $r$-band, normalised in the same way.  The blue dotted line is the best-fit relation for local galaxies from \citet{Kelvinetal2012}.  Grey and slate grey are the median sizes for two mass bins ($\log(M_{*}/M_\odot) \leq10.6$ and $\log(M_{*}/M_\odot)  > 10.6$) respectively.}
  \label{fig_sizewavelength}
\end{figure}

Figure \ref{fig_sizewavelength} shows the change in size with rest-frame wavelength for our sample.  Here we use the light-weighted effective semi-major axis $a_{e}$ from GALFIT, as the galaxy size. We assume every galaxy in the sample is at the cluster redshift.  We select 28 galaxies (out of 36) with no problematic fits in any of the five bands.  The fraction of problematic fits is larger in $i_{775}$ and $Y_{105}$ due to shorter exposure time and lower throughput of the filter, which result in lower S/N.  To facilitate comparison with the literature, the sizes in figure \ref{fig_sizewavelength} are normalised with the median $H_{160}$ sizes of our sample, which is approximately equal to the rest-frame $r$-band size.  We plot the best-fitting relation for local spheroids by \citet{Kelvinetal2012} and the SPIDER cluster sample, normalised in the same way, for comparison. We see a smooth variation of sizes decreasing from $i_{775}$ to $H_{160}$ bands (rest-frame $u$ to $r$).  The reduction in the median size (from $i_{775}$ to $H_{160}$) is $\sim20\%$, which is consistent with the expected decrease across this wavelength range ($\sim19\%$) following the relation of \citet{Kelvinetal2012} and the SPIDER cluster sample \citep{LaBarberaetal2010a}.  The average size gradient of our sample from the best-fit power law is $d \log(a_{e}) / d \log(\lambda) = -0.31 \pm 0.27$.

We also attempt to divide the sample in mass bins as in \citet{Langeetal2015} to investigate the size change with wavelength for different masses.  \citet{Langeetal2015} showed that the size reduction decreases from $\sim13\%$ for local passive galaxies with $\log(M_{*}/M_\odot) = 10.0$ to $\sim11\%$ for those with $\log(M_{*}/M_\odot) = 11.0$.  On the other hand, \citet{Vanderweletal2014} reported no discernible trends with mass in CANDELS.  We split the sample in half at the median mass ($\log(M_{*}/M_\odot) \leq 10.6$ and $\log(M_{*}/M_\odot)  > 10.6$, 14 objects per bin), plotted in grey and slate grey in Figure~\ref{fig_sizewavelength}.  A steeper dependence can be seen for the high mass bins ($d \log(a_{e}) / d \log(\lambda) = -0.57 \pm 0.28$) compared to the whole sample, while the low mass bins ($d \log(a_{e}) / d \log(\lambda) = -0.29 \pm 0.34$) have the same if not slightly shallower wavelength dependence within the uncertainties.  This is the opposite to the finding of \citet{Langeetal2015}.  Nevertheless, the size gradients of the two bins are within 1$\sigma$, a larger sample is needed to confirm the mass dependence.

\subsection{Stellar mass -- light-weighted size relation}
\label{sec:Stellar Mass -- Light-weighted Size Relation}
In this section we show the stellar mass -- $H_{160}$ light-weighted size relation of the cluster XMMUJ2235-2557.  The mass-size relation of this cluster (in $z_{850}$ band, rest-frame UV) has been studied in previous literature \citep{Strazzulloetal2010, Delayeetal2014}.

\begin{figure}
  \includegraphics[scale=0.467]{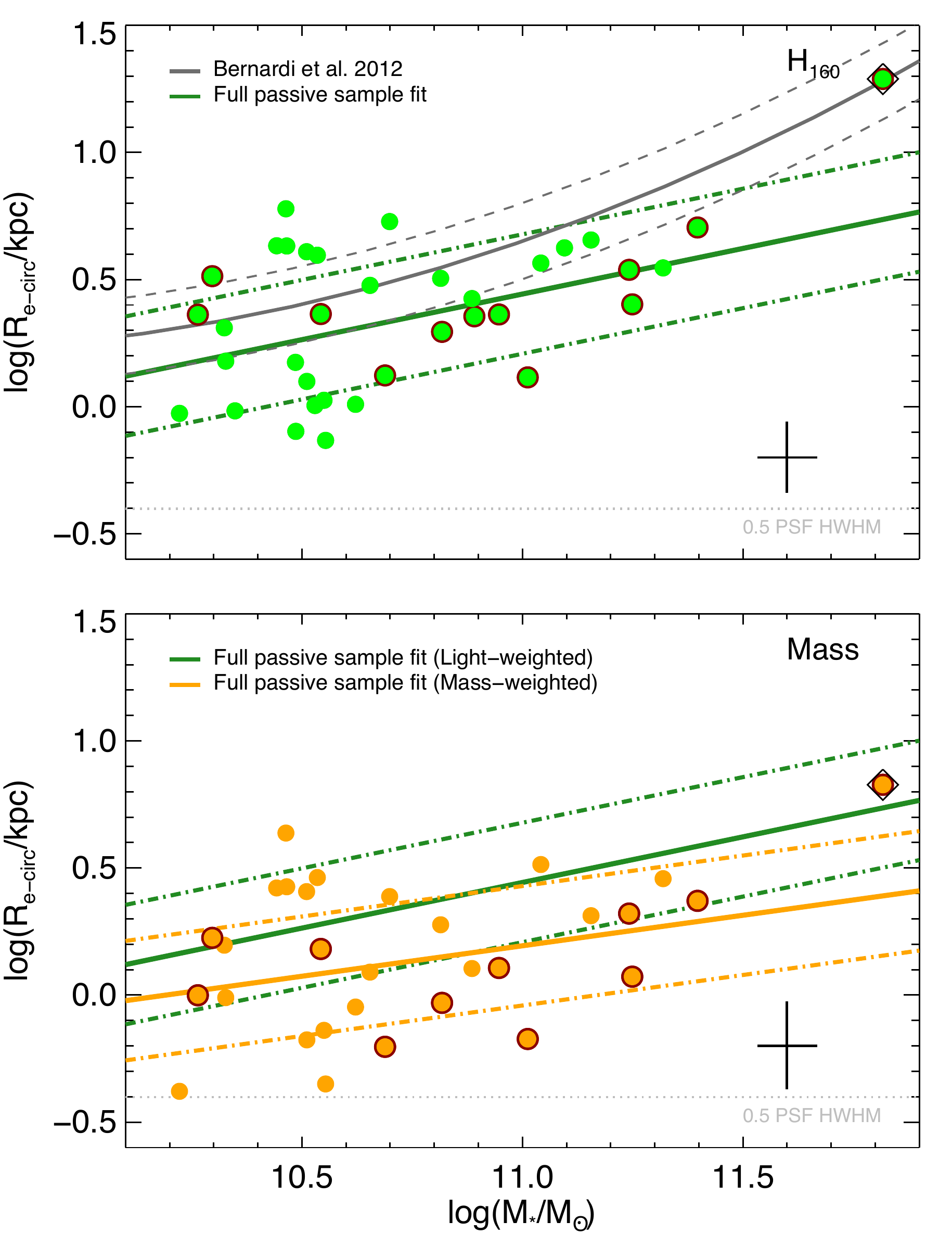}
  \caption{Stellar mass -- size relations of the passive galaxies in XMMUJ2235-2557.  Top: with light-weighted sizes.  Green dots show the sample selected with the passive criteria described in Section~\ref{sec:Object Detection, Sample Selection and Photometry}.  Spectroscopically confirmed objects are circled with dark red.  The green line is a linear fit to the full passive sample (Case A), while the dot-dashed lines represent $\pm 1\sigma$.  The dark grey line corresponds to the local $r$-band mass-size relation from \citet{Bernardietal2012}.  Bottom: with mass-weighted sizes.  Individual objects are shown in orange.  The orange solid line corresponds to the full sample fit (Case A) for the mass-- mass-weighted size relation, and the orange dot-dashed lines represent $\pm 1\sigma$.  The green line is the same linear fit in the top panel for comparison. The BCG is indicated with the black diamond.  The cross shows the typical uncertainty of the sizes and the median uncertainty of the integrated mass in our sample.}
  \label{fig_lightmasssize_mod}
\end{figure}

In the top panel of Figure \ref{fig_lightmasssize_mod} we plot the mass -- size relation for the passive population in the cluster selected from red sequence fitting.  Circled objects are spectroscopically confirmed cluster members \citep{Grutzbauchetal2012}.  The size we use from this point onwards is the circularised effective radius ($R_{e-circ}$), defined as:
\begin{equation}
   R_{e-circ} =  a_{e} \times \sqrt{q}
\end{equation}
where $a_{e}$ is the elliptical semi-major radius and $q = b / a$ is the axis ratio from the best-fit GALFIT S\'ersic profile.  

The integrated stellar masses are derived from the $M_{*}/L$ - colour relation and are scaled with the total GALFIT S\'ersic magnitude (see Section~\ref{sec:Integrated Stellar Masses} for details).  We also plot the local mass-size relation for the SDSS passive sample by \citet{Bernardietal2012} for comparison.  We note that although this relation was derived for galaxies regardless of their local density, a number of studies have established that there is no obvious environmental dependence on passive galaxy sizes in the local universe \citep{Guoetal2009, Weinmannetal2009, Tayloretal2010, Huertascompanyetal2013, Cappellari2013}.  We pick the single S\'ersic fit relation in \citet{Bernardietal2012} for consistency, which is shown to have slightly larger sizes than the two-component exponential + S\'ersic fit relation.

\citet{Hydeetal2009} first demonstrated that the mass-size relation of passive galaxies shows curvature and \citet{Bernardietal2012} fitted the curvature with a second order polynomial; their best-fit values were consistent with \citet{Simardetal2011}.  As we have shown in the last section, size shows wavelength-dependence, hence care has to be taken to ensure the sizes being compared are at around the same rest-frame wavelength. The \citet{Bernardietal2012} local relation is based on the Sloan $r$-band, while our sizes are measured in the $H_{160}$ band at a redshift of 1.39, which roughly corresponds to the same rest-frame band.  As a result, no size-correction is required as the correction to $r$-band is negligible.

The $H_{160}$ band sizes of the passive galaxies in this cluster are on average $\sim40\%$ smaller than expected from the local relation by \citet{Bernardietal2012} with $\langle \log(R_{e-circ} / R_{Bernardi}) \rangle = -0.21$ ($\sim45\%$ smaller for the spectroscopic confirmed members, $\langle \log(R_{e-circ} / R_{Bernardi}) \rangle= -0.25$).  There are also galaxies whose sizes are $\sim70\%$ smaller than those of their local counterparts ($\log(R_{e-circ} / R_{Bernardi}) = -0.56$).  As one can see from Figure~\ref{fig_lightmasssize_mod}, the most massive object in the cluster is the BCG, which also has the largest size ($\sim24$ kpc) and lies on the local relation.  This is consistent with findings from \citet{Stottetal2010, Stottetal2011}, who showed that as a population, BCGs have had very little evolution in mass or size since $z \sim 1$. \citet{Tiretetal2011} suggested that major mergers at $z \geq 1.5$ are required to explain the mass growth of these extremely massive passive galaxies.  Hence below we exclude the BCG when fitting the mass-size relation. We fit the mass-size relation with a Bayesian inference approach using Markov Chain Monte Carlo \citep[MCMC;][]{Kelly2007} with the following linear regression:

\begin{equation}
  \label{eqt-fit}
   \log(R_{e-circ}/kpc) =  \alpha +  \beta~ (\log(M_{*} /M_\odot) - 10.5) + N(0, \epsilon)
\end{equation}
\noindent where $N(0,\epsilon)$ is the normal distribution with mean 0 and dispersion $\epsilon$. The $\epsilon$ represents the intrinsic random scatter of the regression.

The best-fit parameters (the intercept $\alpha$, slope $\beta$ and the scatter $\epsilon$) for both the entire red-sequence selected sample (A) and the spectroscopically confirmed members only (B) are summarised in Table~\ref{tab_bestfit}.  For mass completeness and comparison to previous literature, we also fit only the massive objects with $\log(M_{*}/M_\odot) \geq 10.5$, the limiting mass adopted in \citet{Delayeetal2014} (C \& D).  We notice that the slope of the relation can change by more than 1$\sigma$ depending on the considered mass range.  We also fit the slope using the elliptical semi-major axis $a_{e}$ instead of $R_{e-circ}$), which gives us a significantly flatter slope ($\beta = 0.35 \pm 0.15$).

Our measured slope is consistent at the 1$\sigma$ level with the results of \citet{Delayeetal2014}, who studied the mass-size relation using seven clusters at $0.89 < z < 1.2$ in the rest-frame $B$-band (i.e. $\beta = 0.49 \pm 0.08$ for $\log(M_{*}/M_\odot) > 10.5$).  \citet{Papovichetal2012} measured the sizes of passive galaxies in a cluster at $z = 1.62$ and found that ETG with masses $\log(M_{*}/M_\odot) > 10.48$ have $\langle R_{e-circ} \rangle = 2.0$ kpc with the interquartile percentile range (IQR) of $1.2-3.3$ kpc.  Sizes in XMMUJ2235-2557 are on average $40\%$ larger ($\langle R_{e-circ} \rangle = 2.80$~kpc, IQR $= 1.45-4.38$~kpc).

While the fits are consistent with each other on a 1$\sigma$ level, we notice that the relation in \citet{Delayeetal2014} for this cluster is flatter ($\beta = 0.2 \pm 0.3$) compared to both our full sample fit (A) and massive sample fit (C).  This difference could be due to a combination of a) their mass-size relation is computed in the $z$-band while ours is in $H_{160}$, b) the two red sequence samples are selected differently and c) the masses computed here are scaled with total GALFIT S\'ersic magnitude instead of \texttt{MAG\_AUTO} (the relation is slightly flatter: $\beta = 0.38 \pm 0.27$ instead of $0.43$ if we use the masses scaled with \texttt{MAG\_AUTO}).

A caveat of the above comparison is that we have not considered the effect of progenitor bias \citep{vanDokkumetal2001}.  Correcting the progenitor bias (in age and morphology) has been shown to reduce the magnitude of the observed size evolution \citep{Sagliaetal2010, Valentinuzzietal2010a, Beifiorietal2014}.  Recently, \citet{Jorgensenetal2014} corrected the progenitor bias by removing galaxies that are too young in the Coma cluster to be the descendants of a cluster at $z = 1.27$ and found no size evolution with redshift.

\begin{table}
  \noindent
  \caption{Best-fit parameters of the Stellar mass -- size relations}
  \label{tab_bestfit}
  \addtolength{\tabcolsep}{-2pt}
  \begin{tabular}{@{}clccc@{}}
  \hline
  \hline
  \multicolumn{5}{|l|}{Stellar mass -- light-weighted size relation} \\
  \hline
  Case &  Mass range &  $\alpha \pm \Delta \alpha$  & $\beta \pm \Delta \beta$ & $\epsilon$ \\
  \hline
A & $10.0 \leq M_{*} \leq 11.5$              & $ 0.263 \pm 1.441$  &  $0.359 \pm 0.135$  &  0.235 \\
B & $10.0 \leq M_{*} \leq 11.5$ (spec)  & $ 0.329 \pm 2.096$   &  $0.138 \pm 0.192$  &  0.195 \\
C & $10.5 \leq M_{*} \leq 11.5$              & $ 0.114 \pm  1.876$  &  $0.576 \pm  0.173 $   & 0.195 \\ 
D & $10.5 \leq M_{*} \leq 11.5$ (spec)  & $ 0.149 \pm  2.935$  &  $0.447 \pm  0.268 $   & 0.175 \\    
   \hline
   \hline
   \multicolumn{5}{|l|}{Stellar mass -- mass-weighted size relation} \\
   \hline
   Case &  Mass range &  $\alpha \pm \Delta \alpha$  & $\beta \pm \Delta \beta$ & $\epsilon$ \\
   \hline
 A & $10.0 \leq M_{*} \leq 11.5$              & $ 0.074  \pm  1.733$  &  $0.240 \pm 0.162$  &  0.235 \\
 B & $10.0 \leq M_{*} \leq 11.5$ (spec)  & $ 0.037  \pm  2.467$  &  $0.141 \pm 0.227$  &  0.212 \\
 C & $10.5 \leq M_{*} \leq 11.5$              & $ -0.043 \pm   2.093$  &  $0.477  \pm  0.192 $  & 0.182 \\ 
 D & $10.5 \leq M_{*} \leq 11.5$ (spec)  & $ -0.152  \pm  3.839$  &  $0.411  \pm  0.350 $  & 0.209 \\
  \hline
\end{tabular}
\addtolength{\tabcolsep}{2pt}
\end{table}

\subsection{Colour gradients in the passive cluster galaxies}
\label{sec:Colour Gradients in the Early-type Cluster Galaxies}
\begin{figure}
  \includegraphics[scale=0.457]{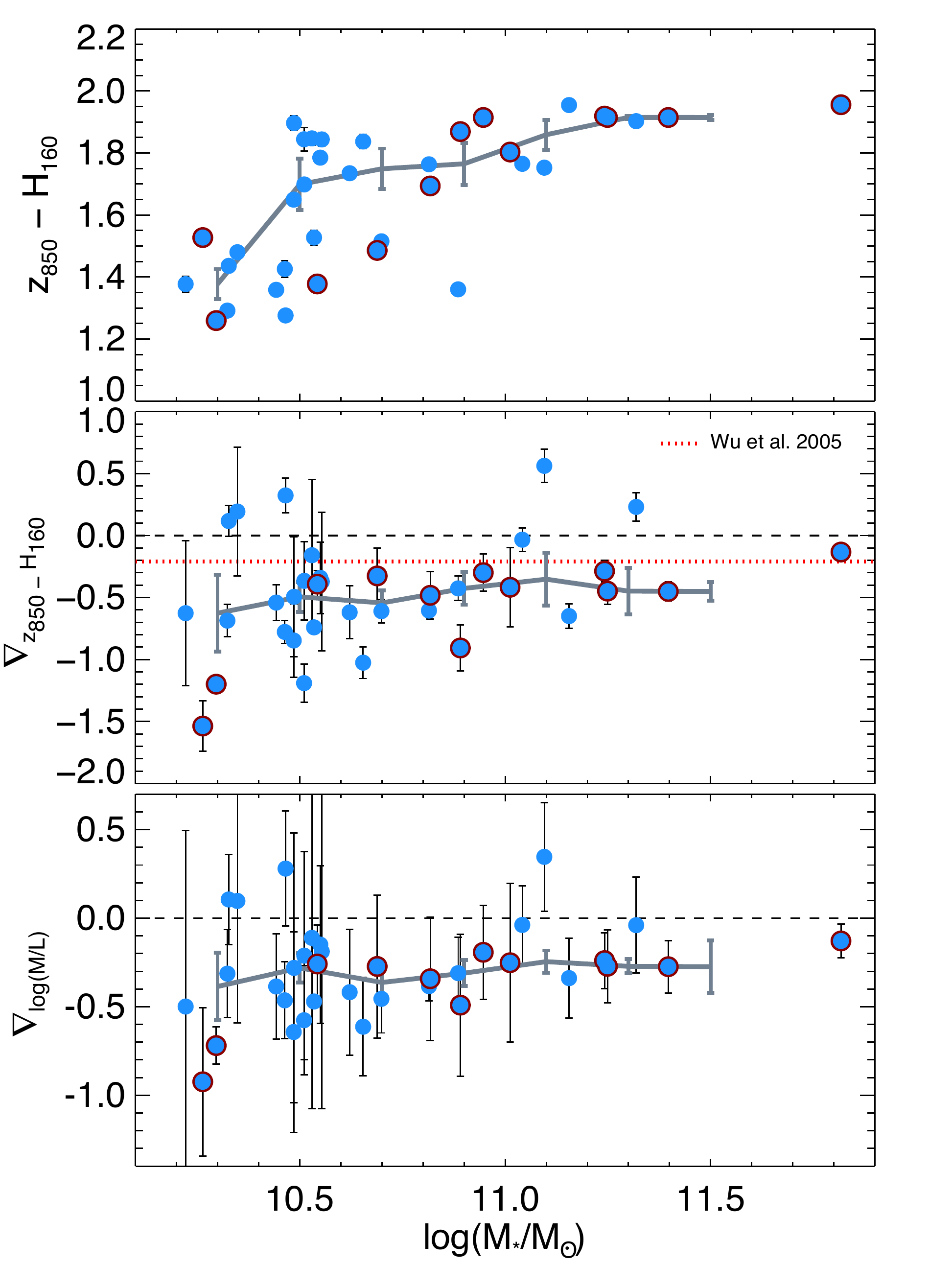}
  \caption{Colour and colour gradients in the cluster XMMUJ2235-2557.  From top to bottom: $z_{850} - H_{160}$ aperture colour ($1''$ in diameter), colour gradient $\nabla_{z_{850} - H_{160}}$ and $\log (M_{*}/L)$ gradient $\nabla_{\log(M/L)}$ as a function of stellar mass.  Spectroscopically confirmed objects are circled in dark red.  At redshift 1.39, this roughly corresponds to rest-frame $(U-R)$ colour gradient.  The black dashed line in each panel shows the reference zero level.  The red dotted line shows the average local $(U-R)$ gradient from \citet{Wuetal2005}.  The grey line in each panel shows the running median and the error bars show the uncertainty of the median in each bin.  When there is only one point in the bin, the uncertainty of the quantity is plotted instead.}
  \label{fig_colourgrad}
\end{figure}

In Figure~\ref{fig_colourgrad} we show the $1''$ aperture colour, colour gradients $\nabla_{z_{850} - H_{160}}$ and $\log (M_{*}/L)$ gradients $\nabla_{\log(M/L)}$ of the passive sample as a function of stellar mass.  The $\log (M_{*}/L)$ gradients are derived from fitting 1D $M_{*}/L$ profiles, which are derived from 1D colour profiles using the $M_{*}/L$-colour relation.  Note that since the $M_{*}/L$-colour relation is essentially a one-to-one mapping, measuring the colour gradient is qualitatively equivalent to measuring the $\log (M_{*}/L)$ gradient.

More massive galaxies appear to have a redder $z_{850} - H_{160}$ colour, as also shown by \citet{Strazzulloetal2010} with HST/NICMOS data.  Redder colour implies a higher median $M_{*}/L$ from the $M_{*}/L$-colour relation.  The passive sample has a range of colour from $\sim1.2 \leq z_{850} - H_{160} \leq 2.0$, which corresponds to a range of $M_{*}/L$ of $-0.37 \leq \log(M_{*}/L) \leq -0.85$.

Most of the galaxies have negative colour gradients.  28 out of 36 galaxies ($\sim 78\%$) show a negative gradient, and 15 out of 36 ($\sim 42\%$) have very steep gradients with $\nabla_{z_{850} - H_{160}} < -0.5$.  The median colour gradient and $1\sigma$ scatter is $\langle \nabla_{z_{850} - H_{160}} \rangle = -0.45 \pm 0.43$ (error on the median 0.09) and the median $\log (M_{*}/L)$ gradient $\nabla_{\log(M/L)}$ is $-0.27 \pm 0.25$ (error on the median 0.05).  This is consistent with previous findings at higher redshift $1.3 < z < 2.5$ \citep{Guoetal2011} which showed that passive galaxies have red cores and bluer stellar population at the outskirts. We find no strong dependence of colour gradients with stellar mass, with a median Spearman's rank correlation coefficient of $\rho \simeq 0.32, p \simeq 0.06$ computed using a bootstrapping method.

At redshift 1.39, the observed $\nabla_{z_{850} - H_{160}}$ colour gradient corresponds to rest-frame $\nabla_{U-R}$.  To ensure rest-frame $\nabla_{U-R}$ matches $\nabla_{z_{850} - H_{160}}$, we compute $\nabla_{U-R}$ from the observed $\nabla_{z_{850} - H_{160}}$ and $z_{850} - H_{160}$ colour using simple stellar population models (SSPs) of \citet{BruzualCharlot2003} as a sanity check.  The details of the methodology are described in Section~\ref{sec:Methodology}.  We confirm that the median gradient $\langle \nabla_{U-R} \rangle = -0.53$ is comparable to $\nabla_{z_{850} - H_{160}}$.  

We overplot the average local $(U-R)$ colour gradient from \citet{Wuetal2005} on Figure~\ref{fig_colourgrad} ($\nabla_{U-R} = -0.21 \pm 0.04$) for comparison.  \citet{Wuetal2005} studied the colour gradients for a sample of $36$ local field early-type galaxies from SDSS and 2MASS.  Due to a lack of deep $U$-band imaging, the $(U-R)$ colour gradient is not available in most local galaxy surveys.  In order to take into account the average age difference between field and cluster passive galaxies \citep[e.g.][]{Thomasetal2005,Thomasetal2010}, we evolve the gradients of \citet{Wuetal2005} for an additional $2$~Gyr with \citet{BruzualCharlot2003} SSP models (assuming an age gradient of $-0.05$ and a metallicity gradient of $-0.2$ consistent with the literature \citep{LaBarberaetal2005, Wuetal2005}).  The extrapolated value ($\nabla_{U-R} = -0.20$) is very close to the one for local ETGs.  The average $(U-R)$ colour gradient at $z\sim1.39$ is found to be $\sim2$ times steeper than colour gradients observed locally.

As a consistency check, we repeat the colour gradient measurements in $\nabla_{Y_{105} - H_{160}}$ ($\sim$ rest-frame $g-r$), and find consistent results with the $U-R$.  We also compare them with the $g-r$ colour gradient in the SPIDER cluster sample and note that the $g-r$ gradients at $z\sim1.39$ (median and $1\sigma$ scatter $\langle \nabla_{g-r} \rangle = -0.16 \pm -0.16$) are steeper than the local $g-r$ gradients ($-0.042 \pm 0.144$), although with a smaller dynamic range.  Details are described in Appendix~\ref{sec:The g-r color gradients and the evolution with redshift}.  Appendix~\ref{sec:The g-r color gradients and the evolution with redshift} also explores the dependence of the local $g-r$ gradients on environment within the full SPIDER sample.  We report to later sections for a discussion on the origin of colour or $M_{*}/L$ gradient in these high redshift passive galaxies.

\subsection{Comparison of light-weighted to mass-weighted structural parameters}
\label{sec:Comparison of Light-weighted to Mass-weighted Structural Parameters}

\begin{figure}
  \includegraphics[scale=0.59]{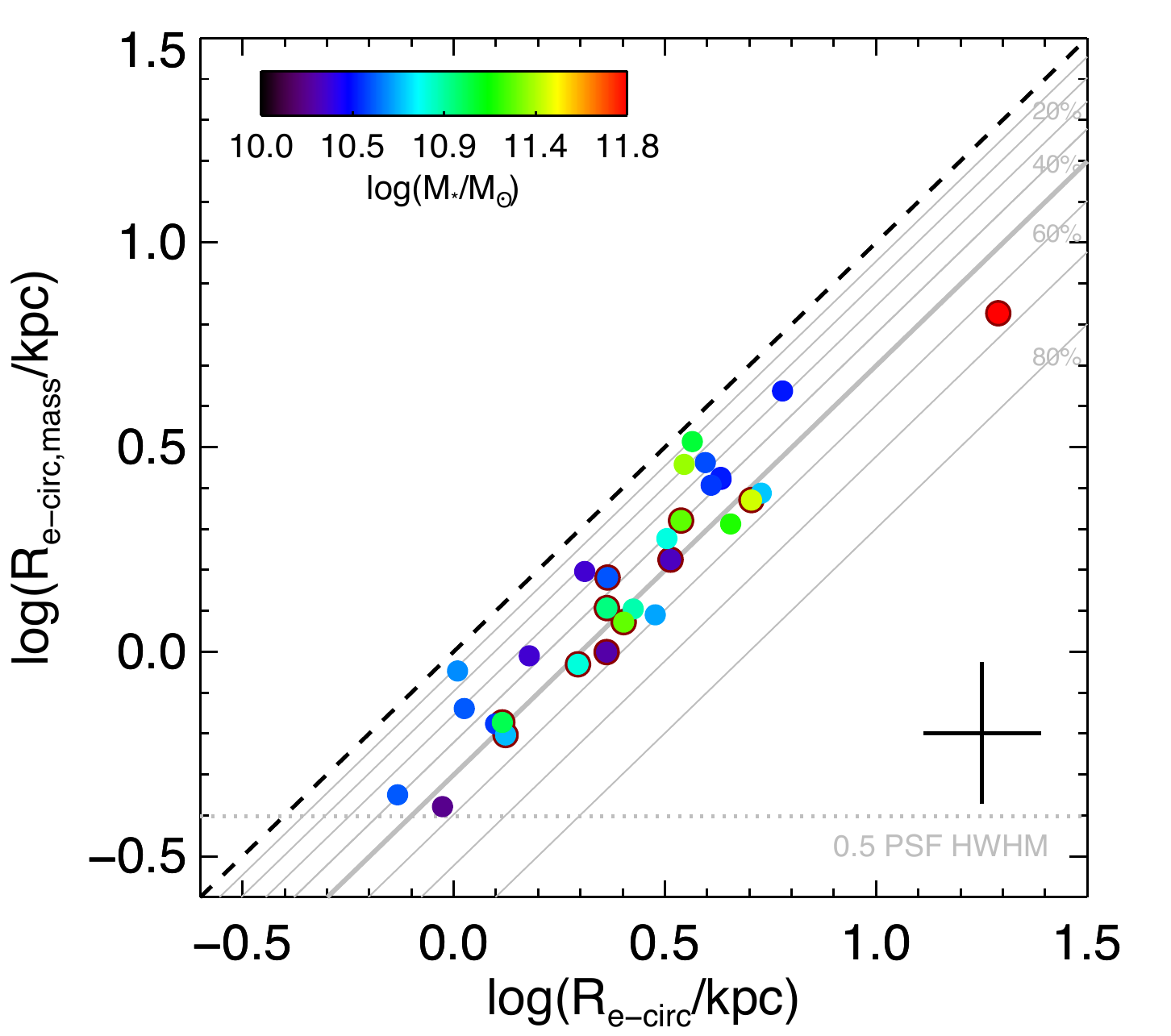}
  \caption{Comparison between mass-weighted size $R_{e-circ,mass}$ and light-weighted size $R_{e-circ}$ of passive galaxies in the cluster XMMUJ2235-2557.  Spectroscopically confirmed objects are circled in dark red.  The dashed line is the one-to-one relation.  Each successive grey line represents a 10\% decrement to the one-to-one relation.  50\% decrement are marked with a thick grey line.  The galaxies are colour coded with their integrated mass. The cross shows the median uncertainty of the light-weighted sizes and mass-weighted sizes.  Note that the two uncertainties are correlated to some extent.}
  \label{fig_lightmasscomp}
\end{figure}

In Figure~\ref{fig_lightmasscomp}, we compare the light-weighted sizes ($R_{e-circ}$) measured in $H_{160}$ band to the mass-weighted sizes (hereafter $R_{e-circ,mass}$) measured from the mass maps.  The mass-weighted sizes are $\sim41\%$ smaller than the $H_{160}$ light-weighted sizes, with a median difference of $\langle \log(R_{e-circ,mass}/ R_{e-circ})\rangle = -0.23$.  The scatter $\sigma_{\log(R_{e-circ,mass}/ R_{e-circ})} $ is $\sim0.11$.  In the most extreme case the mass-weighted size can be up to $\sim60\%$ smaller than its light counterpart (excluding the cluster BCG which is $\sim65\%$ smaller).

The general trend of mass-weighted sizes being smaller is in qualitative agreement with the study at similar redshift by \citet{Szomoruetal2013}, who computed the mass-weighted sizes using radially binned 1D surface brightness profiles for passive field galaxies in CANDELS.  As we will show in the discussion, this is consistent with the colour gradients in high redshift passive galaxies, in the sense that they usually have redder cores and bluer outskirts \citep[see also, e.g.][]{Sagliaetal2000, Wuytsetal2010, Szomoruetal2011,Guoetal2011}.  Negative colour gradients can lead to smaller mass-weighted sizes compared to light-weighted sizes, as a higher ${M_{*}/L}$ ratio at the centre results in a more concentrated mass distribution compared to the light distribution (hence, a smaller $a_e$).

\subsection{Stellar mass -- mass-weighted size relation}
\label{sec:Stellar Mass -- Mass-weighted Size Relation}
In the bottom panel of Figure \ref{fig_lightmasssize_mod} we show the mass -- size relation with the mass-weighted sizes. In the above section we have demonstrated that the mass-weighted sizes are $\sim$41\% smaller than light-weighted sizes.  Here we investigate how using mass-weighted sizes can affect the mass -- size relation.

We fit the stellar mass -- mass-weighted size relation, using equation~\ref{eqt-fit}.  The best-fit parameters are summarised in the second half of Table~\ref{tab_bestfit}. The fact that the mass-weighted sizes are smaller can be seen from the intercept of the fits.  Apart from the intercept, there seems to be a slight change in the slope of the relation if mass-weighted sizes are used.  The best-fitted relation for the full sample has a value $\beta = 0.240$, $34\%$ lower than the light-weighted size -- mass relation, though the two relations are consistent within $1\sigma$.

We check that the change of slope is not due to the discarded objects.  More statistics are required to confirm if there is a shallower mass dependence for mass-weighted sizes with respect to light-weighted sizes.


\begin{figure*}
\centering
  \includegraphics[scale=0.51]{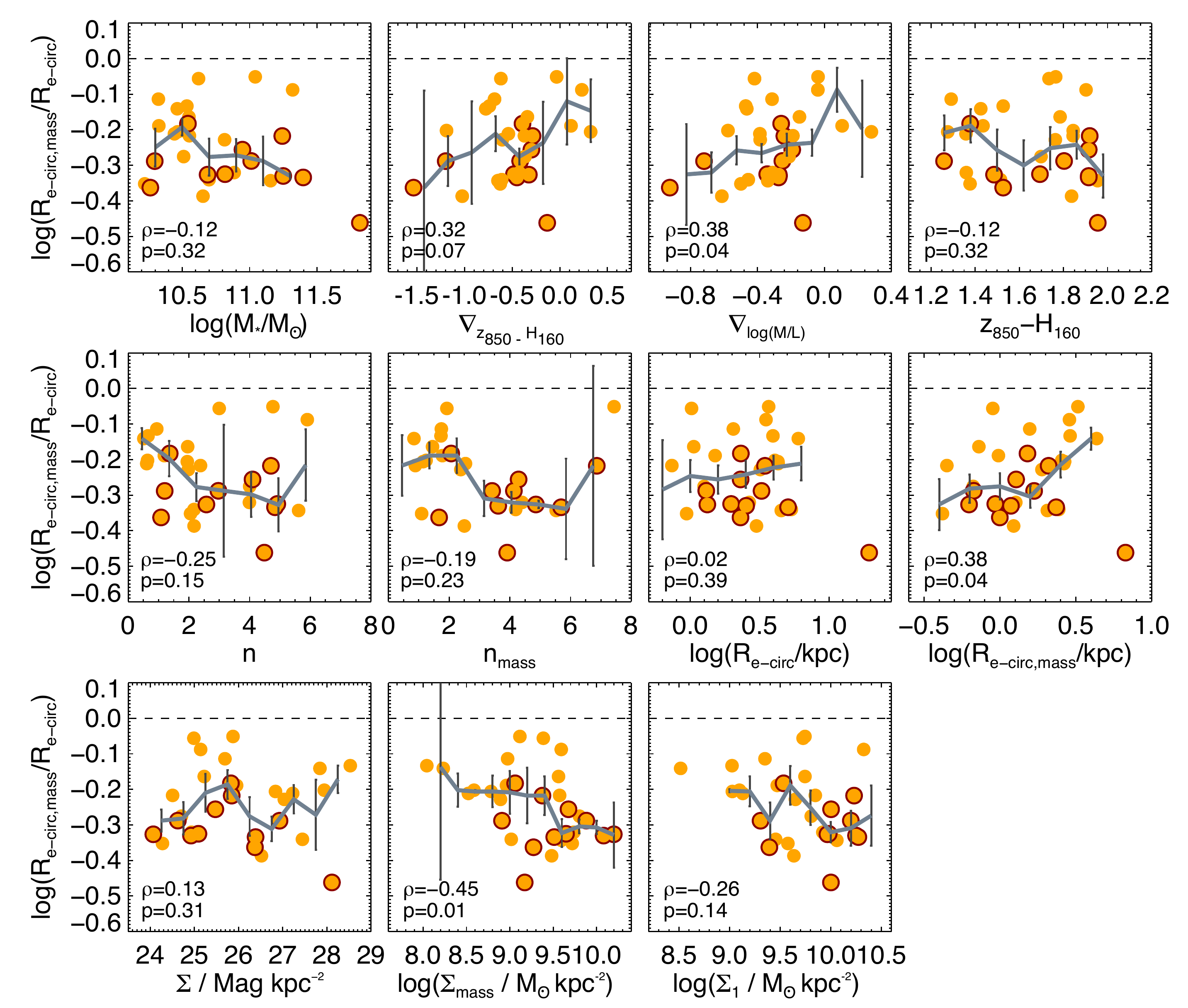}
  \caption{Dependence of ratio of mass-weighted to light-weighted sizes on different galaxy parameters. $H_{160}$ band sizes (rest-frame $r$-band) are used to compute the ratio of mass-weighted to light-weighted sizes ($R_{e-circ,mass}/R_{e-circ}$). From top left to bottom right: stellar mass, colour gradient $\nabla_{z_{850} - H_{160}}$, $M_{*}/L$ gradient $\nabla_{\log(M/L)}$,  $z_{850} - H_{160}$ colour, light-weighted S\'ersic index $n$, mass-weighted S\'ersic index $n_{mass}$, light-weighted effective radius $R_{e-circ}$, mass-weighted effective radius $R_{e-circ,mass}$, mean surface brightness $\Sigma$, mean surface mass density $\Sigma_{mass}$ and mean surface mass density within 1 kpc $\Sigma_{1}$.  Spectroscopically confirmed objects are circled in dark red.  Grey line in each panel shows the running median.  The error bars show the uncertainty of the median in each bin.  When there is only one point in the bin, the uncertainty of the ratio is plotted instead.}
  \label{fig:massmapdep}
\end{figure*}

\section{Discussion}
\label{sec:Discussion}
\subsection{Dependence of ratio of mass-weighted to light-weighted sizes on galaxy properties}
In Section~\ref{sec:Comparison of Light-weighted to Mass-weighted Structural Parameters} we have shown that the mass-weighted sizes are smaller than the corresponding light-weighted sizes and that the majority of the galaxies have negative colour gradients steeper than local passive galaxies.  Intuitively, one might expect some correlation between the ratio of mass-weighted to light-weighted sizes ($\log(R_{e-circ,mass}/R_{e-circ})$, hereafter the size ratio) on the physical parameters that are related to the underlying stellar population, such as colour, stellar mass, and various structural parameters.  

Hence, here we investigate the origin of the size ratio in our cluster by examining the correlation with a number of integrated properties.  In Figure~\ref{fig:massmapdep} we show the correlations between the size ratio with the stellar mass, colour gradient $\nabla_{z_{850} - H_{160}}$,  $M_{*}/L$ gradient $\nabla_{\log(M/L)}$, $z_{850} - H_{160}$ colour, light-weighted / mass-weighted S\'ersic indices, sizes, mean surface brightness $\Sigma$, mean surface mass density $\Sigma_{mass}$ and the mean surface mass density within a radius of 1 kpc $\Sigma_{1}$.  Recent works have shown that $\Sigma_{1}$ is tightly correlated with stellar mass and is closely related to quenching of star formation \citep{Fangetal2013, vanDokkumetal2014, Barroetal2015}.  All parameters plotted are given in Table~\ref{tab_para}. Running median and $1 \sigma$ scatter are over-plotted in each panel.

We search for possible correlations with these physical parameters and compute the correlation coefficients again using the bootstrapping method.  We see a mild dependence with the mass surface density with a $\rho$ value of $-0.45, p \simeq 0.01$.  There is also a weak dependence for the colour gradient and the $M_{*}/L$ gradient, with $\rho \simeq 0.32, p \simeq 0.07$ and $\rho \simeq 0.38, p \simeq 0.04$.  In addition, we see a weak dependence with the mass-weighted size, which has the highest $\rho$ among the light-weighted and mass-weighted structural parameters. In Section~\ref{sec:Stellar Mass -- Mass-weighted Size Relation}, we suspect a difference in the mass dependence for the mass-weighted sizes with respect to the light-weighted sizes, which if genuine, implies a correlation between size ratio and the stellar mass.  Nevertheless, there is no significant correlation with mass.  All other correlations have a $|\rho|$ value $< 0.3$.

In summary, with the exception of mass surface density, most of the parameters do not show significant dependence with the size ratio.  That our measured mass-weighted sizes tend to be significantly smaller than light-weighted sizes can only happen because there are gradients in mass-to-light ratio and colours, as seen previously. Therefore it is encouraging to see that there are (mildly significant) positive correlations between the ratio of sizes and the gradients in colour and ${M_{*}/L}$. That the correlations are not perfect illustrates the contributions of both uncertainties in the data and method, and the fact that our S\'ersic fits are actually quite different from a straightforward linear 1D fit as used to derive the gradients.  A more sophisticated fit to these parameters, and better correlations, require higher S/N and / or a larger sample.

\subsection{Evolution of the ratio of mass-weighted to light-weighted sizes to $z\sim0$}
\label{Evolution of ratio of mass-weighted to light-weighted sizes over redshift}
\begin{figure}
\centering
  \includegraphics[scale=0.468]{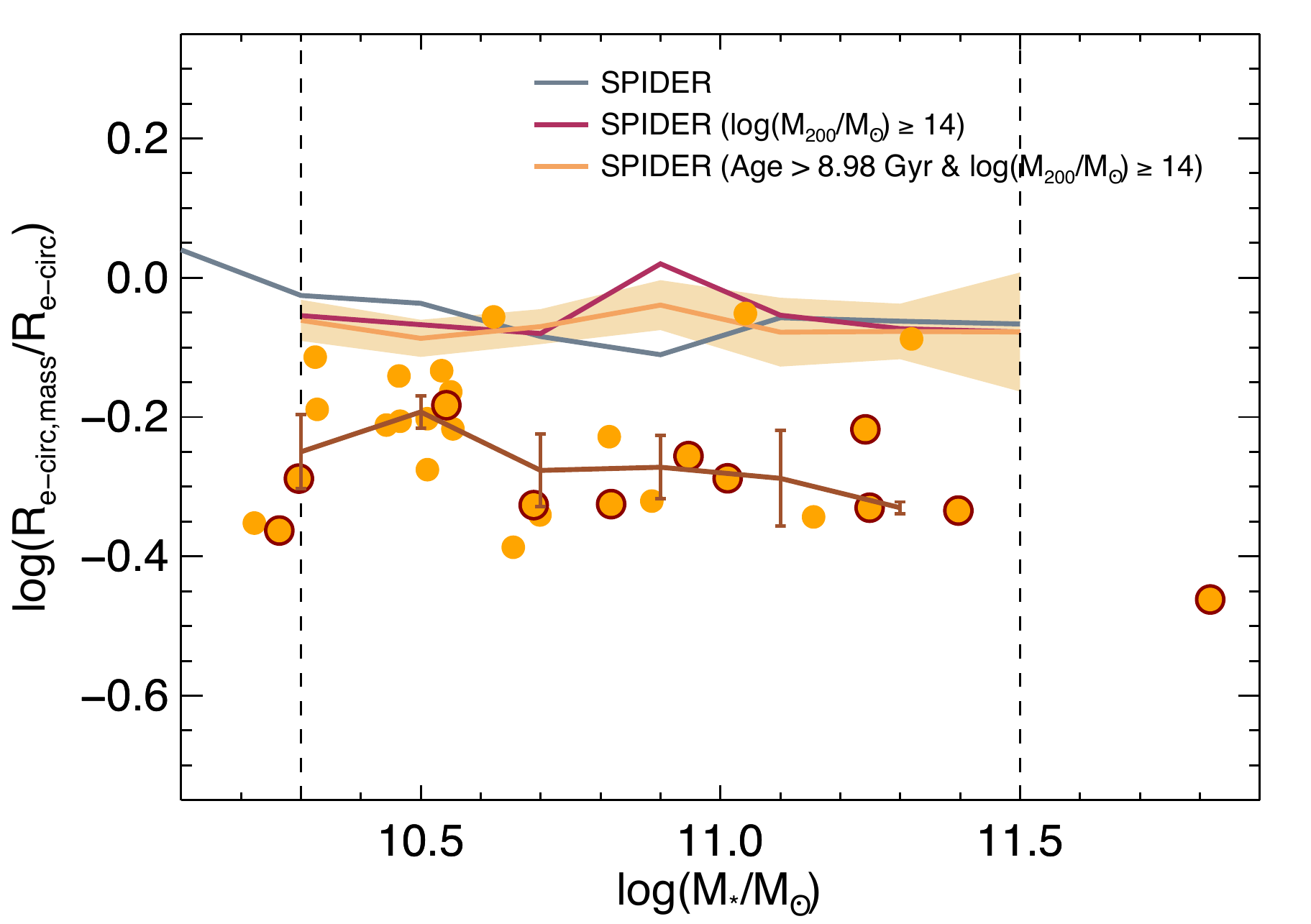}
  \caption{Comparison of the ratio of mass-weighted to light-weighted sizes vs. stellar mass at different redshift.  Same as top leftmost panel of Figure~\ref{fig:massmapdep} but includes the local size ratios from the SPIDER cluster sample.  Spectroscopically confirmed objects are circled in dark red.  The brown line corresponds the running median in mass bins of 0.2 with a window of width 0.3, and the error bars show the uncertainty of the median in each bin. The median size ratio of the SPIDER cluster sample (from mass range $10.2 \leq \log(M_{*}/M_\odot) \leq 11.6$) is plotted as a dark magenta line, while the median size ratio of the SPIDER sample (from mass range $10.0 \leq \log(M_{*}/M_\odot) \leq 11.6$) is plotted as a slate grey line in mass bin of 0.2.  The light brown line and wheat shaded region correspond to the median size ratio and $\pm1\sigma$ error on the median for the progenitor bias corrected SPIDER cluster sample (age $> 8.98$ Gyr and $\log(M_{200}/M_{\odot}) \geq 14$) from mass range $10.2 \leq \log(M_{*}/M_\odot) \leq 11.6$.}
  \label{fig_SPIDER}
\end{figure}

To investigate the evolution of the size ratio,  we compare the size ratio of the cluster sample with the local size ratio computed from a sample of local passive galaxies in high-density environment selected from the SPIDER survey in Figure~\ref{fig_SPIDER} (the SPIDER cluster sample, see Section~\ref{Local Comparison Sample} for details).  We also compare with the SPIDER sample for completeness.  We binned the size ratio of the SPIDER cluster sample (and the SPIDER sample) in mass bins of 0.2, in the mass range $10.2 \leq \log(M_{*}/M_\odot) \leq 11.6$ ($10.0 \leq \log(M_{*}/M_\odot) \leq 11.6$ for the SPIDER sample), to ensure there are sufficient numbers of local galaxies ($>50$) in individual bins. 

We find that the mass-weighed sizes in the SPIDER cluster sample are on average $\sim$ 12\% smaller than the $r$-band sizes with a median $\langle \log(R_{e-circ,mass}/R_{e-circ}) \rangle = -0.055$ ($\sim$13 \% for the SPIDER sample, $\langle \log(R_{e-circ,mass}/R_{e-circ}) \rangle = -0.062$), consistent with previous result \citep{Szomoruetal2013}.

In addition, we find that there is an intriguing offset between the median size ratio of the cluster sample and the SPIDER cluster sample, with a difference of $\langle \log(R_{e-circ,mass,1.39} / R_{e-circ,1.39}) - \log(R_{e-circ,mass,0}/R_{e-circ,0}) \rangle = -0.18$ ($-0.17$ for the SPIDER sample).

A possible issue is the effect of recently quenched galaxies on the size evolution, i.e. the progenitor bias.  It has been shown to have a non-negligible effect on inferred size evolution, and is able to explain part if not all of the observed size evolution \citep[e.g.][]{Sagliaetal2010, Valentinuzzietal2010b, Carolloetal2013, Poggiantietal2013, Jorgensenetal2014, Bellietal2015, Keatingetal2015}.  The effect on the evolution of the size ratio is however unclear, as the newly quenched galaxies may have a range of $M_{*}/L$ gradients that depends on the quenching mechanism involved.

Using age measurements from \citet{LaBarberaetal2010c}, we try to correct the progenitor bias in the size ratio of the SPIDER cluster sample.  An age cut is applied to the SPIDER cluster sample to remove galaxies that are younger than 8.98 Gyr, the time duration from $z\sim1.39$ to $z\sim0$.  The result is shown as a light brown line and wheat band in Figure~\ref{fig_SPIDER} in the mass range $10.2 \leq \log(M_{*}/M_\odot) \leq 11.6$.  Although some changes can be seen, the size ratios of the progenitor bias corrected sample are in general consistent with the SPIDER cluster sample, with median $\langle \log(R_{e-circ,mass}/R_{e-circ}) \rangle = -0.065$.  The median logarithmic size ratios in each bin between the two are within $\pm 0.05$.  The offset between the median size ratio of the cluster sample and the progenitor bias corrected SPIDER cluster sample is $-0.16$.  Hence, the progenitor bias alone does not explain the observed offset.

The smaller size ratio at $z\sim1.39$ suggests that the $M_{*}/L$ gradient is larger (i.e. steeper) in these high redshift passive cluster galaxies compared to the local ones.  This implies an evolution of $M_{*}/L$ gradient with redshift, consistent with our finding that the colour gradient at high redshift cluster passive galaxies is much steeper than the local ones.  In the next section, we try to explore the origin of the colour (and $M_{*}/L$) gradient and the physical processes for the evolution of these passive galaxies.

\subsection{Origin and evolution of colour gradients with redshift}
\label{sec:Evolution of the colour gradients over redshift}
In Section~\ref{sec:Colour Gradients in the Early-type Cluster Galaxies} we have shown that the median colour gradient in our sample is $\sim2$ times steeper than the measured local $(U-R)$ gradient $\nabla_{U-R}$ from \citet{Wuetal2005}.  

The origin of the colour gradients is directly related to how the stellar population in galaxies assembled and evolved.  It is however challenging to segregate the impact of age or metallicity using a small sample of galaxies due to degeneracies between colour, age and metallicity.  In previous studies, colour gradients are mostly interpreted as either age gradients ($\nabla_{age} = d \log(age) / d \log(a)$) at fixed metallicity or metallicity gradients ($\nabla_{Z} = d \log(Z) / d \log(a)$) at fixed age. Some works in clusters at $z\sim0.4$ \citep{Sagliaetal2000} and local clusters \citep[e.g.][]{TamuraOhta2003} showed however that the colour gradients may be preferentially produced by radial variation in metallicity rather than age.  The age gradients in local passive galaxies are consistent with 0 (or slightly positive), while the average metallicity gradient is found to be of $\nabla_{Z} \approx -0.1 $ to $ -0.3$ \citep[see also][]{Mehlertetal2003, Wuetal2005, LaBarberaetal2009b}.  This result is also supported by recent studies with integral field spectroscopy \citep[e.g][]{Kuntschneretal2010, GonzalezDelgadoetal2014, Oliva-Altamiranoetal2015, Wilkinsonetal2015}.

Our $(U-R)$ and $(g-r)$ colour gradient measurements alone unfortunately do not allow us to break the age-metallicity degeneracy.  Nevertheless, with additional colour information at $z\sim0$, by studying the evolution of the colour gradient with redshift we can shed light on the origin of the colour gradients in our sample of early type galaxies in clusters.

\subsubsection{Methodology}
\label{sec:Methodology}
We investigate quantitatively the evolution of colour gradients by modeling them in our cluster sample under different assumptions of the radial variation of stellar population properties.  Simply put, we would like to evolve the observed $z_{850} - H_{160}$ colour gradients $\nabla_{z_{850} - H_{160}}$ in the cluster sample to see under which conditions in age and metallicity gradient will they match the observed $(U-R)$ gradient at $z\sim0$.

For simplicity, here we assume the stellar populations in the passive galaxies are coeval and chemically homogeneous in the regions we considered, hence they can be described by simple stellar populations (SSP) models.  We use the models of \citet{BruzualCharlot2003} (hereafter BC03) and adopt a Chabrier IMF.  The BC03 distribution provides SSP models with metallicities $Z = [0.0001, 0.0004, 0.004, 0.008, 0.02, 0.05]$ from $t = 0$ to the age of the Universe in unequally spaced time steps.  The results below do not strongly dependent on the choice of IMF, since the $U-R$ broad band optical colours under different IMFs (e.g. Chabrier vs. Salpeter) are in reasonable agreement with each other.  In Appendix~\ref{sec:Evolution of colour gradients using exponentially declining tau models} we show that adopting exponentially declining $\tau$-models for this analysis (instead of SSPs), does not change the results.

\begin{figure}
\centering
  \includegraphics[scale=0.42]{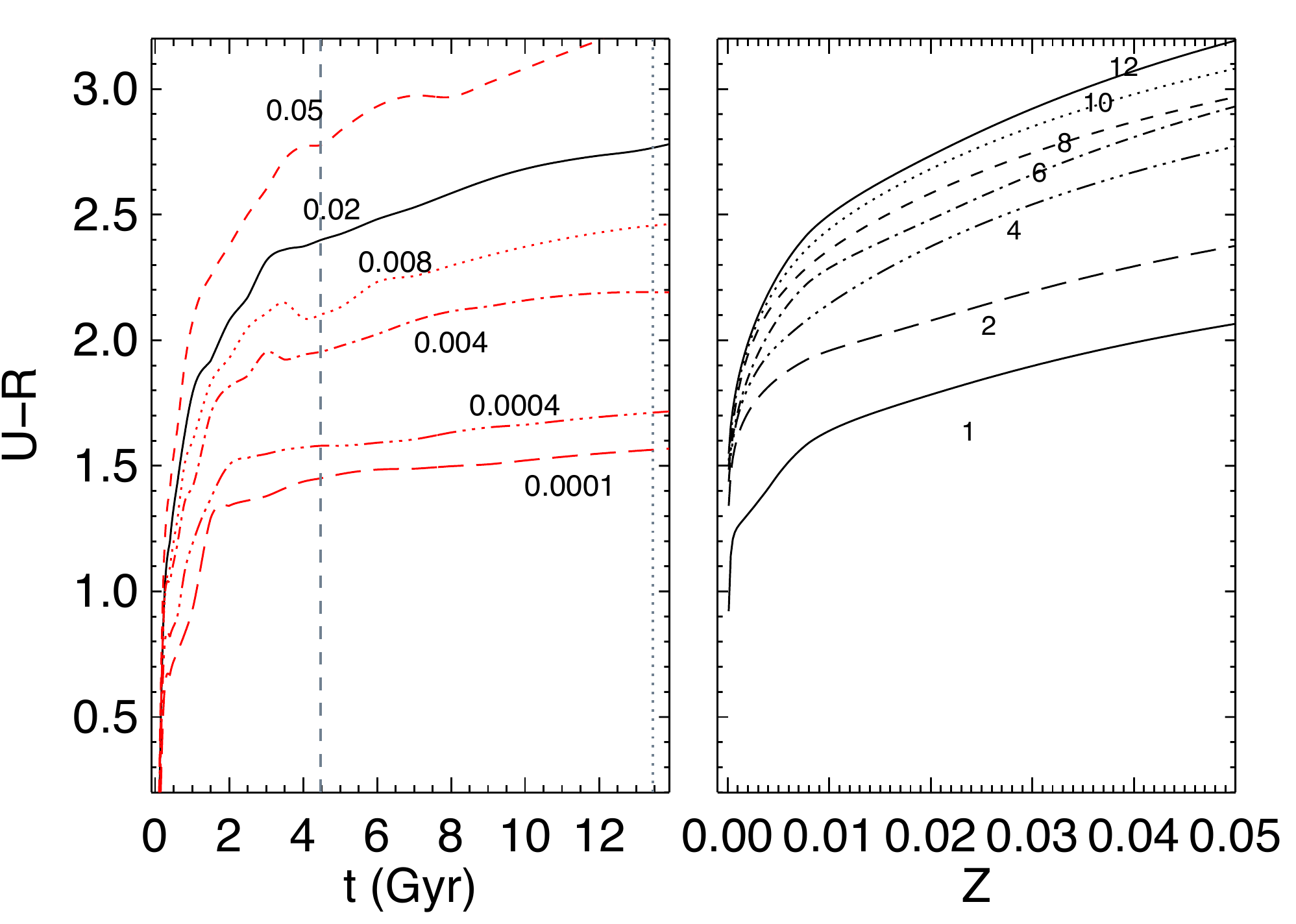}
  \caption{Rest-frame $U-R$ colour of stellar populations with different ages and metallicities.  Left panel: $U-R$ colour-age relations. The black line shows the stellar populations with solar metallicity ($Z=0.02$).  The red lines show populations with different metallicities ($Z=0.0001,0.0004,0.004,0.008,0.05$) as indicated. The grey dotted line shows the current age of the Universe (13.45 Gyr) with our choice of cosmology and the grey dashed line shows the age of Universe at redshift 1.39 (4.465 Gyr).  Right panel: $U-R$ colour-metallicity relations. The black lines shows populations with different ages (in Gyr) as indicated.}
  \label{fig_URcolourgrid}
\end{figure}

We compute the rest-frame $U-R$ colour for individual SSP models with different ages and metallicities by convolving the model SEDs with the $U$ and $R$ filters.  The colours are then interpolated with a cubic spline to obtain an equally spaced colour grid in age and metallicity.  Figure~\ref{fig_URcolourgrid} shows the $U-R$ colour at different ages (left, i.e. the colour-age relations) and metallicities (right, i.e. the colour-metallicity relations) as an example.  Using the same method, we compute a $z_{850} - H_{160}$ colour grid by redshifting the SSP models to $z=1.39$.

Similar to Section~\ref{sec:Colour Gradients in the Early-type Cluster Galaxies}, we have also repeated the above analysis using the $g-r$ colour gradients and find results which are completely consistent (Appendix~\ref{sec:The g-r color gradients and the evolution with redshift}). Since the $g-r$ colour has less dynamic range in the evolution, in the following we will mainly discuss the result of the $U-R$ colour gradients.

To simplify the modeling process, we analyse the evolution of colour at two radii, $0.5 a_e$ and $2 a_e$, representing the inner region and outer region of the galaxy respectively.  In similar studies \citep[e.g.][]{Sagliaetal2000} more central regions are used instead ($0.1-0.2 R_{e}$), but this is not possible at this redshift due to limited resolution.  Nevertheless, our choice of radial range is sufficient for the purpose as the colour gradients are well-fitted by a linear relation in logarithmic radius (see Figure~\ref{fig_colourgradexample}).

Because of the age-metallicity degeneracy, we consider several scenarios with additional assumptions in the age or metallicity gradients.  In this study we explore three possibilities (cases) to interpret the colour gradient evolution:

\begin{itemize}
   \item \textbf{Case I - Pure age-driven gradient evolution} -- In this case we explore the possibility of using a single age gradient to interpret the evolution of colour gradients.  The inner and outer regions are assumed to have identical metallicities (i.e. flat metallicity gradients $\nabla_{Z} = 0$).  Assuming a certain metallicity for the inner regions (and equivalently the outer regions), we derive ages of the stellar population of the inner and outer regions in each galaxy respectively through matching the observed $z_{850} - H_{160}$ colours to the derived $z_{850} - H_{160}$ SSP colours.
   
   \item \textbf{Case II - Pure metallicity-driven gradient evolution} -- In this case we assume that the stellar population in the inner and outer region are coeval (i.e. flat age gradients $\nabla_{age} = 0$).  Assuming a certain metallicity for the inner regions, we derive the inner ages in each galaxy using the same method as I.  The same age is then applied to the outer regions.  With ages and $z_{850} - H_{160}$ colours, metallicities in the outer regions are then derived using the colour-metallicity relations.
   
   \item \textbf{Case III - Age-driven gradient evolution with an assumed metallicity gradient} --  Same as case I,  but assume a fixed metallicity gradient with $\nabla_Z  = -0.2$, which is the mean value observed in local passive galaxies \citep[e.g.][]{TamuraOhta2003, Wuetal2005, Broughetal2007, Redaetal2007} as well as in recent simulations \citep[e.g.][]{Hirschmannetal2015}.  Again assuming a certain metallicity for the inner regions, we derive the inner ages from the $z_{850} - H_{160}$ colour.  The metallicity in the outer regions is then computed according to the assumed gradient; outer ages are then derived with the computed metallicity.
\end{itemize}

In summary, in each case we obtain the ages and metallicities in the inner and outer region of each cluster galaxy.  We then evolve the corresponding SSPs in both regions to $z=0$, and compute the corresponding local $(U-R)$ colour gradients.  We also compute the rest-frame $(U-R)$ gradient for the high-redshift sample for comparison.  For each of the three cases above, we test three scenarios with different assumed metallicity for the inner regions, sub-solar, solar and super-solar ($Z = 0.008, 0.02, 0.05$ or equivalently $[Fe/H] = -0.33, 0.09, 0.56$) \citep{BruzualCharlot2003}.  Assuming metallicities with $Z < 0.008$ or $Z > 0.05$ is unphysical for most galaxies in the cluster sample.

Under different assumed metallicity for the inner region, occasionally the age (or metallicity for case II) determination for some galaxies results in an unphysical age (or metallicity).  With our choice of cosmological parameters, the age of universe at $z=1.39$ is 4.465 Gyr.  Deduced ages that are too old ($> 4.465$ Gyr for $z=1.39$ or $> 13.45$ Gyr for $z=0$ within $1\sigma$ uncertainty) are rejected to avoid drawing incorrect conclusions.  Galaxies that are rejected may simply be unphysical to be modelled with particular metallicity (see, for example in Figure~\ref{fig_URcolourgrid},  a galaxy with $U-R > 2.15$ at $z=1.39$ will result in an unphysical age if one assumes $Z=0.008$) or have a more complicated star formation history, which cannot be well-represented by SSPs.

\begin{figure}
  \includegraphics[scale=0.46]{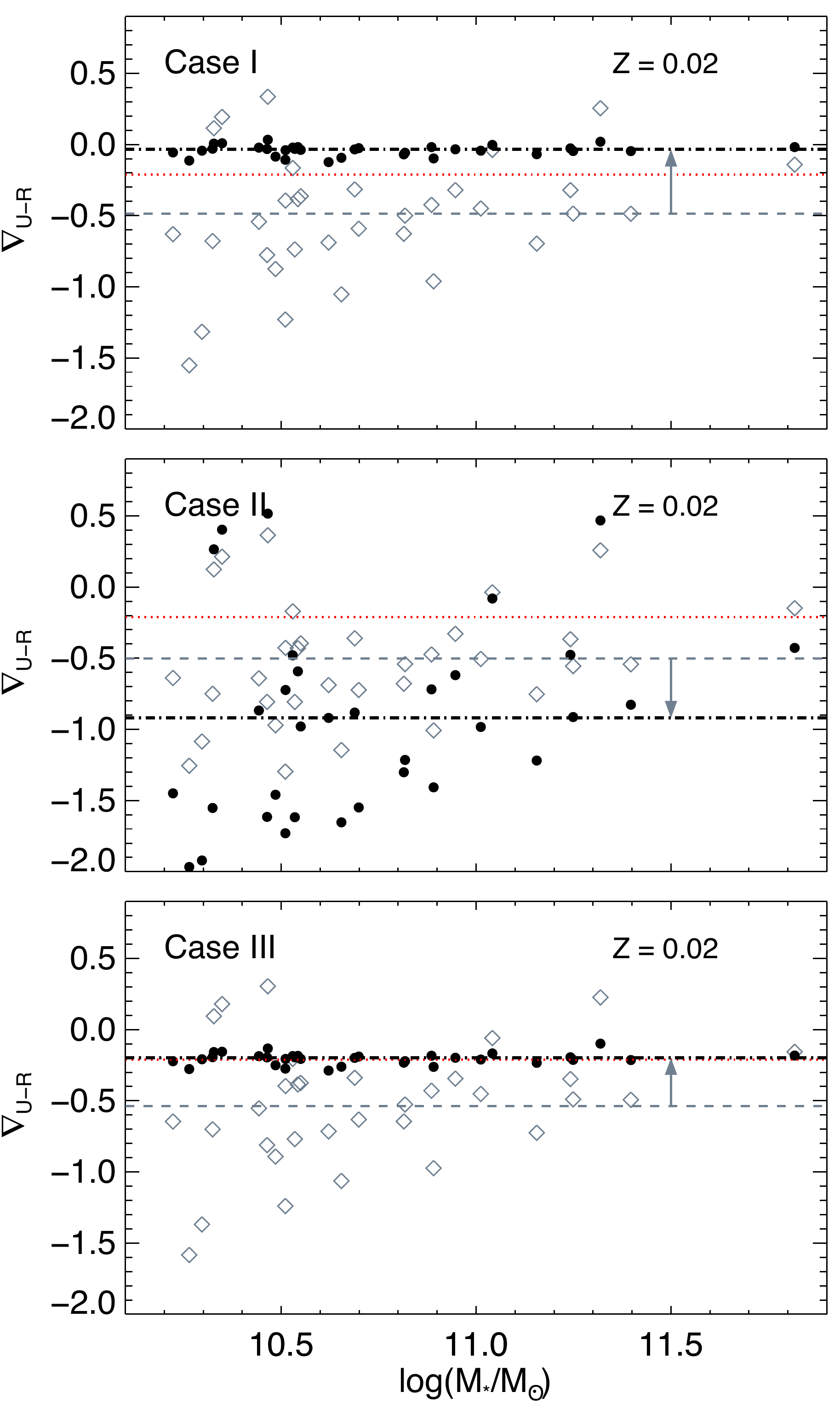}
  \caption{Evolution of colour gradient under different assumptions in the age / metallicity gradient. From top to bottom:  Case I -- Pure age-driven gradient evolution, Case II -- Pure metallicity-driven gradient evolution and Case III -- Age-driven gradient evolution with assumed metallicity gradient. Only the solar ($Z=0.02$) metallicity scenario in each case is shown.  Grey diamonds correspond to the $(U-R)$ gradient at redshift 1.39, with the median plotted as the grey dashed line.  Black circles indicate the predicted $(U-R)$ gradient at redshift 0 of the same galaxy, and the black dot-dashed line indicate the median.  Their masses remain unchanged as we do not consider any mass growth over the period.  The grey arrow in each panel shows the direction of evolution of the median gradient.  The red dotted line corresponds to the observed $(U-R)$ gradient at redshift 0 by \citet{Wuetal2005}.}
  \label{fig_gradevo}
\end{figure}

\subsubsection{Case I -- Pure age-driven gradient evolution}
In the top panel of Figure~\ref{fig_gradevo}, we show the evolution of the rest-frame $(U-R)$ colour gradient from $z=1.39$ to $z=0$ under the assumption of pure age gradient.  We show the scenario with assumed solar metallicity ($Z=0.02$) in the inner region.  We find that although the gradients evolve in the correct direction, the median gradient of the evolved sample is too shallow. The result is almost identical if we assume sub-solar or super-solar metallicity for the inner region instead, with median evolved colour gradient of $(Z, \nabla_{U-R}) = (0.008, -0.036), (0.02, -0.034), (0.05, -0.037)$.  Under the assumption of sub-solar metallicity, 19 out of 36 galaxies have a physical age.  On the other hand, most of the galaxies are retained if we assume a solar (33 out of 36) or super-solar metallicity (36 out of 36).  We conclude that in the reasonable range of metallicity that we covered, a pure age-driven gradient is not able to match the observed evolution of colour gradient.

The reason behind the rapid evolution is the flattening of the SSP colour-age relation over time.  Since we assume identical metallicities for both inner and outer regions, the inner and outer region of an individual galaxy lie on the same colour-age relation in Figure~\ref{fig_URcolourgrid}.  Take the solar metallicity $Z=0.02$ case (black solid line) as an example, the $U-R$ colour increases sharply from 0 to 4 Gyr but flattens after, hence the $(U-R)$ gradient evolves to almost zero at redshift 0.

\subsubsection{Case II -- Pure metallicity-driven gradient evolution}
Instead of using a flat metallicity gradient as case I, the middle panel of Figure~\ref{fig_gradevo} shows the evolution of the $(U-R)$ gradient under the assumption of pure metallicity-driven gradient, or in other words, a flat age gradient $\nabla_{age} = 0$.  Again, we show the scenario with assumed solar metallicity ($Z=0.02$) in the inner region.  Similar to case I, galaxies that have unphysical ages / metallicities are discarded.  21, 33 and 31 out of 36 galaxies are retained in each metallicity scenario ($Z=0.008,0.02,0.05$) respectively. From Figure~\ref{fig_gradevo}, we can see that the median gradient of the evolved sample is even steeper compared to the one at redshift 1.39.  Hence, it is clear that a pure metallicity-driven gradient fails to reproduce the observed gradient.  The median gradients of the evolved sample in the three metallicity scenarios are $(Z, \nabla_{U-R}) = (0.008, -0.691), (0.02, -0.920), (0.05, -0.900)$.

The evolution can be explained using the colour-metallicity relations in the right panel of Figure~\ref{fig_URcolourgrid}.  As the population ages, the $U-R$ colour-metallicity relation steepens (for example, from 2 Gyr to 4 Gyr), which causes the colour gradient to become more negative. The steepening stops at around $\sim4$ Gyr, thus the colour gradient after then remains unchanged.  For the solar metallicity scenario, the median metallicity gradient and $1\sigma$ scatter we found is $\langle \nabla_{Z} \rangle \sim -1.32 \pm 1.22$ at $z=1.39$, much higher than that observed in local galaxies.  Of course in reality metallicity in individual galaxies differs, but mixing galaxies with different metallicity within our metallicity range would not change this conclusion.

\subsubsection{Case III -- Age-driven gradient evolution with assumed metallicity gradient}
The bottom panel of Figure~\ref{fig_gradevo} shows the evolution of the $(U-R)$ gradient with a metallicity gradient as observed in local passive galaxies: $\nabla_{Z} = -0.2$.  Similar to case I,  galaxies with unphysical ages in the inner or outer regions are rejected. 19, 33 and 36 out of 36 galaxies are retained in each metallicity scenario respectively.  The solar metallicity scenario works reasonably well for the majority of the sample with evolved median gradient of $(Z, \nabla_{U-R}) = (0.02, -0.198)$, which is in close agreement with the observed value in the local universe by \citet{Wuetal2005}.  Despite a number of objects have to be discarded due to unphysical age, the median gradient as well as the individual gradients of the evolved samples in the sub-solar metallicity scenario $(Z, \nabla_{U-R}) = (0.008, -0.180)$ is also close to but slightly smaller than the observed local value.  Assuming super-solar metallicity for the inner regions on the other hand, predicts gradients that are slightly too steep $(Z, \nabla_{U-R}) = (0.05, -0.232)$.

Besides the median values, the scatter in the evolved colour gradients is also in excellent agreement to the local value by \citet{Wuetal2005} ($\nabla_{U-R} = -0.21 \pm 0.04$).  For example, for the solar metallicity scenario the scatter reduces from 0.37 at $z\sim1.39$ to 0.06 at $z=0$.

\begin{figure*}
  \includegraphics[scale=0.49]{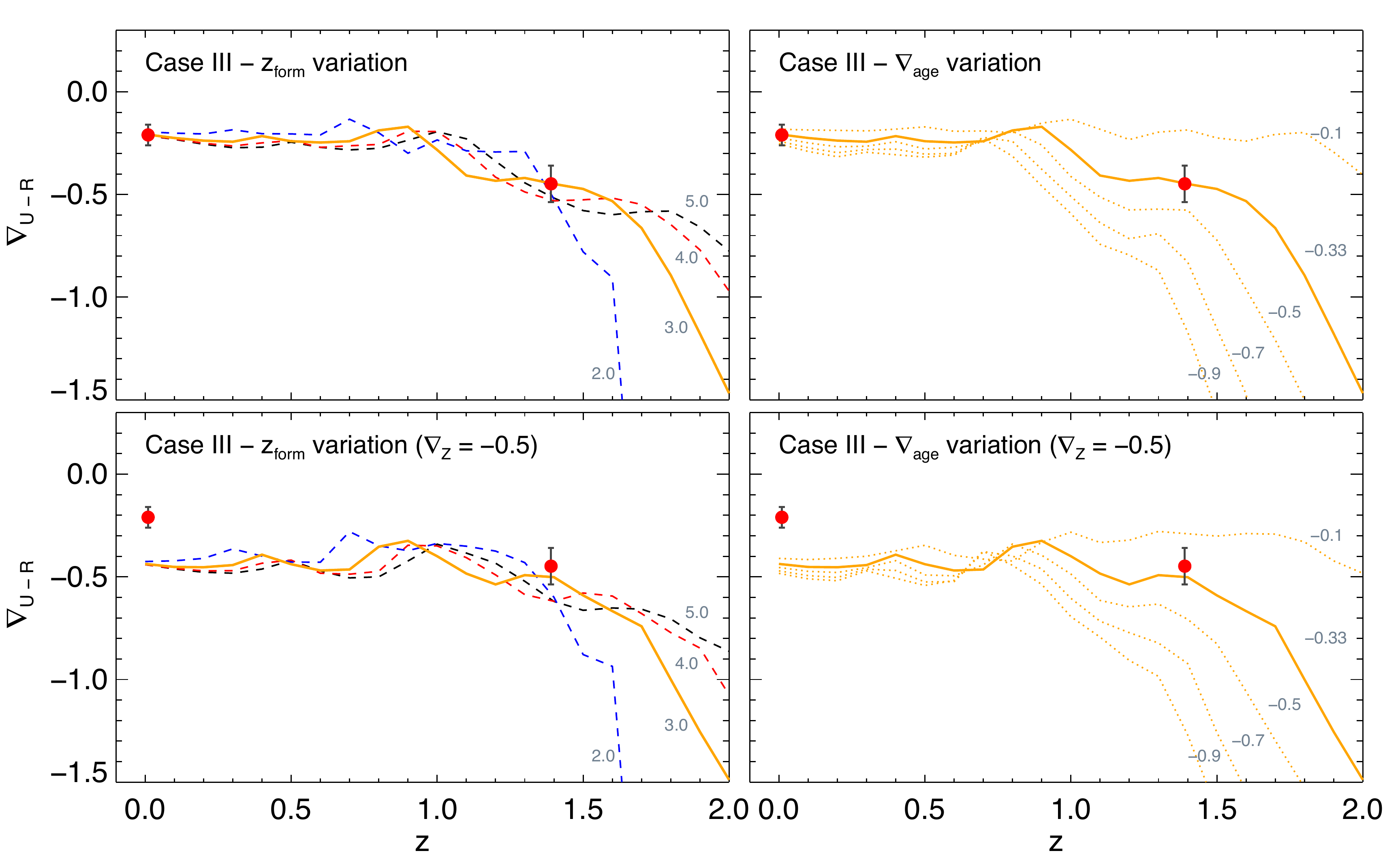}
  \caption{Evolution of colour gradient over redshift in case III (age and metallicity gradient) with assumed solar metallicity $Z=0.02$. The inner region is assumed to have a formation redshift $z_{form} = 3.0$. The initial age gradient at $z=1.39$ is $\nabla_{age} = -0.33$, and the assumed metallicity gradient is $\nabla_{Z} = -0.2$.  Top left: variation in formation redshift $z_{form} = 2.0,4.0,5.0$ as indicated by blue, red and black dashed lines respectively. Top right: variation in age gradients at $z=1.39$. The dotted lines show the evolution with different initial age gradient as indicated ($\nabla_{age} = -0.1, -0.5, -0.7, -0.9$). Bottom left and right: same as top left and right but with an assumed metallicity gradient of $\nabla_{Z} = -0.5$ as predicted by the monolithic collapse formation scenario. Red circles correspond to the median $z_{850}-H_{160}$ gradient of our sample at redshift 1.39, and the observed $(U-R)$ gradient at redshift 0 by \citet{Wuetal2005}. The error bars show the uncertainty of the median.} 
  \label{fig_gradevoT2}
\end{figure*}

\subsubsection{Implications and limitations}
\label{sec:Implications and Limitations}
From the above case study, we find that the presence of an age gradient is a necessary condition for the evolution of the colour gradient, and with metallicity gradient they can sufficiently reproduce the magnitude of the evolution of the colour gradient from $z=1.39$ to $z=0$.

An age-driven gradient evolution with a metallicity gradient close to the local value is the most probable scenario, as it can well-reproduce the observed evolution of the colour gradients over redshift in both median and scatter.  Below we try to understand why this is the case.  Among the three metallicity scenarios, the one with solar metallicity seems to best match the evolution of colour gradients for most galaxies. In this scenario we find a median age gradient and $1\sigma$ scatter of $\langle \nabla_{age} \rangle = -0.33 \pm 0.37$ at $z=1.39$ (i.e. a median age difference $\sim 1.4$ Gyr between the inner and outer regions).

Figure \ref{fig_gradevoT2} shows the evolution of the colour gradient in this case (case III), assuming a formation redshift of the inner regions of $z_{form} = 3.0$ and an age gradient of $\langle \nabla_{age} \rangle = -0.33$ at $z=1.39$.  The top left panel shows the change in the evolution for different formation redshifts $z_{form} = 2.0, 4.0, 5.0$, selected to be consistent with findings in recent spectroscopic studies at similar redshift \citep{Gargiuloetal2012, Bellietal2015}.  The net evolution from $z\sim1.39$ to $z\sim0$ is clearly insensitive to the formation redshift. On the other hand, the path of evolution depends largely on the age gradient; from the top right panel we show that with different initial age gradient at $z=1.39$, a large range of colour gradients at $z=1.39$ can reach similar value at $z\sim0$.  In other words, if the colour gradients in high redshift passive galaxies are mainly due to radial variation in age, this assumption would be able to match the evolution of colour gradient for most galaxies.

This is in agreement with \citet{Gargiuloetal2012}, who investigated the origin of the colour gradient on a sample of early-type galaxies at $0 < z < 1.9$ with spatially resolved colour and global SED fitting.  They found that the colour gradients of $\sim50\%$ of their sample can be reproduced with pure age gradients, while invoking pure metallicity gradients can only explain a small subset of their sample. In addition, extremely steep metallicity gradients are required that are only marginally comparable with those observed in the local Universe.  A similar recent study by \citet{DeProprisetal2015} studied the ratio of galaxy sizes in two bands (as a proxy of the colour gradient) in red sequence galaxies in four clusters with $<z> \sim 1.25$ (including this cluster) also found an indication of negative colour gradients, which they also attribute to due to the presence of age gradients. 

Our result is also not inconsistent with studies on local and intermediate redshift passive galaxies which suggest colour gradients are mainly due to metallicity gradients \citep[e.g.][]{Sagliaetal2000, TamuraOhta2003}.  For example, \citet{Kuntschneretal2010} found a mean $\nabla_{Z} =  -0.25 \pm 0.11$ and mean $\nabla_{age} = 0.02 \pm 0.13$ for galaxies with age $> 8$ Gyr.  Nevertheless, the age gradient (or its presence) is very difficult to constrain in local passive galaxies.  As \citet{Gargiuloetal2012} pointed out, the effect of the age difference in the inner and outer regions is much more enhanced when the stellar population is young (i.e. at high redshift).  Indeed, the age gradient flattens quickly over redshift. For example, with a median age gradient of $\langle \nabla_{age} \rangle = -0.33 \pm 0.37$ at $z=1.39$ (from our best scenario); assuming passive evolution this corresponds to a median age gradient of $\langle \nabla_{age} \rangle = -0.05 \pm 0.06$ at $z=0$, which is consistent with a flat age gradient.

Given the assumption that the cluster passive galaxies at $z=1.39$ have the same metallicity gradient as the local ones, it is implied that the evolution of colour gradients from $z\sim1.4$ to $0$ can be explained simply through passive evolution.  This is consistent with luminosity function studies in clusters \citep[e.g.][]{Andreonetal2008, DeProprisetal2013}.

\subsubsection{Physical processes responsible for the evolution of size and colour gradient}
While passive evolution is a very tempting conclusion, it alone cannot explain the observed size evolution in clusters over redshift.  In Figure \ref{fig_lightmasssize_mod} we have shown that the cluster passive galaxies are $\sim40\%$ smaller than their local counterparts.  Similarly, a large number of previous studies have confirmed that the sizes of passive galaxies in high redshift clusters are smaller than those in the local Universe \citep[e.g.][]{Retturaetal2010, Strazzulloetal2010, Papovichetal2012, Strazzulloetal2013}, although a part of this size evolution \citep[e.g.][]{Sagliaetal2000} or even all \citep{Jorgensenetal2014} may be due to progenitor bias.  If we assume this observed size evolution is genuine, in the sense that it is not completely an effect of progenitor bias, additional physical processes must be in place over redshift to increase the size of the population but not significantly their stellar mass, and at the same time cannot severely disrupt the existing stellar population gradients.

The ``puffing-up'' scenario (or adiabatic expansion) is one of the candidates to explain the size evolution of passive galaxies \citep{Fanetal2008,Fanetal2010}.  While it may work for increasing the size, it is yet unclear whether it can sufficiently explain the observed evolution of colour gradients.  Further detailed investigation with an accurate model is required to test this scenario, but current models \citep[e.g.][]{Fanetal2008} are much simplified.

Minor mergers, on the other hand, seem to be a viable scenario as the effects are primarily on the outer part of the galaxies. \citet{Hilzetal2013} showed from n-body/SPH simulations that for minor (with a mass ratio 1:10) or intermediate (1:5) dry mergers, the inner region of the galaxy remains almost unchanged and the accreted mass assembles predominately in the outer part of the galaxy.  This is also seen in the cosmological simulations of \citet{RodriguezGomezetal2015} where the accreted stars from mergers with lower mass ratio (i.e. merging with smaller galaxies) dominate at outer radius.  Hence, the inner stellar populations of the galaxy can age through passive evolution without major disturbance. The negative age gradients we find here seem to be consistent with this picture given that the minor mergers are dry and the stars accreted are relatively young.

The minor merger scenario has been known to be a viable mechanism in the field.  It is consistent with the observed inside-out growth as seen from the evolution of the stellar mass surface density profiles of passive field galaxies over redshift \citep[e.g.][]{vanDokkumetal2010, Pateletal2013}.  Nevertheless, traditionally this type of merger activity is believed to be suppressed in virialized clusters because of the high velocity dispersion, resulting in high relative velocities between cluster members \citep[e.g.][]{Conroyetal2007, Lotzetal2013}.  The exception being mergers of satellite galaxies onto the BCG due to dynamical friction \citep[e.g][]{Burkeetal2013, Burkeetal2015}, which contribute to the mass growth of BCG and the intracluster light (ICL).  On the other hand, merger events are thought to be very common in galaxy groups where the velocity dispersion is lower, or when the cluster is still assembling.  Recent works have found that sizes of massive passive galaxies are larger in clusters compared to the field at high redshift  \citep[e.g.][]{Cooperetal2012, Zirmetal2012, Papovichetal2012, Lanietal2013, Strazzulloetal2013, Jorgensenetal2013, Delayeetal2014}, but not in the local Universe \citep[e.g.][]{Huertascompanyetal2013, Cappellari2013}, suggesting an accelerated size evolution in high density environments.  This accelerated size evolution is probably due to an enhanced rate of mergers during infall of groups, when the cluster is being assembled \citep{Lotzetal2013, Delayeetal2014, Newmanetal2014}.  While this explains the elevated sizes in cluster compared to the field at fixed redshift, it does not provide an explanation to the subsequent size evolution observed in clusters over redshift $z <1.5$.

Despite the suppression of galaxy merging activity in clusters based on relative velocity arguments, it is clear that clusters themselves and their associated dark matter halos continue to grow by accreting galaxy groups. It is possible that some accretion can still happen to the cluster galaxy population during infall of these group-scale structures.  Simulations of mergers with cluster mass halos have shown that the accreted mass resides mainly in the satellite galaxies and the ICL, but only mildly in the BCG \citep[e.g.][]{Whiteetal2007, Brownetal2008}.  Recent simulations also demonstrate that the size of clusters members can grow significantly via major and minor mergers and the frequency of mergers is sufficient to explain the observed size growth in clusters since $z\sim2$ \citep{Laporteetal2013}.  In addition, the observed merger rate in the cluster galaxy population, excluding the BCG, is poorly constrained.  If this is possible, this kind of gradual mass growth is able to explain at the same time the evolution of both size and colour gradient in clusters.

If we take the size evolution into account and assume the mass growth takes place predominately at the outskirts, the stellar population we considered here in the colour gradient ($a < 3.5 a_{e}$) will correspond to the central population at $R < 1.5 - 2 R_{e-circ}$ in local cluster galaxies.  If the evolution is primarily merger or accretion driven as we suggest above, one would expect that the outer stellar population depends on past merger activity.

Interestingly, there has been some evidence indicating changes in stellar population properties at the outer region of local massive passive cluster galaxies \citep[e.g. NGC 4889 in Coma cluster][]{Coccatoetal2010} as well as in field ellipticals \citep[e.g.][]{Puetal2010}.  More recent studies have extended the age and metallicity measurements to large radii ($\sim 8-10 R_e$, for example in \citealt{LaBarberaetal2012}) and revealed that the outer age or metallicity gradients at $\gtrsim1-2 R_e$ are distinct from those in the inner region \citep[e.g.][]{Greeneetal2013, Pastorelloetal2014, Raskuttietal2014}.  These changes in stellar population gradients are commonly interpreted to be result of mergers.  Nevertheless, these changes can also come from recent quenched galaxies that underwent minor mergers before infalling to the cluster.  A progenitor biased corrected sample is needed to address this issue.

\subsubsection{Monolithic collapse model}
Traditional monolithic models predict very high metallicity gradients in passive galaxies, around $\nabla_Z \sim -0.5$, and a flat if not slightly positive age gradients \citep[e.g.][]{Larson1974, Carlberg1984}.  The fact that we find that on average a negative age gradient is necessary to explain the evolution of the colour gradient is hence inconsistent with the monolithic collapse scenario.  This conclusion is independent of the metallicity gradient we assumed; in case III we assume the metallicity gradient to be $\nabla_{Z} = -0.2$ as observed in local passive galaxies, but using a steeper metallicity gradient would not work.  The bottom two panels in Figure~\ref{fig_gradevoT2} show the effect of assuming a metallicity gradient $\nabla_Z = -0.5$ as predicted by traditional monolithic models.  While the $U-R$ colour gradient at redshift 1.39 is in reasonable agreement with the observed value,  the evolved gradient is too steep (i.e. insufficient evolution) compared to the observed value at redshift 0.

We also compare our observed $z_{850}-H_{160}$ colour gradients at redshift 1.39 with recent simulations based on a revised version of the monolithic model by \citet{Pipinoetal2008a, Pipinoetal2010}.  With semi-cosmological initial conditions, they are able to match the age gradients and metallicity gradients observed in local passive galaxies.  \citet{Tortoraetal2013} computed the colour gradients from the \citet{Pipinoetal2010} simulations using BC03 SSP models, for which we can directly use for the comparison.  Among the four models presented in their work (E1, E2, E3, E4), we find that our observed median colour gradient seems to be in reasonable agreement with models that predict steep metallicity gradient (E2, $\nabla_Z \sim -0.35$ and E4, $\nabla_Z \sim -0.45$) and nearly flat age gradient at $z=0$.  The comparison with the colour gradients in \citet{Gargiuloetal2011, Gargiuloetal2012} at redshift $1<z<2$ also gives a similar result.  However, the local colour gradient at $z \sim 0$, as well as the $F606W - F850LP$ gradients of high-redshift galaxies in \citet{Gargiuloetal2011} favours the other two models (E1, E3) instead, so there is not a single model that can explain the evolution of colour gradients.  Hence, our result cannot be reproduced by the revised monolithic collapse model.

\subsubsection{Effect of dust obscuration}
A complication that we have not considered above is the effect of dust obscuration.  The colour gradient can be affected by the radial variation of dust content.  For local passive galaxies, \citet{WiseSilva1996} pointed out that their colour gradients can be reproduced by a dust gradient, albeit with much higher dust masses than observed \citep{Sagliaetal2000}.  Hence the colour gradient in these galaxies should be driven from the variation of the stellar population.  Nevertheless, the amount of dust can vary with radius;  it is not uncommon to find dust at the centre, for example in \citet{Laueretal2005}, central dust is visible in almost half of the local passive sample.

At high redshift, measuring the radial variation of dust content is even more difficult due to the compact nature of passive galaxies and limited angular resolution.  Several studies suggest that although the effect of dust cannot be completely neglected, it plays only a minor role in driving colour gradients.  \citet{Bellietal2015} showed that their early-type sample at $1.0 < z < 1.6$ have little to no dust extinction \citep[see also][]{Mendeletal2015}.  For the radial variation of dust, \citet{Guoetal2011} demonstrated from spatially resolved annular SED fitting that, for their sample of six $z \sim2$ galaxies dust partly contributes to the observed colour gradients, the inferred dust gradient and global extinction have a value of $d E(B-V)/ d \log(R) \sim -0.07$ and $\langle E(B-V) \rangle \sim0.1$.  On the other hand, among the 11 early-type galaxies at $1.0 < z < 1.9$ in \citet{Gargiuloetal2012}, half of the sample have no dust extinction (i.e. $A_V = 0$) from global SED fit and for most galaxies the main driver of the colour gradients is certainly not the radial variation of dust.

It is not possible to derive reliable dust gradients from our multi-band photometry data.  Hence, for completeness we test whether the effect of dust would affect our conclusion.   Assuming dust mainly affects the central region as in local passive galaxies, we deredden the $z_{850} - H_{160}$ colour in the inner region by a certain magnitude but leaving the outer part unchanged to reduce the observed gradient, then recompute all the evolution under different assumptions. We find that on average a decrease of 0.14 mag in the $z_{850} - H_{160}$ colour at the inner region ($\sim 0.25$ mag at $0.1 a_{e}$) will remove the observed evolution, i.e. the difference of the observed median colour gradient at $z=1.39$ with local passive galaxies.  Assuming the extinction curve by \citet{Calzettietal2000}, this corresponds to a gradient of $d A_V / d \log(a) \sim -0.40$ or $d E(B-V) / d \log(a) \sim -0.10$.  Hence, our conclusion is robust if the dust gradient is less steep than this value.  If the genuine colour gradient is shallower at $z=1.39$, the evolution will be best explained with a shallower age gradient (as seen from the top right panel of Figure \ref{fig_gradevoT2}), and will not change any of our conclusions.  Nevertheless, although unlikely, we cannot rule out the possibility that the observed colour gradients are driven by a large amount of dust located in the central region.

\section{Summary and Conclusions}
\label{sec:Conclusion}
We have presented the structural parameters, resolved stellar mass distribution and colour gradient of a sample of 36 passive galaxies in the red sequence of the cluster XMMUJ2235-2557 at redshift $z\sim1.39$.  With HST/ACS and WFC3 data we derive light-weighted structural parameters independently in five different bands ($i_{775}, z_{850}, Y_{105}, J_{125}, H_{160}$) through 2D S\'ersic fitting.  We compute 1D $z_{850} - H_{160}$ colour profiles for individual galaxies and fit logarithmic gradients $\nabla_{z_{850}-H_{160}}$ in the range of PSF HWHM $< a < 3.5 a_e$ to derive colour gradients.  In addition, we derive resolved stellar mass surface density maps for individual galaxies with an empirical $M_{*}/L$-colour relation and the $z_{850}$ and $H_{160}$ images.  Mass-weighted structural parameters are derived from the resolved stellar mass surface density map.  We find the following:

\begin{enumerate}[--]
  \item From our multi-band light-weighted structural parameter measurements, the passive galaxies in this cluster show a reduction of $\sim20\%$ in sizes from $i_{775}$ to $H_{160}$, consistent with the wavelength-dependence found in local passive galaxies.
  \item The $H_{160}$ band sizes in this cluster are on average $\sim40\%$ smaller than that expected from the local mass-size relation by \citet{Bernardietal2012} at the same rest-wavelength, with a median of $\langle \log(R_{e-circ} / R_{Bernardi}) \rangle = -0.21$.  In the extreme cases the galaxies can be $\sim70\%$ smaller than their local counterparts.
  \item The mass-weighted sizes of the galaxies are $\sim41\%$ smaller than their own light-weighted sizes, with a median $\langle \log(R_{e-circ,mass}/R_{e-circ}) \rangle = -0.23$, in the extreme cases the mass-weighted sizes can be up to $\sim60\%$ smaller.
  \item $78\%$ of the galaxies in our sample show a negative colour gradient $\nabla_{z_{850}-H_{160}}$, with redder colours at the core and bluer colour in the outskirts. $42\%$ have steep gradients with $\nabla_{z_{850}-H_{160}} < -0.5$.  The median colour gradient is $\langle \nabla_{z_{850}-H_{160}} \rangle = -0.45$, two times steeper than the colour gradient found in local passive galaxies in previous studies.
  \item The ratio of mass-weighted to light-weighted size does not show any significant correlation to galaxy properties, and is only mildly correlated to $M_{*}/L$ gradient, mass surface density and mass-weighted size.  The mild correlation with $M_{*}/L$ gradients supports our findings about smaller mass-weighted sizes compared to light-weighted sizes.
 
  \item By using the local SPIDER sample we find that the mass-weighted sizes are on average $\sim13\%$ smaller than the rest frame $r$-band light-weighted sizes, consistent with previous studies. Comparing the cluster sample to the local SPIDER cluster sample, we find an offset in the ratio of mass-weighted sizes to the $H_{160}$ band light-weighted sizes with a median difference of $\langle \log(R_{e-circ,mass,1.39} / R_{e-circ,1.39}) - \log(R_{e-circ,mass,0}/ R_{e-circ,0}) \rangle = -0.18$, which we attribute to an evolution of the $M_{*}/L$ gradient over redshift.  We also find that the progenitor bias cannot explain this observed offset.  This also seems to be consistent with the steeper colour gradient we find in the cluster galaxies compared to those seen in local passive galaxies.  
\end{enumerate}

We then investigate the origin and the evolution of the observed colour gradient by modeling the colour gradients with SSPs under three different assumptions.  We analyse the evolution of the rest-frame (U-R) colour gradient at two radii, $0.5 a_{e}$ and $2 a_{e}$, representing the inner and outer region of the galaxy respectively.  We subdivide each of the assumptions summarised below into three different metallicity scenarios: we fix the metallicity of the inner regions to sub-solar, solar and super-solar $Z=(0.008,0.02,0.05)$:

\begin{enumerate}[--]
   \item \textbf{Case I - Pure age-driven gradient evolution} -- Evolution of colour gradients is solely due to age gradient.  The inner and outer regions of the passive galaxies are assumed to have identical metallicities (i.e. flat metallicity gradients $\nabla_{Z} =0$).  We find that although an age gradient alone is sufficient to reproduce all the evolution in the colour gradient, it over-predicts the evolution over redshift, causing the evolved local gradients to be too shallow compared to the observation.
   
   \item \textbf{Case II - Pure metallicity-driven gradient evolution} -- Evolution of colour gradients is solely due to metallicity gradient.  We assume the inner and outer regions have identical ages (i.e. flat age gradients $\nabla_{age} = 0$).  A metallicity gradient alone cannot explain the observation as the evolution it predicts goes into the wrong direction.
   
   \item \textbf{Case III - Age-driven gradient evolution with an assumed metallicity gradient} --  Evolution of colour gradients is due to a combination of age and metallicity gradients.  The galaxies are assumed to have a fixed metallicity gradient identical to that observed in local passive galaxies, $\nabla_Z \simeq -0.2$.  This model works well, the solar metallicity scenario can well reproduce the observed evolution of the colour gradients from $z\sim1.39$ to $z\sim0$.   
   
   \item We show that the above findings are still robust if any central dust reddening at $0.5 a_{e}$ is $\lesssim 0.14$ mag, or equivalently an extinction gradient $d A_V/d \log(a) \sim 0.40$ or $d E(B-V) /d \log(a) \sim -0.10$.
\end{enumerate}

Our case study indicates that the presence of an age gradient at high redshift is a necessary condition to explain the observed evolution of the colour gradients, while metallicity gradients probably dominate at $z\sim0$.  We also repeat the study using the rest-frame $g-r$ colour gradient and obtain completely consistent results.  This conclusion is partially consistent with other studies \citep{Sagliaetal2000, Guoetal2011, Gargiuloetal2012}.  For the best-matching scenario (Case III with solar metallicity), the median age gradient of our cluster sample is $\langle \nabla_{age} \rangle \sim -0.33 \pm 0.37$, while the metallicity gradient we assumed is $\nabla_Z \simeq -0.2$.  Given the assumption that passive galaxies at z = 1.39 have the same metallicity gradient as the local ones, the evolution of colour gradients from $z\sim1.4$ to $z\sim0$ can be explained by passive evolution.

This general picture is also consistent with a more gradual mass growth mechanism such as via minor mergers, in the sense that the inner region of the galaxies remains undisturbed and the accreted younger material settles at the outskirts.

\section*{Acknowledgments}
The authors would like to thank Francesco La Barbera for sharing the SPIDER structural parameter catalogue for the comparison of our cluster sample with local passive galaxies.  The authors would also like to thank Thorsten Naab for stimulating discussions.  We would also like to thank the anonymous referee for valuable comments that led to an improved presentation.
J.C.C. Chan acknowledges the support of the Deutsche Zentrum f\"ur Luft- und Raumfahrt (DLR) via Project ID 50OR1513.  David J. Wilman and Matteo Fossati acknowledge the support of the Deutsche Forschungsgemeinschaft via Project ID 3871/1-1.  This work was also supported by the Astrophysics at Oxford grants (ST/H002456/1 and ST/K00106X/1) as well as visitors grant (ST/H504862/1) from the UK Science and Technology Facilities Council.  Roger L. Davies acknowledges travel and computer grants from Christ Church, Oxford.  Roger L. Davies is also grateful for support from the Australian Astronomical Observatory Distinguished Visitors programme, the ARC Centre of Excellence for All Sky Astrophysics, and the University of Sydney during a visit.  Ryan C. W. Houghton is supported by the Science and Technology Facilities Council [STFC grant numbers ST/H002456/1, ST/K00106X/1 \& ST/J002216/1].  John P. Stott gratefully acknowledges support from a Hintze Research Fellowship.

\bibliography{ms}

\appendix

\section[]{Details of the simulations}
\label{Details of the simulations}
 
We perform extensive simulations with a set of 50000 simulated galaxies with surface brightness profiles described by a Se\'rsic profile on the ACS $z_{850}$ and WFC3 $H_{160}$ band images to give a more realistic estimate of the uncertainty of the photometry, light-weighted structural parameters and mass-weighted structural parameters. Here we describe the set-up and the procedures involved in these simulations.

\subsection[]{Constructing the set of simulated galaxies}
\label{Constructing the set of simulated galaxies}
The simulated galaxies (hereafter SGs) are uniformly distributed within a magnitude range of $19.0 < H_{160} < 25.0$ and a colour range of $0.4 <  z_{850} -H_{160} <2.2$ (hence a range of $z_{850}$ magnitudes $19.4 < z_{850} < 27.2$).  This selected magnitude range is a good representation of the cluster sample, and is also the range where our $M_{*}/L$ - colour relation is calibrated as we explained in Section~\ref{sec:Stellar Mass-to-light Ratio-Colour relation}.  Each galaxy is described by a S\'ersic profile with input structural parameters randomly drawn from Gaussian distributions with means and dispersions taken from the real galaxies distributions in the $H_{160}$ band.

The input structural parameters are identical in the two bands. The means and dispersions of the S\'ersic indices $n$, effective semi-major axis $a_e$ and axis ratio $q$ are $(\langle n \rangle, \sigma_{n}) = (3.19, 2.18) $, $(\langle a_e \rangle, \sigma_{a_e}) = (6.07, 5.16 (pixel))$ and $(\langle q \rangle, \sigma_{q}) = (0.67, 0.20) $.  The position angle $P.A.$ is uniformly distributed within $0^{\circ} < P.A. < 180^{\circ}$. To ensure the simulated profiles are physical, we further apply the following constraints:  $n > 0.2$, $a_e > 0.3$, $0.01 < q \leq 1$.

The SGs were then convolved with the adopted PSFs. \citet{Morishitaetal2014} pointed out that there are differences in the central part of the S\'ersic profiles produced by \texttt{IRAF gallist} and \texttt{mkobjects} compared to those produced by GALFIT, which possibly originates with the PSF convolution procedure. In our case, we produce our simulated galaxies using a custom-built IDL routine\footnote{Interactive Data Language, Exelis Visual Information Solutions, Boulder, Colourado} that over-samples the central part of the S\'ersic profiles before resampling it onto a 2D grid.

To check whether the S\'ersic profiles we generated are consistent with those used in GALFIT, we first fit the noise-free SGs with GALFIT and examine the residual maps to compare the S\'ersic profiles. Without PSF convolution, we notice there are residuals at the centre in the residuals map output by GALFIT, although the difference is negligible ($< 0.005\%$ of the flux).  Including the PSF convolution does not noticeably increase this difference.

We then inject the SGs one by one uniformly to the sky regions of both the WFC3 $H_{160}$ images and the PSF-matched and resampled ACS $z_{850}$ images at the same location (i.e. 50000 set of $z_{850}$ and $H_{160}$ images).  The segmentation maps from SExtractor are used as a reference to avoid direct overlap with existing objects in the field.

\subsection[]{Photometry uncertainty test}
\label{Details of the simulations-light}
We run SExtractor on the SG images using the same setting as for the science sample.  We then assess the detection rate of the SGs in different magnitudes, as well as investigate the uncertainties of the galaxy magnitude $\tt{MAG\_AUTO}$ and the $z_{850} - H_{160}$ colour derived from $1''$ aperture magnitudes.

\subsection[]{Light-weighted structural parameter uncertainty test}
\label{Details of the simulations-lightstruct}
We assess the accuracy of our light-weighted structural parameter measurements by measuring the structural parameters of the SGs with the same GALAPAGOS routine. 

\begin{figure}
  \includegraphics[scale=0.46]{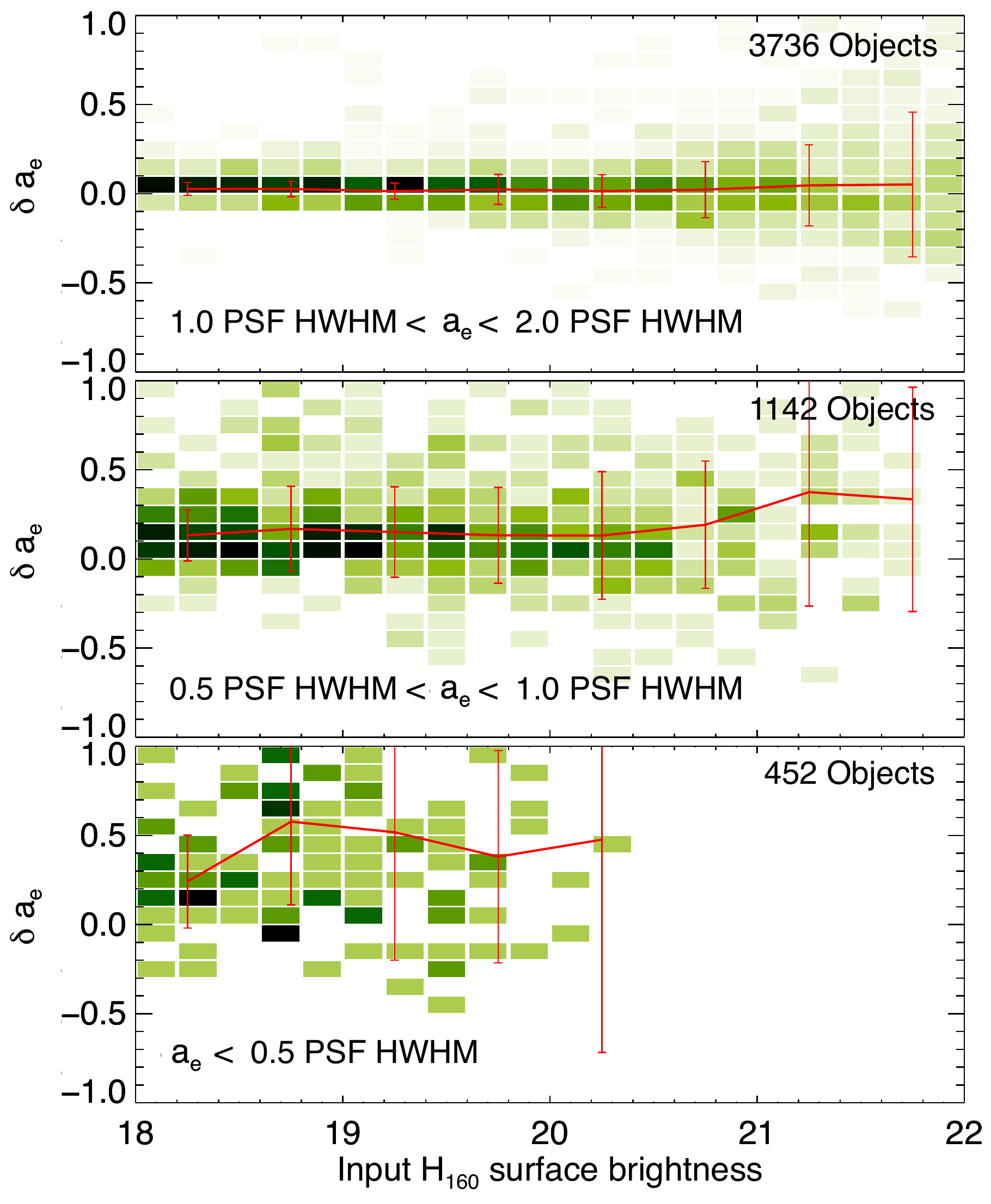}
  \caption{Differences between input and recovered effective semi-major axes by GALFIT $\delta a_e = (a_{e-out} - a_{e-in})/a_{e-in}$ in function of input mean $H_{160}$ surface brightness. From top to bottom: simulated galaxies with different ranges of input effective radius,  $1.0$ PSF HWHM $< a_e < 2.0 $ PSF HWHM, $0.5$ PSF HWHM $< a_e < 1.0 $ PSF HWHM and $a_e < 0.5$ PSF HWHM.  The red line indicates the median and $1\sigma$ dispersion in different bins (0.5 mag arcsec$^{-2}$ bin width) and the green-shaded 2D histogram in each panel shows the number density distribution of the simulated galaxies.}
  \label{fig:lightuncertaintysub}
\end{figure}

In Section~\ref{sec:Quantifying the uncertainties on the photometry} we mention there is a bias in the recovered effective radius at high mean surface brightness ($< 19$ mag arcsec$^{-2}$) due to unresolved SGs in our simulations. Here we expand the discussion on this.  Figure \ref{fig:lightuncertaintysub} shows the difference between input and recovered structural parameters by GALFIT for three subpopulations of SGs with descending range of input $a_e$ in terms of the size of PSF. We find that the bias between input and recovered effective radii increases sharply for SGs with input $a_e <$ PSF HWHM. For SGs with input $a_e$ in range of $1.0$ PSF HWHM $< a_e < 2.0 $ PSF HWHM, the average bias is typically limited to $1-2\%$, while it increases to $\sim10\%$ for SGs with $0.5$ PSF HWHM $< a_e < 1.0 $ PSF HWHM.  For those with $a_e < 0.5$ PSF HWHM, the average bias rises sharply to $\sim50\%$.  We conclude that our method is unable to measure sizes reliably from galaxies with $a_e < 0.5$ PSF HWHM.  However, since the sizes of high-redshift galaxies can indeed be very small, we do not exclude this small-sized population from our set of simulated galaxies.

None of the galaxies in our sample have light-weighted sizes smaller than $0.5$ PSF HWHM. Nevertheless, in the case of mass-weighted sizes, three of the objects at the low-mass end have sizes $< 0.5$ PSF HWHM and are hence discarded in the subsequent analyses.

\subsection[]{Mass-weighted structural parameter uncertainty test}
\label{Details of the simulations-massstruct}
We also perform a similar test to investigate the biases and uncertainties of our mass-weighted structural parameter measurements.  We start with the $z_{850}$ and $H_{160}$ postage stamps output from the light uncertainty test. These images are converted into mass maps with the pipeline described in Section~\ref{sec:Resolved Stellar Mass Surface Density Maps}.  The resultant stacked mass maps are then fitted with GALFIT.

As mentioned in Appendix~\ref{Constructing the set of simulated galaxies}, the galaxies we inject have identical initial structural parameters in both $z_{850}$ and $H_{160}$ bands, there is no internal colour gradient within the set of simulated galaxies. Nevertheless, this allows us to assess the accuracy of the output mass-weighted structural parameters, as the retrieved parameters should be in theory, exactly the same as the input (light) structural parameters. As the Voronoi binning and stacking take a certain time, we perform the mass map conversion for a subsample of galaxies ($\sim1500$). This sample is sufficient to provide an uncertainty estimates on the mass-weighted sizes in different bins of surface brightness.

\begin{figure}
  \includegraphics[scale=0.46]{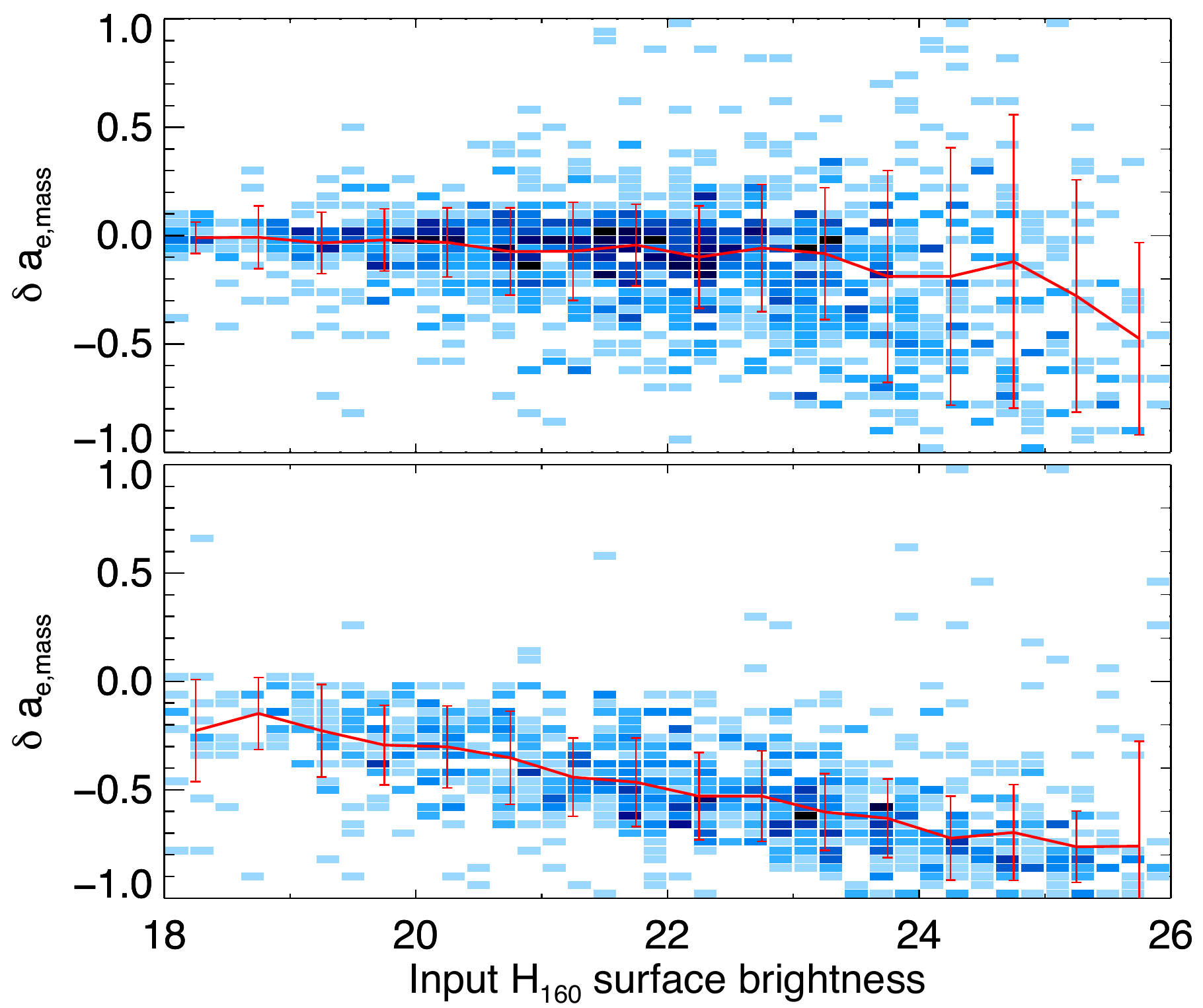}
  \caption{Differences between input and recovered effective semi-major axes by GALFIT $\delta a_e = (a_{e-out} - a_{e-in})/a_{e-in}$ in function of input mean $H_{160}$ surface brightness. Top: sizes derived from mass maps with the extrapolation applied, same as the middle panel in Figure \ref{fig:massuncertainty}. Bottom: sizes derived from mass maps without applying the extrapolation scheme. Red line indicates the median and $1\sigma$ dispersion in different bins (0.5 mag arcsec$^{-2}$ bin width) and blue-shaded 2D histogram in each panel shows the number density distribution of the simulated galaxies.}
  \label{fig:massuncertaintysub}
\end{figure}

In deriving the mass maps, we implement an extrapolation scheme to determine the $M_{*}/L$ in regions with insufficient signal-to-noise (described in Section~\ref{sec:From Colour to Stellar Mass Surface Density}). The low S/N or sky regions are problematic as the colours (and hence $M_{*}/L$) are not well determined. Converting mass directly on these regions will induce a huge scatter of mass in the background, which in turn have serious effects on the structural parameter measurements. Our extrapolation scheme can preserve the sky noise and at the same time provide a reasonable $M_{*}/L$ estimate to these regions.  We illustrate this effect in Figure \ref{fig:massuncertaintysub}. The top panel shows the differences between input and recovered sizes by GALFIT with extrapolation, while the bottom panel shows the differences without applying the extrapolation. In the absence of extrapolation, a huge bias can be seen in all bins of surface brightness. Sizes are more underestimated in galaxies with low surface brightness. 

Previous studies use a different method to solve this issue, for example in \citet{Langetal2014} who derived mass maps for galaxies in CANDELS, these low S/N regions are assigned the average $M_{*}/L$ of the three nearest Voronoi bins. The data we used in this study is not as deep as the CANDELS HST imaging. We find that averaging the nearest three Voronoi bins does not work as well as our annular average extrapolation. For star-forming galaxies which have sub-structures such as star forming clumps and spiral arms \citep[see, e.g.][]{Wuytsetal2012}, it might be more suitable to use the nearest neighbour extrapolation as in \citet{Langetal2014}. Our method, on the other hand, works well for early-type galaxies which usually have smooth(er) surface brightness profiles. 

\section{Details of PSF matching}
\label{app:psfmatching}
We first stack the unsaturated stars in the $z_{850}$ and $H_{160}$ bands to obtain characteristic PSFs respectively. The matching kernel is generated using the \texttt{psfmatch} task in IRAF. Cosine bell tapering is applied to filter the high frequency component of the input $z_{850}$ PSF, which is presumably induced by noise, to clean the output kernel.  In the \texttt{psfmatch} task, there are some free parameters that can be tweaked (e.g. kernel sizes, highest cosine bell frequencies and apodize), a systemic search is performed to find the best parameters to match the PSFs.

\begin{figure}
  \includegraphics[scale=0.46]{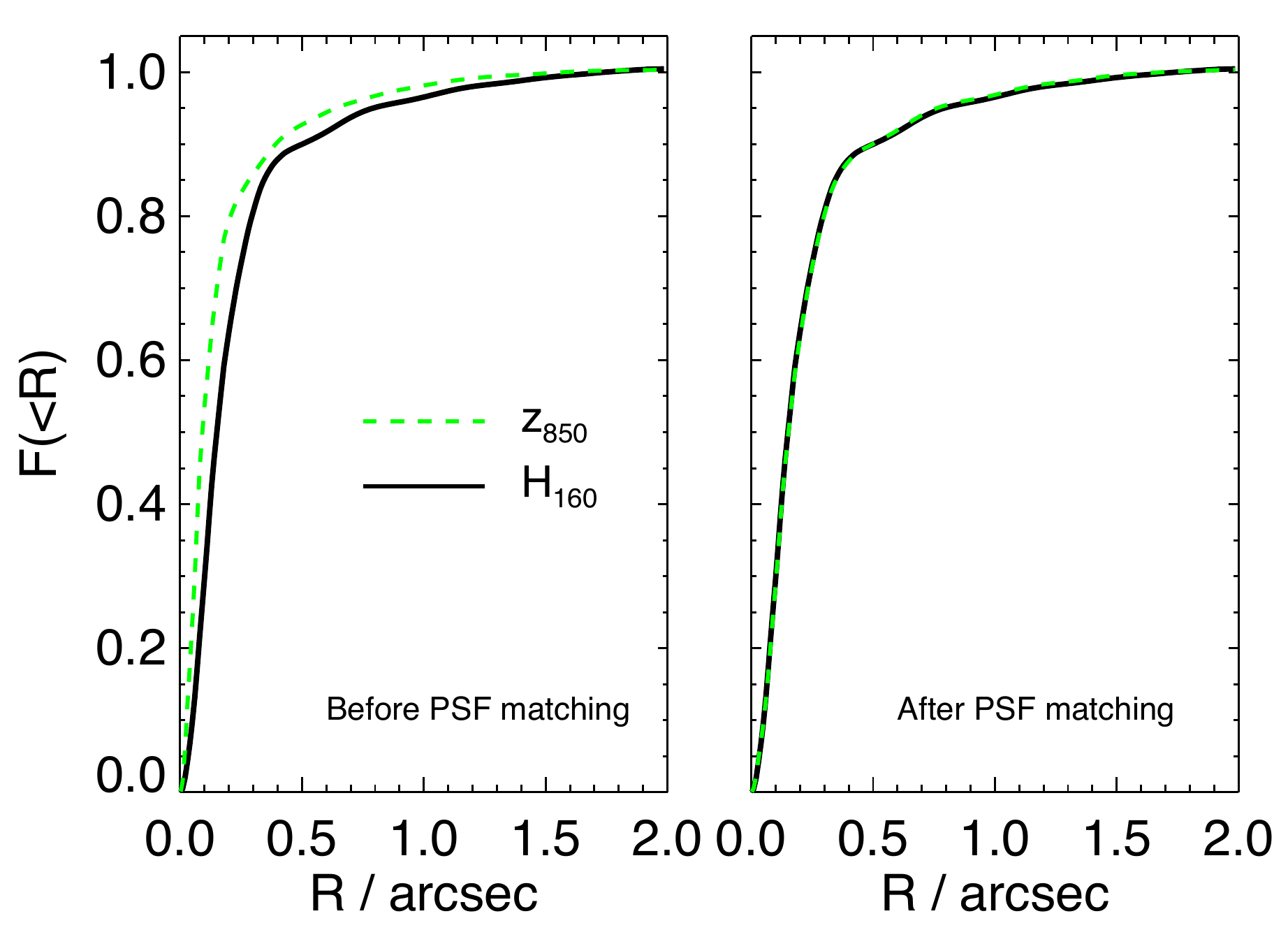}
  \caption{Fractional encircled energy of the $z_{850}$ and $H_{160}$ PSFs. The green dashed line corresponds to the $z_{850}$ PSF while the black solid line corresponds to the $H_{160}$ PSF. Left: before PSF matching. Right: after PSF matching.}
  \label{fig:psfcompare}
\end{figure}

We assess the accuracy by comparing the fractional encircled energy of the $z_{850}$ PSF before and after the procedure to the $H_{160}$ PSF. The convolved $z_{850}$ PSF matches almost perfectly to the $H_{160}$ PSF with only tiny difference in the wing ($<1\%$).  We also re-constructed a new $z_{850}$ PSF from the PSF matched $z_{850}$ images to assess the result.  Figure \ref{fig:psfcompare} shows the fractional encircled energy of the PSFs constructed from images before and after PSF matching.  The ratios of their growth curves deviate by $< 2.5\%$ from unity.

\section[]{Comparison of masses derived using $M_{*}/L$ - colour relation with SED fitting}
\label{app:masscomparison}
In this study, the stellar masses of the galaxies are derived from an empirical $M_{*}/L$ - colour relation and the total $H_{160}$ luminosity from 2D GALFIT S\'ersic fitting.  Other studies usually estimate the stellar masses through spectral energy distribution fitting of multiple photometric bands \citep[e.g.][]{Strazzulloetal2010, Delayeetal2014}.  The advantage of using $M_{*}/L$ - colour relation over SED fitting is that it does not require a number of photometric bands, hence is a relatively inexpensive mass indicator.  The accuracy of the stellar mass estimates of then depends on how well constrained the $M_{*}/L$ - colour relation is, which in turn depends on the colour used \citep[see discussion in e.g.,][]{BelldeJong2001, Belletal2003}.  Here we assess whether our mass estimates is biased.

For this cluster, \citet{Delayeetal2014} estimated stellar masses of our galaxies through SED fitting with four bands (HST/ACS $i_{775}$, $z_{850}$, HAWK-I, $J$, $K_s$) with BC03 models, exponential declining SFHs and a \citet{Chabrier2003} IMF.  The setting is almost identical to the stellar masses from the NMBS catalogue we picked to construct the $M_{*}/L$ - colour relation, thus can be compared directly.

\begin{figure}
\centering
  \includegraphics[scale=0.55]{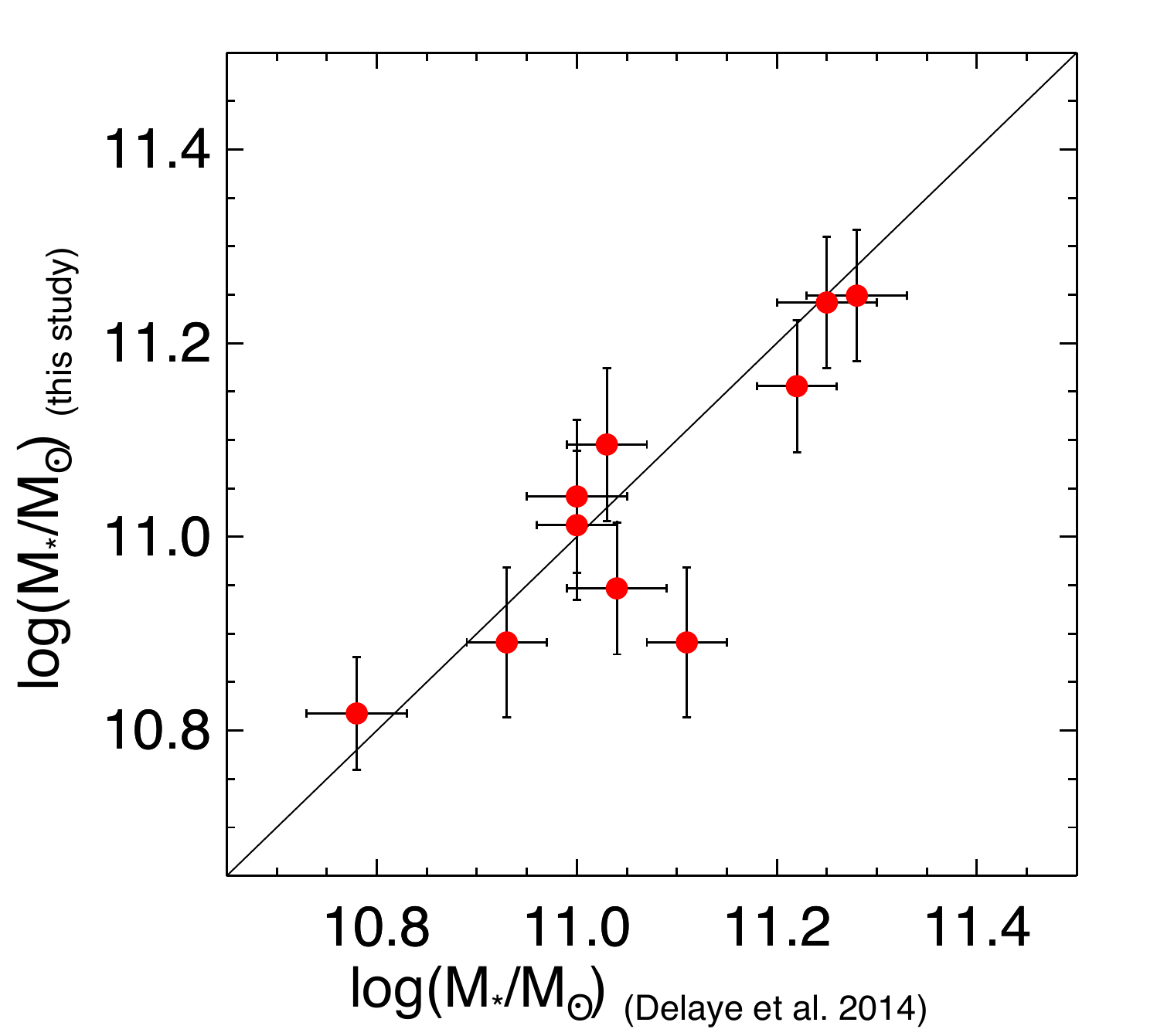}
  \caption{Comparison of masses derived using $M_{*}/L$ - colour relation with SED fitting from \citet{Delayeetal2014}.  The red circles are the stellar mass estimates of 10 galaxies that are common in our sample to \citet{Delayeetal2014}.  The stellar masses in \citet{Delayeetal2014} are derived from SED fitting with four bands, while our masses are from $M_{*}/L$ - colour relation. The solid black line is the one-to-one relation.  The error bars are the $1\sigma$ uncertainties from both methods.}
  \label{fig:massdelaye}
\end{figure}

Figure \ref{fig:massdelaye} shows a direct comparison of the mass estimated using our $M_{*}/L$ - colour relation and SED fitting from \citet{Delayeetal2014}.  Our sample covers 10 out of 13 early-type galaxies in their sample.  The remaining 3 galaxies are out of the field of view of our WFC3 images (but are in the FOV of the ACS $z_{850}$ image), thus are not included in our sample.  The mass estimates from the two methods are consistent with each other, with a median difference and $1\sigma$ scatter of $0.03 \pm 0.09$ dex.  The object that deviates from the one-to-one relation the most (at $\log(M_{*}/M_\odot) = 11.11$) is a galaxy close to the core of cluster with a very close neighbour (ID 368), which probably affect the mass estimates in both methods.  Removing this object reduces the median difference to $0.01 \pm 0.07$ dex.  Therefore we conclude that the masses derived using $M_{*}/L$ - colour relation are not biased.

As we mentioned in Section~\ref{sec:Integrated Stellar Masses}, the typical uncertainty of the mass estimates from $M_{*}/L$ - colour relation is $\sim 0.1$ dex, which is comparable to the uncertainties obtained from SED fitting.  These uncertainties are correct for the relative masses of multiple galaxies. Note that the uncertainty of the absolute stellar masses is larger as in the case of SED fitting, depending on the details of NMBS SED fitting and choice of IMF.

\section[]{Evolution of colour gradients using exponentially declining tau models}
\label{sec:Evolution of colour gradients using exponentially declining tau models}
In Section~\ref{sec:Evolution of the colour gradients over redshift} we discuss the evolution of colour gradients assuming the stellar population in the inner and outer regions can be well described by simple stellar population models (SSPs).  Here we present the result of using exponentially declining tau models with different $\tau$ ($\tau = [0.2, 0.4, 0.6]$) instead of SSPs.

\begin{figure*}
\centering
  \includegraphics[scale=0.45]{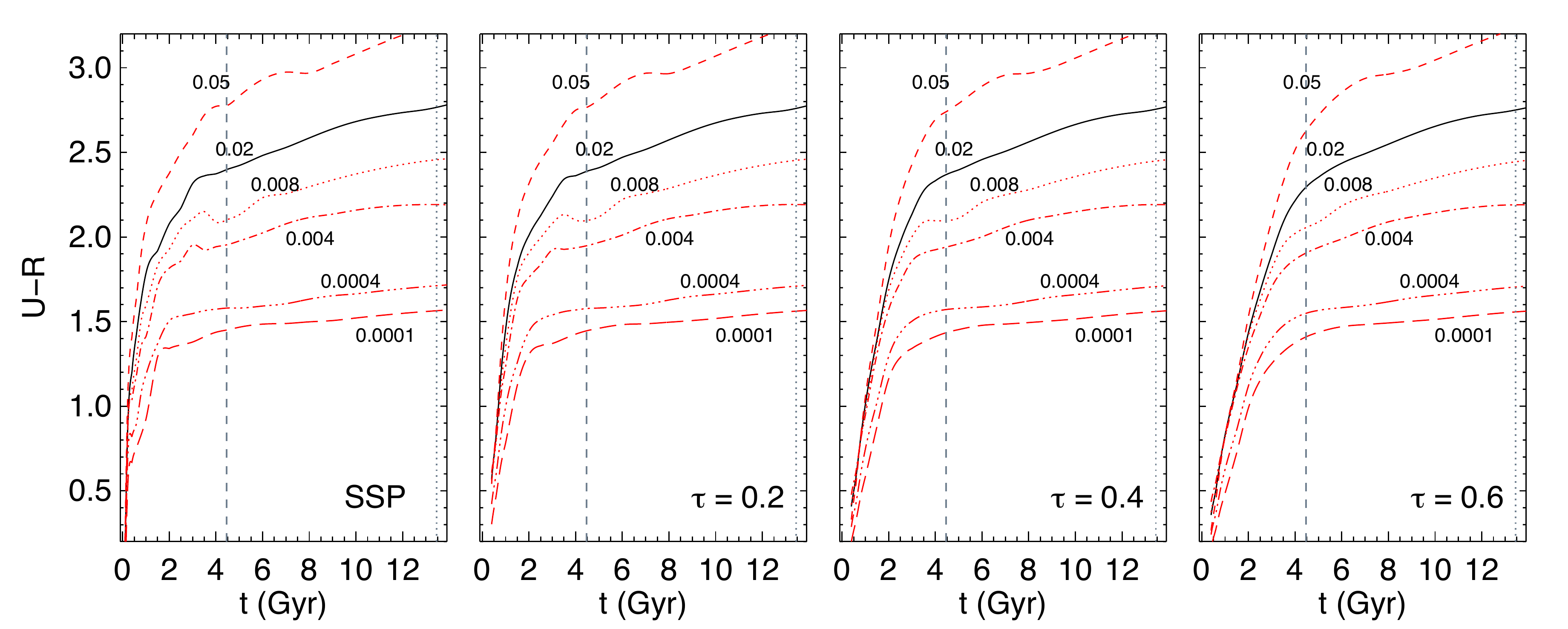}
  \caption{Rest-frame $U-R$ colour of stellar populations with different ages and metallicities with different values of $\tau$s (from left to right: SSP, $\tau = 0.2, 0.4, 0.6$).  The black line shows the stellar populations with solar metallicity ($Z=0.02$).  The red lines show populations with different metallicities ($Z=0.0001,0.0004,0.004,0.008,0.05$) as indicated. The grey dotted line shows the current age of the Universe (13.45 Gyr) with our choice of cosmology and the grey dashed line shows the age of Universe at redshift 1.39 (4.465 Gyr). Note that age refers to the time passed when the populations start forming stars.}
  \label{fig_URcolour_tau}
\end{figure*}

Figure~\ref{fig_URcolour_tau} shows the $U-R$ colour at different ages (left, i.e. the colour-age relations) from models with various $\tau$ using BC03, similar to Figure~\ref{fig_URcolourgrid}.  The methodology is described in Section~\ref{sec:Methodology}.  Comparing to SSPs, due to the continual star formation the $(U-R)$ colour is bluer for a given age, which is more pronounced if $\tau$ is larger.  We stop at $\tau = 0.6$ as otherwise the $(U-R)$ colour would be too blue to account for the observations.  Using $\tau$ models also has the effect of reducing the colour differences between different metallicities when the galaxy is young (see the trend before the grey dashed line), while the evolution is very similar for later ages when the contribution of young stars falls off.

\begin{figure}
\centering
  \includegraphics[scale=0.45]{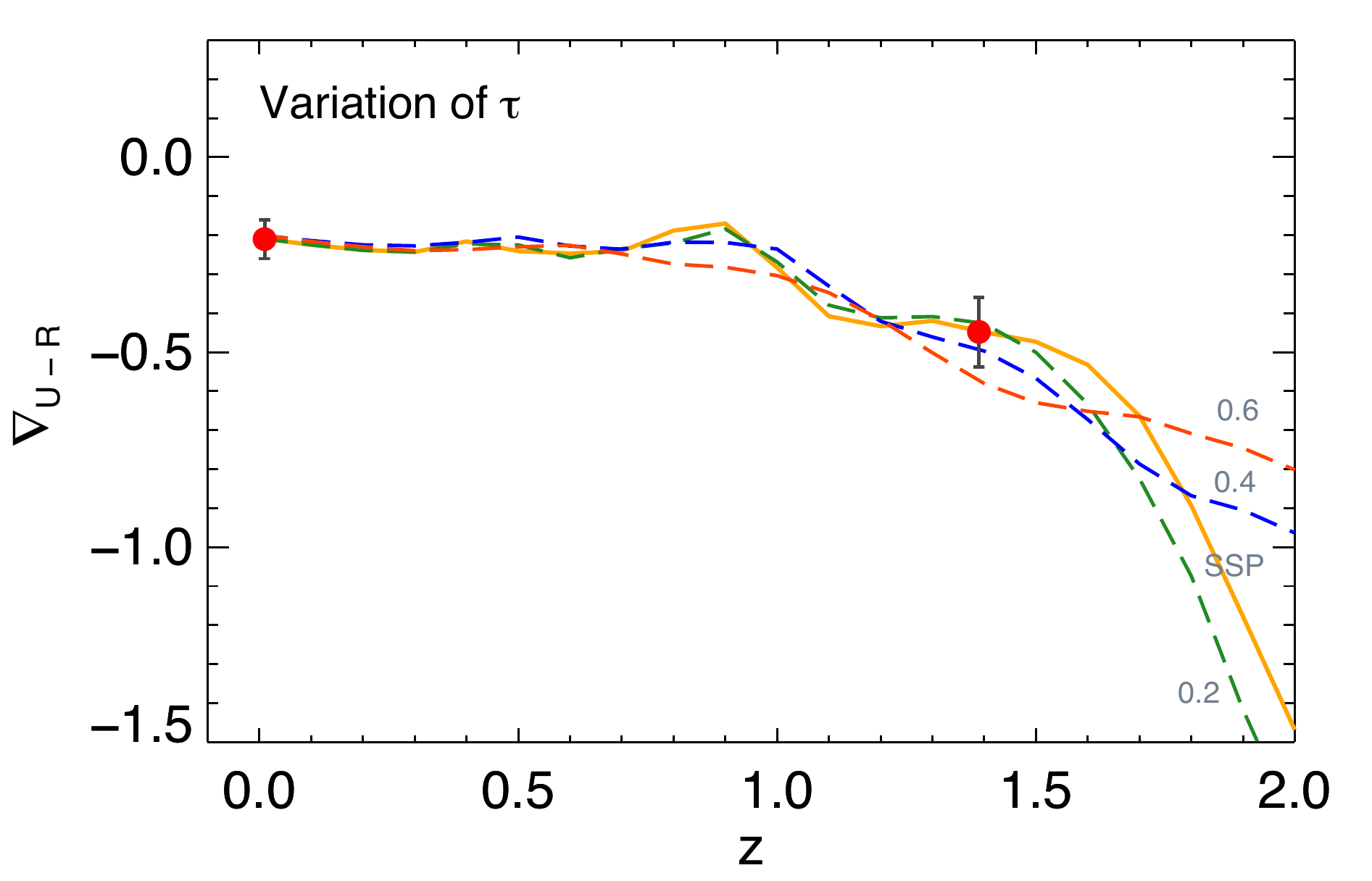}
  \caption{Evolution of colour gradient over redshift in case III (age and assumed metallicity gradient $\nabla_{Z} = -0.2$) using SSP and tau models with $\tau = 0.2, 0.4, 0.6$ with assumed solar metallicity $Z=0.02$. The solid yellow line shows the result with SSP models and is identical to one in the top left panel of Figure~\ref{fig_gradevoT2}.  Green, blue and red lines show the result with tau models with $\tau = 0.2, 0.4, 0.6$ respectively.  The derived initial age gradient at $z=1.39$ as well as formation redshift for each $\tau$ can be found in Table~\ref{tab_age}.  Red circle corresponds to the median $z_{850}-H_{160}$ gradient of our sample at redshift 1.39, and the observed $(U-R)$ gradient at redshift 0 by \citet{Wuetal2005}. The error bars show the uncertainty of the median.}
  \label{fig_gradevoTtau}
\end{figure}

Using $\tau$ models instead of SSPs does not change our main conclusion.  Since the evolution of the colour gradient at later times is very similar to SSPs, we find that an age-driven gradient evolution with a metallicity gradient close to the local value (case III) remains the best scenario to explain the colour gradients independent of the $\tau$ used.  Figure~\ref{fig_gradevoTtau} shows the best-fit scenario of the evolution of the colour gradients with different $\tau$ (case III with solar metallicity).  Within the range of $\tau$ the evolution can still be well modelled.  Nevertheless, due to the change in $(U-R)$ colour over time (see Figure~\ref{fig_URcolour_tau}), the resulting age gradient, age difference of the inner and outer population and formation redshift for the best-fit scenario varies by a certain amount with $\tau$.  With an increasing $\tau$, a flatter but still significant age gradient (e.g. $\langle \nabla_{age} \rangle = -0.21$ for $\tau = 0.4$) is needed to explain the colour gradient.  This strengthens our result that an age gradient is a necessary component in the colour gradient at high-redshift.  The values of the median age gradient, the evolved age gradient at $z=0$ and the formation redshift for the best-fit scenario with different $\tau$s can be found in Table~\ref{tab_age}.

\begin{table}
  \caption{The derived median age gradient at $z=1.39$, $z=0$ and formation redshift with exponentially declining $\tau$-models (Case III)}
  \label{tab_age}
  \begin{tabular}{cccc}
  \hline
  \hline
  $\tau$ & $\nabla_{age}$ & Evolved $\nabla_{age}$ & Formation redshift \\
              &  at $z=1.39$     &  at $z = 0$ &  z \\
  \hline
  SSP & $-0.33 \pm 0.37$  &  $-0.05 \pm 0.06$ &  3.0  \\
  0.2   & $-0.29 \pm 0.29$  &  $-0.04 \pm 0.06$ &  3.5  \\  
  0.4   & $-0.21 \pm 0.22$  &  $-0.04 \pm 0.04$ &  4.0  \\  
  0.6   & $-0.19 \pm 0.19$  &  $-0.05 \pm 0.04$ &  7.0  \\  
  \hline
\end{tabular}
\end{table}

\section[]{\MakeLowercase{$g-r$} color gradients and their evolution with redshift}
\label{sec:The g-r color gradients and the evolution with redshift}
Here we present the measurements of the $Y_{105} - H_{160}$ colour gradient in XMMUJ2235-2557 and the $g-r$ colour gradient in the local SPIDER cluster sample.  To compute the $g-r$ colour gradient from the SPIDER sample,  we again make use of the structural parameters in $g$-band and $r$-band of the publicly available multiband structural catalogue from \citet{LaBarberaetal2010b}.  We generate 2D S\'ersic model images in the two bands with fitted parameter from the structural catalogue, then convert the 2D image in both bands into 1D radial surface brightness profiles, similar to the procedure described in Section~\ref{sec:Elliptical aperture photometry and color gradients}.  This allows us to derive 1D $g-r$ colour profiles and measure the colour gradients of the galaxies in the SPIDER sample by fitting the logarithmic slope of their $g-r$ profiles along the major axis.  The sample is split into low density and high density environment with a halo mass cut ($\log(M_{200}/M_{\odot}) < 14$ and $\log(M_{200}/M_{\odot}) \geq 14$).  The detail selection is described in Section~\ref{Local Comparison Sample}.  We also apply the age cut (age > 8.98 Gyr) using age measurements from \citet{LaBarberaetal2010a} to correct for the progenitor bias in the SPIDER sample.  The median $g-r$ gradient and $1\sigma$ scatter in the local SPIDER cluster sample is $\nabla_{g-r} = -0.042 \pm 0.144$ (error on the median 0.008), while the median gradient in the low density sample is $\nabla_{g-r} = -0.060 \pm 0.158$ (error on the median 0.008), consistent with \citet{LaBarberaetal2005}.

\begin{figure}
  \includegraphics[scale=0.457]{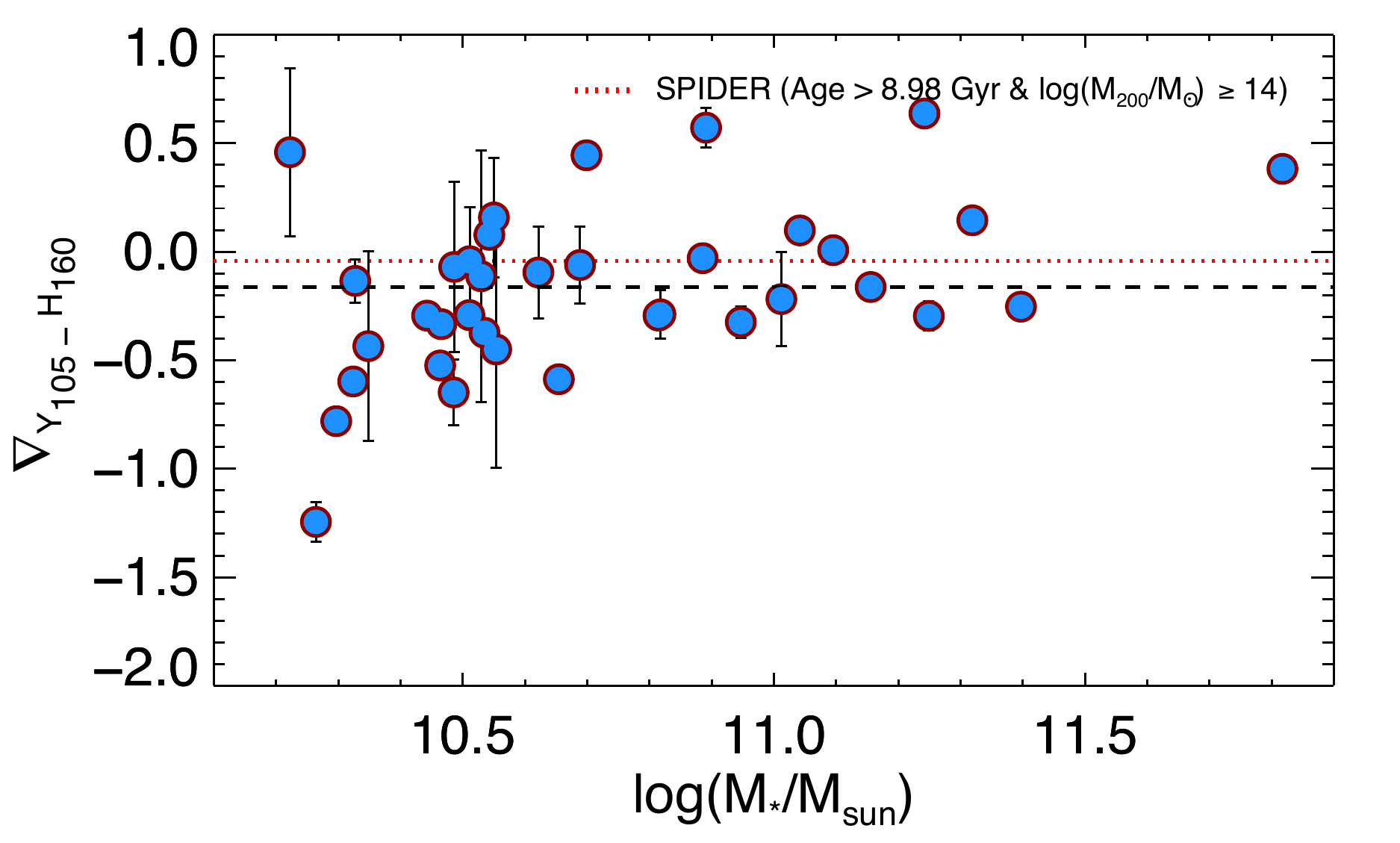}
  \caption{$Y_{105} - H_{160}$ colour gradients in the cluster XMMUJ2235-2557.  At redshift 1.39, this roughly corresponds to rest-frame $g-r$ colour gradient.  The red dotted line shows the median local $g-r$ gradient from the SPIDER cluster sample. The black dashed line shows the median $Y_{105} - H_{160}$ colour gradients.}
  \label{fig_yhcolourgrad}
\end{figure}

We derive the $Y_{105} - H_{160}$ colour gradient in XMMUJ2235-2557 with structural parameters of the $Y_{105}$ and  $H_{160}$ bands.  This is because the above $g-r$ colour gradients are intrinsic gradients without PSF convolution, hence for better comparison and consistency we use the same method as above.  Figure~\ref{fig_yhcolourgrad} shows the $Y_{105} - H_{160}$ colour gradients, roughly corresponds to rest-frame $g-r$.  $\nabla_{Y_{105} - H_{160}}$ is less steep compared to $\nabla_{z_{850} - H_{160}}$, with a median and $1\sigma$ scatter of $\langle \nabla_{Y_{105} - H_{160}} \rangle = -0.16 \pm -0.16$  (error on the median 0.08).  Hence, the $g-r$ colour gradient at $z=1.39$ is also much steeper than the local sample.

We repeat the analysis described in Section~\ref{sec:Methodology} to model the $g-r$ colour gradients under different assumptions in the radial variation of stellar population properties.  The $g-r$ colour is less sensitive to age variation than $(U-R)$ colour, as the $g$-band is on the 4000\AA~break.  Hence, the evolution in the $g-r$ colour gradient is less pronounced than the $(U-R)$ gradients.

\begin{figure}
  \includegraphics[scale=0.45]{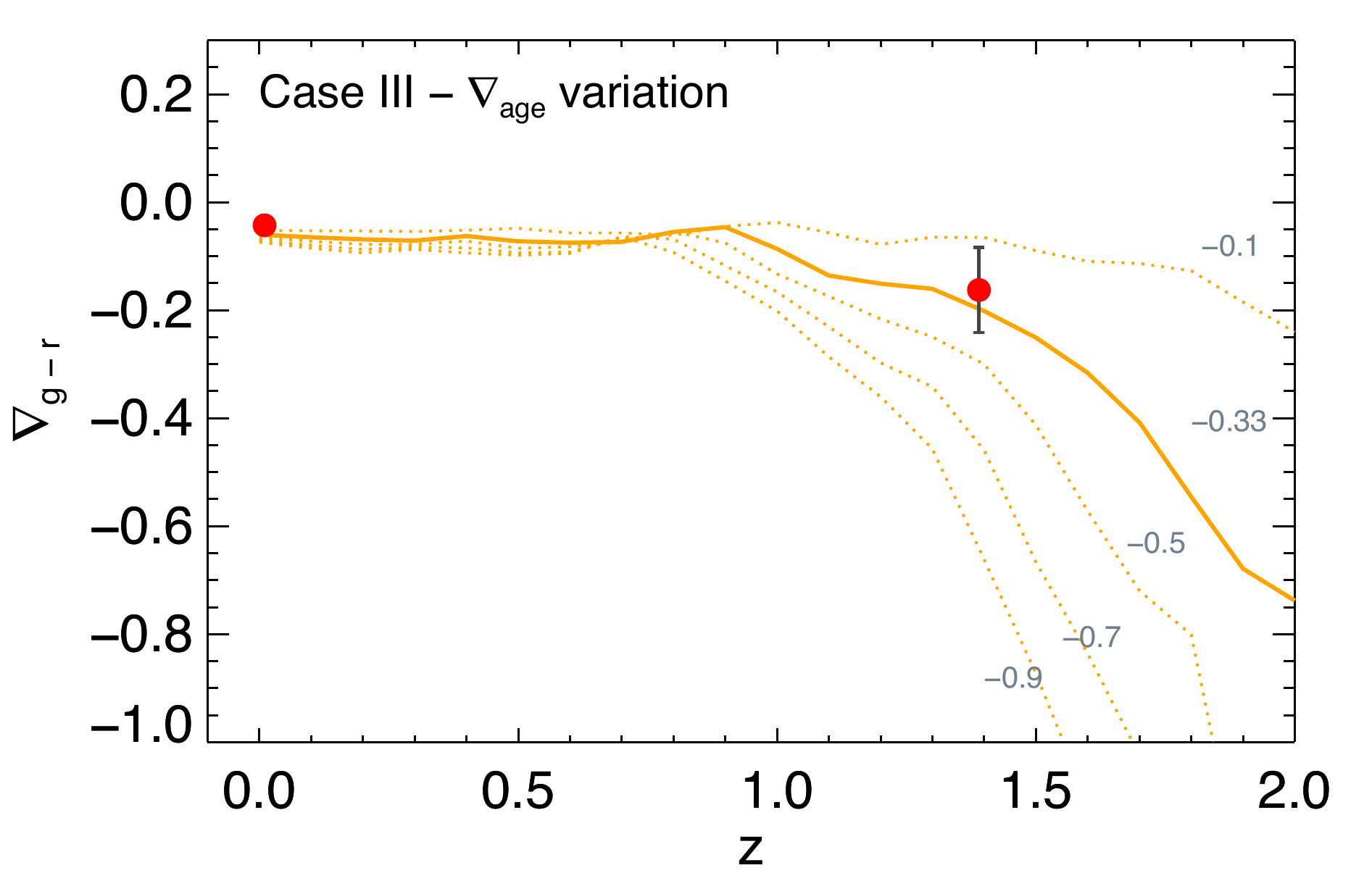}
  \caption{Evolution of $g-r$ colour gradient over redshift in case III (age and metallicity gradient) with assumed solar metallicity $Z=0.02$. The inner region is assumed to have a formation redshift $z_{form} = 3.0$. The initial age gradient at $z=1.39$ is $\nabla_{age} = -0.33$, and the assumed metallicity gradient is $\nabla_{Z} = -0.2$.  The dotted lines show the evolution with different initial age gradient as indicated ($\nabla_{age} = -0.1, -0.5, -0.7, -0.9$). Red circles correspond to the median $Y_{105}-H_{160}$ gradient of our sample at redshift 1.39, and the observed local $g-r$ gradient from the SPIDER cluster sample. The error bars show the uncertainty of the median.} 
  \label{fig_gradevoTgr}
\end{figure}

Figure~\ref{fig_gradevoTgr} shows the evolution of the $g-r$ colour gradients (case III with solar metallicity).  Despite the lack in dynamic range, the result with the $g-r$ colour gradient is completely consistent with the $(U-R)$ colour gradients, in the sense that an age-driven gradient evolution with a metallicity gradient close to local value (case III) is the best scenario to explain its evolution.  A pure age gradient would predict  $g-r$ gradients that are too shallow at $z=0$, while a pure metallicity gradient would predict gradients that are too steep.  In addition, we find that the derived median age gradient is in good agreement with the one derived from $(U-R)$ gradients.  The evolution can be well described with an age gradient of $\nabla_{age} = -0.33$, identical to the one we found from $(U-R)$ gradients. The consistent result from $g-r$ colour gradients reinforces our conclusion that age gradient is necessary to explain the colour gradient at high-redshift.

\section[]{Galaxy properties and best-fit light weighted and mass-weighted structural parameters}
\label{parametertablesection}

\newpage 
\onecolumn

\clearpage

\begin{landscape}
\begin{center}
\begin{table}
  \setlength\tabcolsep{2pt}

\centering 
\caption{Galaxy parameters}
    \label{tab_para}
  \begin{tabular}{@{}lcccccccccccccccc@{}}
  \hline
  \hline
  ID\textsuperscript & R.A. & Decl. & $\log M_{*}\textsuperscript{a} $ & $z_{850}-H_{160}$\textsuperscript{b} & $\nabla_{z_{850}-H_{160}}$\textsuperscript{c} & $\nabla_{\log(M/L)}$\textsuperscript{c} & $a_{e}$ & $n$ & $q$ & $a_{e,mass}$ & $n_{mass}$ & $q_{mass}$ & $\Sigma$\textsuperscript{d} & $\log(\Sigma_{mass})$\textsuperscript{d} & $\log(\Sigma_{1})$\textsuperscript{e} \\
    & (J2000)  & (J2000) & ($M_\odot$) & (AB mag) &  &  & (kpc) &  &  & (kpc) & & & (mag kpc$^{-2})$ & $(M_\odot$ kpc$^{-2})$ & $(M_\odot$ kpc$^{-2})$ \\
  \hline
   36& 338.829468& $ -25.974178$ & $ 11.04\pm 0.08$ & $ 1.765\pm 0.010$ & $ -0.033\pm 0.096$ & $ -0.038\pm 0.220$ & $ 4.12\pm 0.41$ & $ 4.77\pm 0.48$ & $ 0.79\pm 0.03$ & $ 3.66\pm 0.85$ & $ 7.43\pm 1.11$ & $ 0.79\pm 0.05$ & $ 25.87\pm 0.09$ & $ 9.12\pm 0.20$ & $ 9.74\pm 0.03 $\\
 61& 338.815826& $ -25.971949$ & $ 10.44\pm 0.05$ & $ 1.359\pm 0.018$ & $ -0.540\pm 0.145$ & $ -0.386\pm 0.296$ & $ 5.16\pm 1.08$ & $ 0.61\pm 0.12$ & $ 0.69\pm 0.06$ & $ 3.64\pm 1.07$ & $ 2.54\pm 0.73$ & $ 0.52\pm 0.06$ & $ 27.23\pm 0.18$ & $ 8.52\pm 0.25$ & $ 9.13\pm 0.10 $\\
 77& 338.818085& $ -25.972015$ & $ 10.51\pm 0.08$ & $ 1.844\pm 0.037$ & $ -1.190\pm 0.154$ & $ -0.576\pm 0.309$ & $ 5.47\pm 1.48$ & $ 0.64\pm 0.17$ & $ 0.55\pm 0.06$ & $ 3.65\pm 1.12$ & $ 1.24\pm 0.42$ & $ 0.49\pm 0.08$ & $ 27.94\pm 0.23$ & $ 8.59\pm 0.27$ & $ 9.10\pm 0.14 $\\
 148& 338.833374& $ -25.967278$ & $ 10.53\pm 0.07$ & $ 1.527\pm 0.024$ & $ -0.740\pm 0.017$ & $ -0.470\pm 0.015$ & $ 9.27\pm 3.22$ & $ 0.65\pm 0.21$ & $ 0.18\pm 0.03$ & $ 7.03\pm 3.44$ & $ 1.73\pm 0.83$ & $ 0.17\pm 0.03$ & $ 28.53\pm 0.30$ & $ 8.04\pm 0.42$ & $ 9.02\pm 0.18 $\\
 159& 338.825348& $ -25.968397$ & $ 10.30\pm 0.07$ & $ 1.259\pm 0.015$ & $ -1.199\pm 0.048$ & $ -0.719\pm 0.105$ & $ 4.09\pm 0.68$ & $ 1.20\pm 0.19$ & $ 0.64\pm 0.04$ & $ 1.97\pm 0.47$ & $ 3.42\pm 0.68$ & $ 0.73\pm 0.07$ & $ 26.93\pm 0.15$ & $ 8.91\pm 0.21$ & $ 9.31\pm 0.06 $\\
 170& 338.836761& $ -25.961046$ & $ 11.82\pm 0.07$ & $ 1.956\pm 0.009$ & $ -0.133\pm 0.038$ & $ -0.129\pm 0.094$ & $ 24.62\pm 6.62$ & $ 4.49\pm 1.18$ & $ 0.62\pm 0.07$ & $ 8.39\pm 2.56$ & $ 3.91\pm 1.31$ & $ 0.64\pm 0.10$ & $ 28.11\pm 0.23$ & $ 9.17\pm 0.26$ & $ 10.00\pm 0.21 $\\
 198& 338.824432& $ -25.936956$ & $ 10.22\pm 0.05$ & $ 1.377\pm 0.025$ & $ -0.626\pm 0.585$ & $ -0.499\pm 0.995$ & $ 1.02\pm 0.07$ & $ 2.05\pm 0.28$ & $ 0.86\pm 0.05$ & $ 0.71\pm 0.10$ & $ 1.09\pm 0.15$ & $ 0.34\pm 0.02$ & $ 24.28\pm 0.06$ & $ 9.72\pm 0.13$ & $ 9.58\pm 0.02 $\\
 220& 338.845001& $ -25.940239$ & $ 10.82\pm 0.06$ & $ 1.694\pm 0.013$ & $ -0.482\pm 0.190$ & $ -0.342\pm 0.348$ & $ 2.34\pm 0.19$ & $ 4.91\pm 0.50$ & $ 0.71\pm 0.03$ & $ 1.53\pm 0.27$ & $ 8.64\pm 1.69$ & $ 0.37\pm 0.02$ & $ 25.09\pm 0.07$ & $ 9.65\pm 0.15$ & $ 9.98\pm 0.06 $\\
 239& 338.824738& $ -25.942131$ & $ 10.51\pm 0.06$ & $ 1.699\pm 0.018$ & $ -0.365\pm 0.314$ & $ -0.212\pm 0.587$ & $ 1.37\pm 0.10$ & $ 4.00\pm 0.43$ & $ 0.84\pm 0.03$ & $ 0.90\pm 0.15$ & $ 8.38\pm 1.72$ & $ 0.55\pm 0.04$ & $ 24.71\pm 0.06$ & $ 9.80\pm 0.14$ & $ 9.81\pm 0.04 $\\
 296& 338.839996& $ -25.957035$ & $ 10.54\pm 0.05$ & $ 1.377\pm 0.011$ & $ -0.392\pm 0.110$ & $ -0.259\pm 0.220$ & $ 3.00\pm 0.28$ & $ 1.37\pm 0.14$ & $ 0.59\pm 0.02$ & $ 2.20\pm 0.50$ & $ 2.07\pm 0.30$ & $ 0.48\pm 0.03$ & $ 25.83\pm 0.08$ & $ 9.06\pm 0.20$ & $ 9.53\pm 0.05 $\\
 308& 338.832703& $ -25.957813$ & $ 10.89\pm 0.05$ & $ 1.360\pm 0.006$ & $ -0.426\pm 0.098$ & $ -0.310\pm 0.204$ & $ 2.99\pm 0.19$ & $ 3.99\pm 0.34$ & $ 0.79\pm 0.02$ & $ 1.50\pm 0.24$ & $ 4.40\pm 0.67$ & $ 0.72\pm 0.04$ & $ 24.93\pm 0.06$ & $ 9.74\pm 0.14$ & $ 9.92\pm 0.04 $\\
 343& 338.840668& $ -25.959299$ & $ 10.55\pm 0.08$ & $ 1.844\pm 0.022$ & $ -0.371\pm 0.558$ & $ -0.189\pm 0.887$ & $ 1.18\pm 0.07$ & $ 2.38\pm 0.24$ & $ 0.39\pm 0.01$ & $ 1.23\pm 0.18$ & $ 0.89\pm 0.15$ & $ 0.13\pm 0.02$ & $ 24.51\pm 0.05$ & $ 9.57\pm 0.13$ & $ 9.85\pm 0.03 $\\
 352& 338.836304& $ -25.962229$ & $ 11.24\pm 0.07$ & $ 1.919\pm 0.009$ & $ -0.287\pm 0.086$ & $ -0.240\pm 0.157$ & $ 4.58\pm 0.45$ & $ 4.70\pm 0.47$ & $ 0.57\pm 0.02$ & $ 3.42\pm 0.78$ & $ 6.86\pm 1.02$ & $ 0.37\pm 0.02$ & $ 25.84\pm 0.08$ & $ 9.38\pm 0.20$ & $ 10.23\pm 0.06 $\\
 357& 338.829987& $ -25.959751$ & $ 10.33\pm 0.05$ & $ 1.436\pm 0.014$ & $ 0.117\pm 0.124$ & $ 0.105\pm 0.254$ & $ 2.38\pm 0.22$ & $ 1.19\pm 0.13$ & $ 0.40\pm 0.02$ & $ 1.92\pm 0.44$ & $ 1.78\pm 0.27$ & $ 0.26\pm 0.02$ & $ 25.96\pm 0.08$ & $ 8.96\pm 0.20$ & $ 9.47\pm 0.04 $\\
 365& 338.833984& $ -25.960121$ & $ 10.53\pm 0.08$ & $ 1.847\pm 0.021$ & $ -0.158\pm 0.612$ & $ -0.111\pm 0.964$ & $ 1.38\pm 0.10$ & $ 3.98\pm 0.47$ & $ 0.53\pm 0.02$ & $-$ & $-$ & $-$ & $ 24.91\pm 0.06$ & $-$ & $-$\\
 368& 338.837128& $ -25.959915$ & $ 10.89\pm 0.08$ & $ 1.869\pm 0.013$ & $ -0.906\pm 0.184$ & $ -0.491\pm 0.401$ & $ 2.41\pm 0.24$ & $ 5.78\pm 0.61$ & $ 0.89\pm 0.03$ & $-$ & $-$ & $-$ & $ 25.25\pm 0.09$ & $-$ & $-$\\
 385& 338.837341& $ -25.959743$ & $ 10.62\pm 0.08$ & $ 1.735\pm 0.013$ & $ -0.619\pm 0.213$ & $ -0.418\pm 0.356$ & $ 1.73\pm 0.11$ & $ 3.00\pm 0.28$ & $ 0.35\pm 0.01$ & $ 1.65\pm 0.27$ & $ 1.92\pm 0.31$ & $ 0.29\pm 0.02$ & $ 24.99\pm 0.06$ & $ 9.39\pm 0.14$ & $ 9.73\pm 0.04 $\\
 407& 338.836243& $ -25.960421$ & $ 11.25\pm 0.07$ & $ 1.915\pm 0.008$ & $ -0.448\pm 0.108$ & $ -0.272\pm 0.206$ & $ 3.03\pm 0.20$ & $ 4.85\pm 0.43$ & $ 0.70\pm 0.02$ & $ 1.53\pm 0.25$ & $ 3.62\pm 0.56$ & $ 0.60\pm 0.03$ & $ 24.92\pm 0.06$ & $ 10.08\pm 0.14$ & $ 10.25\pm 0.04 $\\
 433& 338.829346& $ -25.964228$ & $ 10.95\pm 0.07$ & $ 1.915\pm 0.013$ & $ -0.299\pm 0.150$ & $ -0.193\pm 0.266$ & $ 2.76\pm 0.23$ & $ 4.10\pm 0.40$ & $ 0.70\pm 0.02$ & $ 1.72\pm 0.35$ & $ 4.29\pm 0.67$ & $ 0.55\pm 0.03$ & $ 25.47\pm 0.07$ & $ 9.68\pm 0.18$ & $ 10.00\pm 0.04 $\\
 478& 338.853668& $ -25.943596$ & $ 11.16\pm 0.07$ & $ 1.955\pm 0.013$ & $ -0.649\pm 0.099$ & $ -0.339\pm 0.225$ & $ 5.14\pm 0.67$ & $ 5.61\pm 0.69$ & $ 0.78\pm 0.03$ & $ 2.41\pm 0.46$ & $ 5.53\pm 0.75$ & $ 0.73\pm 0.04$ & $ 26.37\pm 0.11$ & $ 9.59\pm 0.16$ & $ 10.06\pm 0.05 $\\
 516& 338.840942& $ -25.952995$ & $ 10.32\pm 0.07$ & $ 1.292\pm 0.011$ & $ -0.686\pm 0.130$ & $ -0.313\pm 0.248$ & $ 2.33\pm 0.22$ & $ 0.94\pm 0.09$ & $ 0.77\pm 0.03$ & $ 1.89\pm 0.43$ & $ 1.74\pm 0.26$ & $ 0.69\pm 0.04$ & $ 25.69\pm 0.08$ & $ 8.97\pm 0.20$ & $ 9.35\pm 0.05 $\\    
 534& 338.840820& $ -25.953827$ & $ 10.26\pm 0.07$ & $ 1.527\pm 0.021$ & $ -1.536\pm 0.203$ & $ -0.924\pm 0.419$ & $ 2.51\pm 0.30$ & $ 1.08\pm 0.13$ & $ 0.84\pm 0.04$ & $ 1.25\pm 0.24$ & $ 1.67\pm 0.25$ & $ 0.64\pm 0.04$ & $ 26.37\pm 0.10$ & $ 9.27\pm 0.16$ & $ 9.39\pm 0.04 $\\
 538& 338.831543& $ -25.945869$ & $ 10.49\pm 0.06$ & $ 1.649\pm 0.018$ & $ -0.846\pm 0.296$ & $ -0.643\pm 0.566$ & $ 1.65\pm 0.14$ & $ 5.37\pm 0.63$ & $ 0.82\pm 0.03$ & $-$ & $-$ & $-$ & $ 25.10\pm 0.07$ & $-$ & $-$\\
 552& 338.838593& $ -25.953201$ & $ 10.46\pm 0.05$ & $ 1.426\pm 0.026$ & $ -0.777\pm 0.092$ & $ -0.464\pm 0.217$ & $ 6.69\pm 1.80$ & $ 0.51\pm 0.14$ & $ 0.80\pm 0.09$ & $ 5.21\pm 1.59$ & $ 0.84\pm 0.28$ & $ 0.69\pm 0.11$ & $ 27.84\pm 0.23$ & $ 8.23\pm 0.26$ & $ 8.52\pm 0.16 $\\
 558& 338.839447& $ -25.949474$ & $ 11.10\pm 0.08$ & $ 1.753\pm 0.009$ & $ 0.562\pm 0.133$ & $ 0.346\pm 0.306$ & $ 4.93\pm 0.49$ & $ 4.77\pm 0.48$ & $ 0.73\pm 0.03$ & $-$ & $-$ & $-$ & $ 26.11\pm 0.09$ & $-$ & $-$\\
 562& 338.859161& $ -25.945955$ & $ 11.32\pm 0.07$ & $ 1.903\pm 0.008$ & $ 0.231\pm 0.116$ & $ -0.040\pm 0.271$ & $ 3.66\pm 0.32$ & $ 5.90\pm 0.57$ & $ 0.92\pm 0.03$ & $ 2.91\pm 0.60$ & $ 9.04\pm 1.41$ & $ 0.97\pm 0.05$ & $ 25.14\pm 0.08$ & $ 9.59\pm 0.18$ & $ 10.33\pm 0.07 $\\
 571& 338.857452& $ -25.946079$ & $ 10.35\pm 0.05$ & $ 1.480\pm 0.016$ & $ 0.194\pm 0.518$ & $ 0.097\pm 0.688$ & $ 1.16\pm 0.08$ & $ 6.44\pm 0.84$ & $ 0.68\pm 0.03$ & $-$ & $-$ & $-$ & $ 24.42\pm 0.06$ & $-$ & $-$\\
 576& 338.841461& $ -25.949100$ & $ 11.01\pm 0.08$ & $ 1.803\pm 0.008$ & $ -0.417\pm 0.320$ & $ -0.252\pm 0.447$ & $ 2.18\pm 0.11$ & $ 2.97\pm 0.24$ & $ 0.36\pm 0.01$ & $ 1.47\pm 0.21$ & $ 4.13\pm 0.57$ & $ 0.21\pm 0.01$ & $ 24.62\pm 0.04$ & $ 9.88\pm 0.13$ & $ 10.20\pm 0.03 $\\
 585& 338.856934& $ -25.949547$ & $ 10.49\pm 0.08$ & $ 1.896\pm 0.023$ & $ -0.494\pm 0.483$ & $ -0.281\pm 0.762$ & $ 1.05\pm 0.06$ & $ 2.63\pm 0.28$ & $ 0.58\pm 0.03$ & $-$ & $-$ & $-$ & $ 24.50\pm 0.05$ & $-$ & $-$\\
 588& 338.830658& $ -25.948841$ & $ 10.81\pm 0.08$ & $ 1.764\pm 0.015$ & $ -0.605\pm 0.035$ & $ -0.384\pm 0.084$ & $ 5.42\pm 0.91$ & $ 1.98\pm 0.31$ & $ 0.35\pm 0.02$ & $ 3.65\pm 0.86$ & $ 2.38\pm 0.47$ & $ 0.27\pm 0.03$ & $ 27.04\pm 0.15$ & $ 8.89\pm 0.21$ & $ 9.65\pm 0.07 $\\
 599& 338.856018& $ -25.947937$ & $ 11.40\pm 0.07$ & $ 1.915\pm 0.008$ & $ -0.451\pm 0.076$ & $ -0.274\pm 0.148$ & $ 7.05\pm 0.86$ & $ 4.84\pm 0.57$ & $ 0.52\pm 0.02$ & $ 3.50\pm 0.66$ & $ 5.69\pm 0.75$ & $ 0.45\pm 0.03$ & $ 26.39\pm 0.11$ & $ 9.51\pm 0.16$ & $ 10.27\pm 0.05 $\\
 611& 338.857452& $ -25.949520$ & $ 10.47\pm 0.07$ & $ 1.276\pm 0.012$ & $ 0.324\pm 0.141$ & $ 0.280\pm 0.325$ & $ 4.70\pm 0.80$ & $ 1.95\pm 0.31$ & $ 0.83\pm 0.05$ & $ 2.77\pm 0.66$ & $ 1.14\pm 0.23$ & $ 0.93\pm 0.09$ & $ 26.84\pm 0.15$ & $ 8.78\pm 0.21$ & $ 9.03\pm 0.10 $\\
 617& 338.858368& $ -25.948902$ & $ 10.70\pm 0.07$ & $ 1.515\pm 0.016$ & $ -0.610\pm 0.094$ & $ -0.455\pm 0.193$ & $ 6.86\pm 1.47$ & $ 2.17\pm 0.44$ & $ 0.61\pm 0.05$ & $ 2.77\pm 0.82$ & $ 4.19\pm 1.23$ & $ 0.78\pm 0.09$ & $ 27.45\pm 0.19$ & $ 9.02\pm 0.26$ & $ 9.46\pm 0.10 $\\
 618& 338.823639& $ -25.948795$ & $ 10.65\pm 0.08$ & $ 1.837\pm 0.023$ & $ -1.025\pm 0.128$ & $ -0.613\pm 0.278$ & $ 3.37\pm 0.42$ & $ 2.17\pm 0.27$ & $ 0.79\pm 0.04$ & $ 1.54\pm 0.29$ & $ 2.50\pm 0.35$ & $ 0.64\pm 0.04$ & $ 26.52\pm 0.11$ & $ 9.48\pm 0.16$ & $ 9.64\pm 0.04 $\\
 637& 338.844788& $ -25.951603$ & $ 10.69\pm 0.05$ & $ 1.486\pm 0.008$ & $ -0.325\pm 0.223$ & $ -0.273\pm 0.403$ & $ 1.46\pm 0.06$ & $ 2.58\pm 0.23$ & $ 0.82\pm 0.02$ & $ 0.70\pm 0.10$ & $ 4.86\pm 0.77$ & $ 0.80\pm 0.03$ & $ 24.07\pm 0.04$ & $ 10.20\pm 0.12$ & $ 9.97\pm 0.03 $\\
 642& 338.842316& $ -25.951626$ & $ 10.55\pm 0.08$ & $ 1.785\pm 0.019$ & $ -0.341\pm 0.289$ & $ -0.149\pm 0.445$ & $ 1.71\pm 0.14$ & $ 1.96\pm 0.20$ & $ 0.38\pm 0.02$ & $ 1.25\pm 0.25$ & $ 1.45\pm 0.23$ & $ 0.34\pm 0.02$ & $ 25.22\pm 0.07$ & $ 9.56\pm 0.18$ & $ 9.74\pm 0.03 $\\ 
  \hline
  \end{tabular}
\end{table}

\begin{minipage}[t][0pt][t]{20.5cm}
Notes:\\
\textsuperscript{a} Total stellar masses are estimated using the $M_{*}/L$-colour relation, the $z_{850}-H_{160}$ aperture colours and the total luminosity $L_{H_{160}}$ from the best-fit S\'ersic models.\\
\textsuperscript{b}  $1''$ aperture magnitudes from SExtractor.\\
\textsuperscript{c}  Colour gradients $\nabla_{z_{850}-H_{160}}$ and $M_{*}/L$ gradients $\nabla_{\log(M/L)}$ are defined as $d(z_{850} - H_{160}) / d \log(a)$ and $d(\log(M/L)) / d \log(a)$ respectively.  The gradient is fitted in the radial range of PSF HWHM $< a < 3.5~a_{e}$. \\
\textsuperscript{d} Mean surface brightness $\Sigma$ and mean surface mass density $\Sigma_{mass}$ are defined as mag $+ 2.5 \log(2\pi a_e^{2} )$ and $\log(M_{*}/2\pi a_{e,mass}^{2})$. \\
\textsuperscript{e} Mean surface mass density within a radius of 1 kpc $\Sigma_{1}$ is defined as $\log(M_{*} $(<1kpc)$ /\pi $(1kpc)$^{2})$, derived from integrating the fitted 2D mass profiles. \\
\end{minipage}
\end{center}
\end{landscape}


\newpage
%

\label{lastpage}
\end{document}